\documentclass[11pt,a4paper]{article}
\usepackage{jheppub}
\usepackage{epsf}
\usepackage{amsmath}
\usepackage{amssymb}
\usepackage{amsfonts}
\usepackage{graphicx}
\usepackage{graphics}
\usepackage{epstopdf}
\usepackage{url}

\newcommand{\beq}{\begin{equation}}
\newcommand{\eeq}{\end{equation}} 
\newcommand{\beqa}{\begin{eqnarray}} 
\newcommand{\eeqa}{\end{eqnarray}}

\newcommand{\ycrit}{y^{\rm crit}}

\def\bi{\begin{itemize}}
\def\ei{\end{itemize}}
\def\be{\begin{equation}}
 \def\ee{\end{equation}}
\def\ben{\begin{equation*}}
 \def\een{\end{equation*}}
 \def\bea{\begin{eqnarray}}
 \def\eea{\end{eqnarray}}
 \def\bean{\begin{eqnarray*}}
 \def\eean{\end{eqnarray*}}
\newcommand{\ie}{{\it i.e.}}  \newcommand{\eg}{{\it e.g.}}

\def\xtwo{x_{_2}}

\newcommand{\morder}[1]{{\cal O}\left(#1 \right)}
\newcommand{\eq}[1]{(\ref{#1})}
\newcommand{\ed}{\end{document}}

\newcommand{\com}[2]{\left[{#1},{#2}\right]}

\newcommand{\Ep}{E_\mathrm{p}}
\newcommand{\mprot}{m_\mathrm{p}}
\newcommand{\mT}{M_\perp}
\newcommand{\thard}{t_{\mathrm{hard}}}
\newcommand{\tauoctet}{\tau_{\mathrm{octet}}}
\newcommand{\jpsi}{{\mathrm J}/\psi}
\newcommand{\xf}{x_{\mathrm{F}}}
\newcommand{\xfcrit}{\xf^{\rm crit}}
\newcommand{\pt}{p_{_\perp}}

\newcommand{\pp}{p--p}
\newcommand{\pA}{p--A}

\newcommand{\piA}{\ensuremath{\pi}--A}
\newcommand{\pN}{p--N}
\newcommand{\hi}{A--A}
\newcommand{\dd}{{\rm d}}
\newcommand{\half}{\frac{1}{2}}  
\newcommand{\lsim}{\lesssim} \newcommand{\gsim}{\gtrsim}

\newcommand{\qzero}{\hat{q}_0}

\newcommand{\gevsqfm}{GeV$^2$/fm}
\newcommand{\dsigpp}{\dd\sigma_{\rm pp}^\psi/\dd\xf}

\newcommand{\ndf}{{\rm ndf}}
\newcommand{\Leff}{L_{\rm eff}}

 \def\esim{\,\mathrel{\rlap{\lower0.2em\hbox{$-$}}\raise0.15em\hbox{\footnotesize $\hskip0.04em\sim$}}\,}
 \def\gsim{\mathrel{\rlap{\lower0.2em\hbox{$\sim$}}\raise0.2em\hbox{$>$}}}
 \def\ksim{\mathrel{\rlap{\lower0.2em\hbox{$\sim$}}\raise0.2em\hbox{$<$}}}

\def\xf{x_{_F}}

\newcommand{\sqrts}{\sqrt{s}}

\def\pt{p_{_\perp}}

\title{Heavy-quarkonium suppression in \pA\ collisions from parton energy loss in cold QCD matter}

\author[a]{Fran\c{c}ois Arleo}
\author[b]{and St\'ephane Peign\'e}

\affiliation[a]{Laboratoire d'Annecy-le-Vieux de Physique Th\'eorique (LAPTh)\\ UMR5108, Universit\'e de Savoie, CNRS, BP 110, 74941 Annecy-le-Vieux cedex, France}
\affiliation[b]{SUBATECH, UMR 6457, Universit\'e de Nantes, Ecole des
Mines de Nantes, IN2P3/CNRS \\ 4 rue Alfred Kastler, 44307 Nantes cedex 3, France}

\emailAdd{arleo@lapp.in2p3.fr}
\emailAdd{peigne@subatech.in2p3.fr}

\abstract{
The effects of parton energy loss in cold nuclear matter on heavy-quarkonium suppression in \pA\ collisions are studied. It is shown from first principles that at large quarkonium energy $E$ and small production angle in the nucleus rest frame, the medium-induced energy loss scales as $E$. Using this result, a phenomenological model depending on a single free parameter is able to reproduce $\jpsi$ and $\Upsilon$ suppression data in a broad $\xf$-range and at various center-of-mass energies. These results strongly support energy loss as the dominant effect in heavy-quarkonium suppression in \pA\ collisions. Predictions for $\jpsi$ and $\Upsilon$ suppression in p--Pb collisions at the LHC are made. It is argued that parton energy loss scaling as $E$ should generally apply to hadron production in \pA\ collisions, such as light hadron or open charm production.}


\begin{document} 

\maketitle
\setcounter{footnote}{0}
\renewcommand{\thefootnote}{\arabic{footnote}} 	

\section{Introduction}

The spectacular quenching of hadrons produced at large $\pt$ in Pb--Pb collisions at the LHC~\cite{Aamodt:2010jd,CMS:2012aa}, as well as the jet imbalance reported in those collisions~\cite{Collaboration:2010bu,Chatrchyan:2012ni}, find a natural explanation in terms of parton energy loss in a quark-gluon plasma (QGP). For light hadron production at mid-rapidity and sufficiently large $\pt$, the parton energy loss is dominantly radiative, in average of the form $\Delta E \sim \alpha_s \, \hat{q}_{\rm hot} \, L^2$ \cite{Baier:1996kr,Zakharov:1997uu}, with $L$ the distance travelled by the parton through the hot medium and $\hat{q}_{\rm hot}$ the rate per unit length of transverse momentum broadening in the medium. The strength of jet-quenching can be explained if the 
transport coefficient $\hat{q}$ in a QGP is larger, by one to two orders of magnitude, than its estimate 
in cold nuclear matter, $\hat{q}_{\rm cold} \sim 0.045 \, {\rm GeV}^2/{\rm fm}$~\cite{Baier:2001yt}.
This is why jet-quenching is considered as a prominent QGP signal. However, despite the wealth of data accumulated so far at RHIC and LHC, the in-depth understanding of energy loss processes in a QGP remains far from complete (see~\cite{Armesto:2011ht} for a discussion).

Drastic nuclear suppression effects are not only seen in \hi\ but also in \pA\ collisions, at least for some processes and in some kinematical conditions. For instance, quarkonium~\cite{Leitch:1999ea} but also light hadron~\cite{Arsene:2004ux,Adler:2004eh} production at large longitudinal momentum fraction $\xf$ (or large rapidity) is strongly suppressed in \pA\ as compared to \pp\ collisions. Understanding nuclear suppression in cold nuclear matter, a well-controlled medium as opposed to an 
expanding QGP, should be a prerequisite in order to interpret quantitatively nuclear suppression in heavy-ion collisions. However, it is striking that there is no consensus yet on the origin of $\jpsi$ suppression at large rapidity/$\xf$ in \pA\ collisions, from SPS to RHIC~\cite{Badier:1983dg,Leitch:1999ea,Adare:2010fn}, despite many theoretical attempts~(see~\cite{Frawley:2008kk} for a review). 

Recently, new scaling properties have been identified for the induced gluon radiation spectrum $\dd I/\dd \omega$, and associated energy loss $\Delta E$, of hard processes where a color charge undergoes small angle scattering through a static medium (cold matter or QGP) \cite{Arleo:2010rb}. In the present work we address the phenomenological consequences of these results on $\jpsi$ and $\Upsilon$ nuclear suppression in \pA\ and \piA\ collisions, parametrized by the attenuation factor (in the following we use the generic notations ``$\psi$'' and ``\pA'')
\be
\label{RpA}
R_{\mathrm{pA}}^{\psi}\left(\xf, \sqrt{s} \right) = \frac{1}{A} \, {\frac{\dd\sigma_{\mathrm{pA}}^{\psi}}{\dd \xf} \left(\xf, \sqrt{s}\right) \biggr/ \frac{\dd\sigma_{\mathrm{pp}}^{\psi}}{\dd \xf} \left(\xf, \sqrt{s} \right)} \, .
\ee

\begin{figure}[t]
\centering
\includegraphics[scale=0.35]{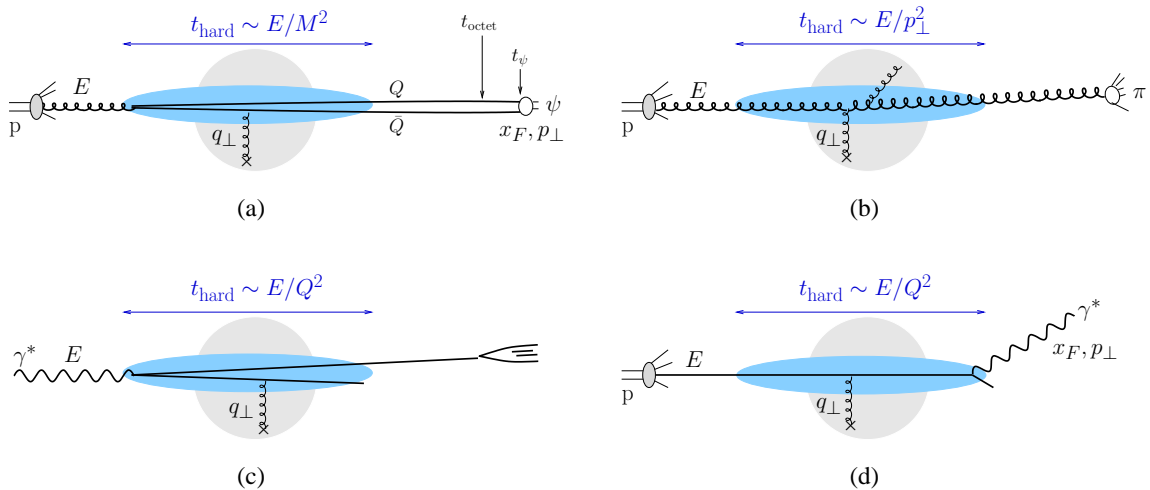}
\caption{Generic processes of (a) heavy-quarkonium hadroproduction (b) light hadron hadroproduction (c) deep inelastic scattering and (d) Drell-Yan production, at large $E$ in the target nucleus rest frame. The ellipse represents the hard subprocess occurring within the time $\thard$. Cases (a) and (b) are similar to small angle scattering of an asymptotic charge.}
\label{fig:generic-processes}
\end{figure}

We will show that the available large-$\xf$ quarkonium suppression data in \pA\ collisions can be explained by parton energy loss in cold matter. Although $\hat{q}$ in cold matter is small, a strong nuclear attenuation arises due to the specific parametric behaviour $\Delta E \propto E$ at sufficiently large $E$, where $E$ is the quarkonium energy in the target nucleus rest frame. As discussed in Ref.~\cite{Arleo:2010rb} and reviewed in the present paper (Section~\ref{sec2}), this behaviour holds when the 
hard 
partonic subprocess can be viewed, in the nucleus rest frame, as the small angle scattering of a color charge. In the following we focus on quarkonium hadroproduction (see Fig.~\ref{fig:generic-processes}a), where the heavy quark mass provides the hard scale allowing for a perturbative QCD description, and for which p--A suppression data are quite abundant. Our discussion should however apply more generally to hadron hadroproduction, for instance to light hadron production in \pA\ collisions (provided the light hadron $p_\perp$ plays the role of the hard scale, \ie, $p_\perp \gsim 1 \, {\rm GeV}$), see Fig.~\ref{fig:generic-processes}b. Light hadron nuclear suppression due to parton energy loss will be addressed in a future work. 

As is well-known, the quarkonium hadroproduction mechanism in elementary \pp\ collisions 
is still under debate\footnote{This is especially true for low $p_\perp$ production, $p_\perp \lsim M$, considered in the present study. The recent findings of Refs.~\cite{Kang:2011mg,Nayak:2005rt} apply specifically to the large $p_\perp \gg M$ domain.} (see for instance Ref.~\cite{Lansberg:2006dh} for a review). In order to study quarkonium nuclear suppression in the most model-independent way, in the present study we will only 
assume that the heavy-quark $Q \bar{Q}$ pair of mass $M$ is produced, within the perturbative proper time scale $\tau_{Q \bar{Q}} \sim 1/M$, in a compact {\it color octet} state, and remains color octet for a time $\tauoctet \gg \tau_{Q \bar{Q}}$. In quarkonium production models where color neutralization is a soft, non-perturbative process, $\tauoctet$ coincides with the quarkonium hadronization time $\tau_{\psi}$, and this assumption holds at any $\xf$. In the Color Singlet Model (CSM), the gluon emission required for color neutralization of the $Q \bar{Q}$ pair is constrained at large enough $\xf$ to become softish and thus to occur late, leading to $\tau_{Q \bar{Q}} \ll \tauoctet \lsim \tau_{\psi}$. Thus, the assumption of a color octet $Q \bar{Q}$ pair living longer than the perturbative time scale $\sim 1/M$ holds quite independently of the quarkonium production model.\footnote{Strictly speaking, the assumption is independent of the quarkonium production model only at large $\xf$. At small $\xf$, the assumption becomes invalid in the CSM. We will briefly come back to this point in our final discussion.}

Working in the nucleus rest frame and considering the limit $E \gg M \gsim p_\perp$, quarkonium hadroproduction looks like small angle scattering of an ``asymptotic'' color charge, \ie, prepared in the ``far past'' and propagating in the ``far future'' as compared to the perturbative time scale $\thard = \tau_{Q \bar{Q}} \cdot (E/M) \sim E/M^2$. This is illustrated in Fig.~\ref{fig:generic-processes}a for the generic $gg \to Q \bar{Q}$ partonic subprocess, viewed in the nucleus rest frame as the splitting $g \to Q \bar{Q}$ of the incoming gluon, followed by a rescattering in the target.\footnote{If the gluon $q_\perp$ in Fig.~\ref{fig:generic-processes}a is relatively large, $q_\perp \sim \morder{M}$, then it should be interpreted as part of the hard subprocess $gg \to Q \bar{Q} + g$ (or $gq \to Q \bar{Q} + q$).} 

In the present study we will assume for simplicity that the octet $Q \bar{Q}$ pair arises dominantly from the splitting of an incoming {\it gluon}. This should be a valid assumption for all \pA\ data considered in this paper, except at {\it very} large values of $\xf$ ($\xf \gsim 0.8$), where quark-induced processes come into play. (We will further comment on this point in Section \ref{sec:procedure}.)

Within those assumptions,
the associated gluon radiation with large formation time, $t_{\mathrm{f}} \gg \thard$, is similar to (non-abelian) Bethe-Heitler radiation off a fast color octet 
undergoing an effective transverse momentum kick $q_\perp$. The typical $q_\perp$ is expected to be larger in \pA\ than in \pp\ collisions due to transverse momentum broadening 
being proportional to the target size $L$,  
$\Delta q_\perp^2 \equiv \ell_\perp^2 \simeq \hat{q}\,L$.\footnote{In our study the broadening $\Delta q_\perp^2$ equals the amount of soft rescattering $\ell_\perp^2$. In other words it is defined with respect to an ``ideal'' target where soft rescatterings are absent, see Section \ref{sec:resolved-charge}.}  
As a result, the {\it medium-induced} radiation spectrum is similar to the Bethe-Heitler spectrum \eq{asym-spec}, 
up to the replacement of the total momentum transfer
by the broadening $\ell_{\perp {\rm A}}^2$ through the nucleus A 
(see \eq{our-spectrum-3} and Section \ref{sec2}),
\be
\label{our-spectrum}
\omega \frac{\dd I}{\dd \omega} = \frac{N_c \alpha_s}{\pi} \left\{ \ln{\left(1+\frac{\ell_{\perp {\rm A}}^2 E^2}{M_\perp^2 \omega^2}\right)} - \ln{\left(1+\frac{\Lambda_{\rm p}^2 E^2}{M_\perp^2 \omega^2}\right)} \right\} \, \Theta(\ell_{\perp {\rm A}}^2 - \Lambda_{\rm p}^2) \, ,
\ee
with $M_\perp=(M^2+\pt^2)^{\half}$ the transverse mass of the $Q \bar{Q}$ pair and $\Lambda_{\mathrm p}^2 = {\rm max}(\Lambda_{_\mathrm{QCD}}^2,\ell_{\perp {\rm p}}^2)$. This leads to an average loss $\Delta E \propto E$. When $\Lambda_{\mathrm p}^2 < \ell_{\perp {\rm A}}^2 \ll M_\perp^2$ we have
\be
\Delta E \equiv \int_0^E \dd\omega \,\omega \frac{\dd I}{\dd \omega} \simeq  N_c \, \alpha_s \, \frac{\sqrt{\ell_{\perp {\rm A}}^2} - \Lambda_{\mathrm p}}{M_\perp} \, E \, .
\label{mean-delta-E}
\ee

The spectrum \eq{our-spectrum} is at the basis of the phenomenological study presented here, whose main results can already be found in Ref.~\cite{Arleo:2012hn}. The scaling $\Delta E \propto E$ in quarkonium hadroproduction was first postulated in~\cite{Gavin:1991qk} (also revisited in~\cite{Kopeliovich:2005ym}) yet this assumption was not motivated and the parametric dependence on $L$ and $M$ arbitrary (and different from \eq{mean-delta-E}). In Ref.~\cite{Brodsky:1992nq}, an {\it energy-independent} bound on $\Delta E$ was derived, but in a specific setup where the nuclear broadening of the final tagged particle was neglected (we will briefly comment on this point in Section \ref{sec2}).

We stress that the spectrum \eq{our-spectrum} is 
{\it coherent}.
This can easily be seen in a calculation using physical polarizations for the radiated gluon (see Section \ref{sec2}). With this choice the 
medium-induced
radiation spectrum indeed arises from the interference between the initial and final state emission amplitudes. The 
energy loss \eq{mean-delta-E} is thus neither a purely initial nor final state effect, and is distinct from gluon radiation resummed in leading-twist parton distribution and fragmentation functions. Being process-dependent 
(\eg, it is expected in $\jpsi$ production in \pA\ collisions but {\it not} in Drell-Yan production, see below) 
and suppressed by a power of the hard scale $M_\perp$, it is naturally interpreted as a higher-twist effect. Nevertheless it plays a crucial role in a broad $\xf$ or rapidity interval, as we shall see.

To make the physics under consideration clear, let us mention that the spectrum \eq{our-spectrum} is not expected in quarkonium (real or virtual) photoproduction, nor in inclusive deep inelastic scattering (DIS) off nuclei, where the incoming energetic particle participating to the hard subprocess is colorless (see Fig.~\ref{fig:generic-processes}c for the DIS case).\footnote{This is to be distinguished from {\it resolved} photoproduction, which at the partonic level is similar to hadroproduction (Fig.~\ref{fig:generic-processes}a), and where we thus expect $\Delta E \propto E$.} In those cases radiation with $t_{\mathrm{f}} \gg \thard$ can only arise from final state radiation. The latter (DGLAP-like) radiation is independent of the medium properties and cancels in the medium-induced spectrum. Quite remarkably, at large $z$ the 
quarkonium production data in deep inelastic muon scattering~\cite{Amaudruz:1991sr} exhibit no nuclear suppression (but instead a slight enhancement), in sharp contrast to hadroproduction. 

Drell-Yan (DY) production off nuclei is similar to quarkonium photoproduction, since by definition the energetic particle produced perturbatively (the Drell-Yan photon of mass $Q$) is colorless, see Fig.~\ref{fig:generic-processes}d. Radiation with $t_{\mathrm{f}} \gg \thard \sim E/Q^2$ must arise from initial state and does not contribute to the medium-induced spectrum. Thus, neither the spectrum \eq{our-spectrum}, nor the energy loss $\Delta E \propto E$ of the type \eq{mean-delta-E}, is expected in DY production. This is qualitatively consistent with the 
much milder nuclear suppression of DY production~\cite{Vasilev:1999fa} when compared to $\jpsi$ hadroproduction in the same kinematical range. 

In Section~\ref{sec2} we discuss the medium-induced gluon radiation spectrum, derived in~\cite{Arleo:2010rb}, which is at the basis of the model detailed in Section~\ref{sec:model}. Phenomenological applications are presented in Section~\ref{sec:phenomenology} and we conclude the paper by a discussion (Section~\ref{sec:discussion}).

\section{Revisiting energy loss scaling properties}
\label{sec2}

In this section we justify the expression \eq{our-spectrum} for the gluon radiation spectrum associated to quarkonium hadroproduction, in a less heuristic way than in Ref.~\cite{Arleo:2010rb}. 

\subsection{Asymptotic charge}
\label{sec:asym-charge}

\begin{figure}[h]
\centering
\includegraphics[scale=0.40]{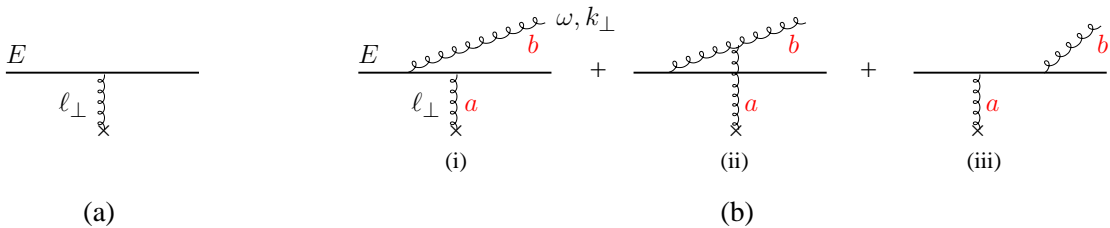}
\caption{(a) Elastic scattering amplitude ${\cal M}_{el}$ of an asymptotic light quark via single gluon exchange. (b) Induced gluon radiation amplitude ${\cal M}_{rad}$.}
\label{fig:GBamplitude}
\end{figure}

We first consider the case of an on-shell (``asymptotic") parton of energy $E$ undergoing an elastic scattering and exchanging a gluon with transverse momentum $\vec{\ell}_\perp$ with a nuclear target, see Fig.~\ref{fig:GBamplitude}a. As is well-known, scattering can induce radiation, provided the quantum state of the charge is perturbed. In QED, this happens when the scattering angle $\theta_s \simeq \ell_\perp/E$ is non-zero. In QCD, radiation can occur even in the limit $\theta_s \to 0$, due to the incoming  parton {\it color rotation} in the elastic scattering. 

These features are illustrated by the expression of the gluon radiation amplitude induced by the elastic scattering, given by the diagrams of Fig.~\ref{fig:GBamplitude}b. We denote by $\omega$ and $\vec{k}_{\perp}$ the radiated gluon energy and transverse momentum. We focus on soft ($\omega \ll E$) and small angle 
($k_{\perp} \ll \omega$) gluon radiation. For an on-shell light quark the radiation amplitude reads~\cite{Gunion:1981qs}
\bea
\label{GB}
\frac{{\cal M}_{rad}}{{\cal M}_{el}} \propto \left[ T^a T^b \frac{\vec{\theta}}{\theta^2} + \com{T^b}{T^a} \frac{\vec{\theta}''}{\theta''^{\,2}} - T^b T^a \frac{\vec{\theta}'}{\theta'^{\,2}} \right] \cdot \vec{\varepsilon}_\perp \, ,  \hskip 2cm \\
{\rm where} \ \ \  \vec{\theta} \equiv \frac{\vec{k}_{\perp}}{\omega} \ \ ; \ \ \vec{\theta}' = \vec{\theta} - \vec{\theta}_s \ \ ; \ \ \vec{\theta}''= \vec{\theta} - \vec{\theta}_g \ \ ; \ \ \vec{\theta}_s \equiv \frac{\vec{\ell}_\perp}{E} \ \ ;\ \ \vec{\theta}_g \equiv \frac{\vec{\ell}_\perp}{\omega} \, , &&
\eea
with $\vec{\varepsilon}$ the physical polarization vector of the radiated gluon, which will be implicit in the following. The first two terms of \eq{GB} correspond to initial state radiation (diagrams (i) and (ii) of Fig.~\ref{fig:GBamplitude}b), and the last (diagram (iii) of Fig.~\ref{fig:GBamplitude}b) to final state radiation. 

In the abelian case, the second diagram is absent, as well as color factors, and the radiation spectrum reads
\be
\omega \left. \frac{\dd{I}}{\dd\omega} \right|_{\rm QED} = \frac{\alpha}{\pi^2} \int \dd^2\vec{\theta} \,\left( \frac{\vec{\theta}}{\theta^2} - \frac{\vec{\theta}'}{\theta'^2}  \right)^2 = \frac{\alpha}{\pi^2} \int \dd^2\vec{\theta} \, \frac{\theta_s^2}{\theta^2 \theta'^2} =   \frac{2\alpha}{\pi} \int^{\theta_s^2} \frac{\dd \theta^2}{\theta^2}\, .
\ee 
The collinear divergence at $\theta \to 0$ is screened by the electron mass $m$. Keeping the latter, the radiation spectrum can be expressed as\footnote{Although the mass dependence of the QED spectrum \eq{qed-spectrum-asym} is not exact, it provides the correct parametric expressions in the two limits $\ell_\perp \gg m$ and $\ell_\perp \ll m$~\cite{Peigne:2008wu,Arleo:2010rb}. This remark also applies to the QCD expression \eq{qcd-spectrum-asym}.}
\be 
\label{qed-spectrum-asym}
\omega \left. \frac{\dd{I}}{\dd\omega} \right|_{\rm QED} =   \frac{2\alpha}{\pi} \int_0^{\theta_s^2} \frac{\dd \theta^2}{\theta^2 + \theta_m^2} = \frac{2\alpha}{\pi} \ln{\left( 1 +\frac{\theta_s^2}{\theta_m^2}  \right)} = \frac{2\alpha}{\pi} \ln{\left( 1 +\frac{\ell_\perp^2}{m^2}  \right)} \, ,
\ee
where $\theta_m \equiv m/E$. The radiation spectrum vanishes when $\ell_\perp = 0$, but also in the formal limit $\theta_s \to 0$ (at fixed $\theta_m$), as expected.

In QCD we single out the purely non-abelian contribution to the radiation spectrum off a light quark by focussing on the $\theta_s \to 0$ limit of Eq.~\eq{GB}, 
\be
\label{GB-qcd}
\frac{{\cal M}_{rad}}{{\cal M}_{el}} \propto \com{T^a}{T^b}  \left[ \frac{\vec{\theta}}{\theta^2} - \frac{\vec{\theta}''}{\theta''^{\,2}} \right]  \propto \com{T^a}{T^b}  \left[ \frac{\vec{k}_\perp}{\vec{k}_\perp^{\, 2}} - \frac{\vec{k}_\perp-\vec{\ell}_\perp}{( \vec{k}_\perp-\vec{\ell}_\perp )^2} \right]  \, . 
\ee
Squaring the r.h.s. of \eq{GB-qcd}, we recover the well-known Gunion-Bertsch spectrum~\cite{Gunion:1981qs},
\be
\omega \left. \frac{\dd{I}}{\dd\omega \, \dd^2 \vec{k}_\perp} \right. \ \sim \ \alpha_s \, \frac{\ell_\perp^2}{k_\perp^2 (\vec{k}_\perp-\vec{\ell}_\perp )^2} \, .
\ee
Integrating over $d^2\vec{k}_\perp$, or equivalently over $d^2\vec{\theta}$, and inserting the quark mass $M$ dependence as in the abelian case above, 
we obtain\footnote{The color factor in \eq{qcd-spectrum-asym} is obtained by summing $|{\cal M}_{rad}|^2$ over initial and final color indices, and normalizing by the color factor associated to $|{\cal M}_{el}|^2$. The resulting factor $N_c$ is independent of the type (quark or gluon) of the energetic color charge.}
\be
\label{qcd-spectrum-asym}
\omega \left. \frac{\dd{I}}{\dd\omega} \right|_{\rm QCD} =  \frac{N_c \, \alpha_s}{\pi} \int_{\frac{\Lambda^2}{\omega^2}}^{\theta_g^2} \frac{\dd \theta^2}{\theta^2 + \theta_M^2}  \, ,
\ee
where $\theta_M \equiv M/E$ and the lower cutoff arises from the constraint $k_\perp > \Lambda$ (with $\Lambda \equiv \Lambda_{_\mathrm{QCD}}$), 
put by hand for the consistency of our perturbative QCD treatment (we also assumed $\ell_\perp > \Lambda$). We obtain the soft radiation spectrum off an on-shell quark of mass $M$~\cite{Peigne:2008wu,Arleo:2010rb},
\be
\label{asym-spec}
\omega \frac{\dd{I}}{\dd\omega} = \frac{N_c \, \alpha_s}{\pi} \left\{ \ln{\left(1+\frac{\ell_\perp^2 E^2}{M^2 \omega^2}\right)} - \ln{\left(1+\frac{\Lambda^2 E^2}{M^2 \omega^2}\right)} \right\} \, . 
\ee
The average energy loss of a heavy quark, $M \gg \ell_\perp $, is dominated by $\omega \sim (\ell_\perp /M)\,E \ll E$ and is proportional to $E$,
\be
\label{deltaE-asym}
\Delta E \equiv \int_0^E \dd\omega \ \omega \frac{\dd{I}}{\dd\omega} \  \mathop{\simeq}_{M \gg \ell_\perp} \  N_c \, \alpha_s \, \frac{\ell_\perp - \Lambda}{M} \, E \, .
\ee

\subsection{Color charge resolved in a hard process}
\label{sec:resolved-charge}

\begin{figure}[h]
\centering
\includegraphics[scale=0.40]{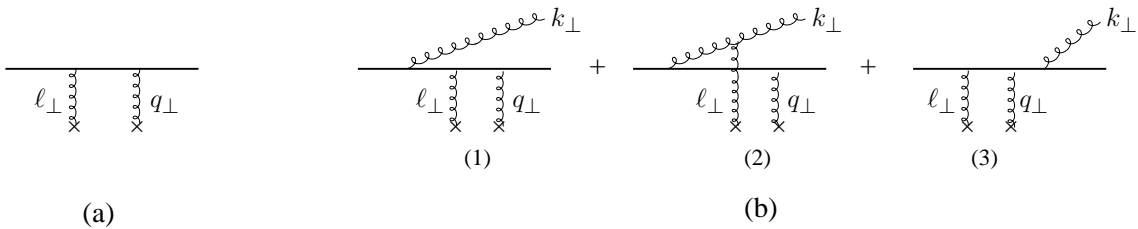}
\caption{(a) Model for hard scattering $q_\perp$ supplemented by transverse momentum broadening $\ell_\perp \ll q_\perp$. (b) Associated gluon radiation amplitude in the large formation time and soft gluon limit.}
\label{fig:GBamplitude-rescatt}
\end{figure}

The case of a color charge resolved in a hard process can be simply illustrated by inserting a hard exchange $q_\perp \gg \ell_\perp$ in the light quark scattering process, see Fig.~\ref{fig:GBamplitude-rescatt}a. The transfer $\ell_\perp$ now plays the role of nuclear momentum broadening, $\ell_\perp^2 \equiv \Delta q_\perp^2$. Since some radiation is released even when $\ell_\perp = 0$ 
due to the presence of the hard exchange, the relevant quantity is the {\it medium-induced} radiation spectrum,
\be
\label{induced-def}
\left. \omega \frac{\dd{I}}{\dd\omega} \right|_{\mathrm{ind}} 
\equiv  \omega \frac{\dd{I}}{\dd\omega} (q_\perp; \, \ell_\perp)- \omega \frac{\dd{I}}{\dd\omega} (q_\perp; \, \ell_\perp =0)  \, .
\ee

\begin{figure}[h]
\centering
\includegraphics[scale=0.43]{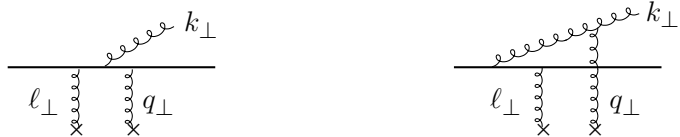}
\caption{Diagrams for the radiation amplitude which are suppressed at large formation time and in the approximation $k_\perp, \ell_\perp \ll q_\perp$.}
\label{fig:negligible}
\end{figure}

We concentrate on soft radiation as compared to the hard process, $\omega \ll E$, $k_\perp \ll q_\perp$, and on the limit of large gluon formation time. It can indeed be checked that the domain $t_{\mathrm{f}} \gg {\rm max}(L,\thard)$ contributes most to the medium-induced loss (see Section~\ref{sec:appl-hadro}). The radiation amplitude is thus dominated by the diagrams of Fig.~\ref{fig:GBamplitude-rescatt}b. Diagrams of the type shown in Fig.~\ref{fig:negligible} can be neglected. The diagram of Fig.~\ref{fig:negligible}a, relevant when $t_f  \lsim L$, contributes to the energy loss as $\Delta E \propto L^2$,  which is negligible compared to the contribution $\Delta E \propto E$ we focus on throughout our study. The diagram of Fig.~\ref{fig:negligible}b is suppressed due to $\ell_\perp \ll q_\perp$.

Note that in the limit $t_{\mathrm{f}} \gg L$, the induced radiation cannot probe the nuclear size $L$, but is however sensitive to the {\it total} amount of soft rescattering, $\ell_\perp^2 = \hat{q}L$, transferred to the fast color charge, independently of the actual number of soft rescatterings. Modeling soft rescatterings by a single {\it effective} 
(semi-hard) 
scattering with $\ell_\perp^2 = \hat{q}L$ should thus allow addressing the main features of medium-induced radiation.

The radiation amplitude is similar to the amplitude \eq{GB} in absence of hard scattering, however with modified color factors and scattering angle $\vec{\theta}_s$,
\bea
\label{GB-rescatt}
\frac{{\cal M}_{rad}}{{\cal M}_{el}} \propto  C_1 \frac{\vec{\theta}}{\theta^2} + C_2 \frac{\vec{\theta}''}{\theta''^{\,2}} - C_3 \frac{\vec{\theta}'}{\theta'^{\,2}}  \, ,  \hskip 2cm \\
\vec{\theta}' = \vec{\theta} - \vec{\theta}_s \ \ ; \ \ \vec{\theta}''= \vec{\theta} - \vec{\theta}_g \ \ ; \ \ \vec{\theta}_s \equiv \frac{\vec{\ell}_\perp + \vec{q}_\perp}{E} \simeq \frac{\vec{q}_\perp}{E} \ \ ;\ \ \vec{\theta}_g \equiv \frac{\vec{\ell}_\perp}{\omega} \, . &&
\eea

In the abelian case, the second term of \eq{GB-rescatt} is absent. Neglecting $\ell_\perp$ compared to $q_\perp$ in the definition of $\vec{\theta}_s$, the radiation amplitude becomes independent of $\ell_\perp$, and the induced spectrum \eq{induced-def} vanishes. As argued by Brodsky and Hoyer~\cite{Brodsky:1992nq}, this means that only radiation with small formation time $t_{\mathrm{f}} \lsim L$ can contribute, resulting in some ($E$-independent) bound on medium-induced energy loss. This statement holds in an abelian model (such as that considered in Ref.~\cite{Brodsky:1992nq}), but fails in QCD, as we shortly recall now.\footnote{In the following we maintain the approximation $\left. \theta_s \right|_{\rm pA} \simeq \left. \theta_s \right|_{\rm pp} = q_\perp/E$, which allows to extract the purely non-abelian medium-induced spectrum \eq{our-spectrum}. This approximation corresponds to an experimental setup where the \pp\  and \pA\ $\xf$ distributions are measured in the same $p_\perp$-bin. If the $\xf$ distributions are averaged over $p_\perp$, we expect $\left. \theta_s \right|_{\rm pA} > \left. \theta_s \right|_{\rm pp}$, leading to some abelian-like radiation arising from $t_{\mathrm{f}} \gg L$. However, the associated loss is suppressed by a power of the hard scale 
as compared to the purely non-abelian loss.} 
 
\begin{figure}[t]
\centering
\includegraphics[scale=0.38]{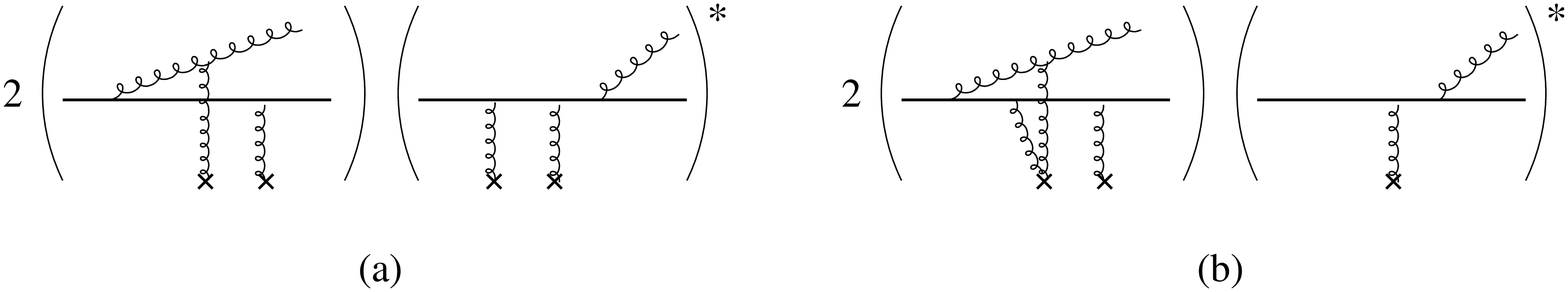}
\caption{Interference term (a) and associated virtual correction (b) contributing to the medium-induced spectrum \eq{qcd-spectrum-qperp}.}
\label{fig:interferencevirtual}
\end{figure} 

In QCD the radiation spectrum is obtained by squaring \eq{GB-rescatt} and putting aside purely initial/final state radiation, which cancels in the medium-induced spectrum. The latter thus arises from the interference between the second and third diagrams of Fig.~\ref{fig:GBamplitude-rescatt}b, see Fig.~\ref{fig:interferencevirtual}a, and reads 
\be
\omega \left. \frac{\dd{I}}{\dd\omega} \right|_{\rm ind} \sim - \alpha_s \int \dd^2\vec{\theta} \, \left. \frac{\vec{\theta}''}{\theta''^{\, 2}} \cdot \frac{\vec{\theta}'}{\theta'^{\,2}}  \right|_{\rm ind} \sim - \alpha_s \int \dd^2\vec{\theta} \,  \left[ \frac{\vec{\theta} - \vec{\theta}_g}{(\vec{\theta} - \vec{\theta}_g)^2} - \frac{\vec{\theta}}{\vec{\theta}^{\,2}} \right] \cdot \frac{\vec{\theta} - \vec{\theta}_s}{(\vec{\theta} - \vec{\theta}_s)^2} \, .
\ee 
The azimuthal integral yields~\cite{Arleo:2010rb} 
\be
\label{kperp-range}
\omega \left. \frac{\dd{I}}{\dd\omega} \right|_{\rm ind} \sim  \alpha_s \int_{\theta_s^2}^{(\vec{\theta}_s - \vec{\theta}_g)^2} \frac{d \theta^2}{\theta^2} \sim \alpha_s \int_{{\rm max}(x^2 q_\perp^2, \Lambda^2)}^{(x \vec{q}_\perp - \vec{\ell}_\perp)^2} \frac{d k_\perp^2}{k_\perp^2}  \, ,
\ee 
where $x \equiv \omega/E$ and we added the constraint  
$k_\perp > \Lambda$
 in the $k_\perp$-integral. Approximating $(x \vec{q}_\perp - \vec{\ell}_\perp)^2 \sim x^2 q_\perp^2 + \ell_\perp^2$ and ${\rm max}(x^2 q_\perp^2, \Lambda^2) \sim x^2 q_\perp^2 + \Lambda^2$ we obtain
\be
\label{qcd-spectrum-qperp}
\omega \left. \frac{\dd{I}}{\dd\omega} \right|_{\rm ind} = \frac{F_c \, \alpha_s}{\pi} \left\{ \ln{\left(1+\frac{\ell_\perp^2 E^2}{q_\perp^2 \omega^2}\right)} - \ln{\left(1+\frac{\Lambda^2 E^2}{q_\perp^2 \omega^2}\right)} \right\} \, .
\ee
It is easy to check that for the scattering of a fast {\it octet} charge (the case of interest), the color factor $F_c$ in \eq{qcd-spectrum-qperp} reads $F_c = N_c$. 
Note that the virtual correction shown in Fig.~\ref{fig:interferencevirtual}b contributes a factor 2 in \eq{qcd-spectrum-qperp}.\footnote{For a quick way to derive color factors and the relative contributions of virtual corrections, see for instance Ref.~\cite{Baier:1998kq}.}

We can also verify that with a parton mass $M \neq 0$ we get the same expression as \eq{qcd-spectrum-qperp}, up to the change in the hard scale $q_\perp \to M_\perp = (M^2 + q_\perp^2)^{\half}$~\cite{Arleo:2010rb}. The spectrum off an energetic color octet charge thus reads
\be
\label{our-spectrum-2}
\omega \left. \frac{\dd{I}}{\dd\omega} \right|_{\rm ind} = \frac{N_c \, \alpha_s}{\pi} \left\{ \ln{\left(1+\frac{\ell_\perp^2 E^2}{M_\perp^2 \omega^2}\right)} - \ln{\left(1+\frac{\Lambda^2 E^2}{M_\perp^2 \omega^2}\right)} \right\} \, \Theta(\ell_\perp^2 - \Lambda^2) \, ,
\ee
where the $\Theta$-function reminds us that only $\ell_\perp > \Lambda$ can induce the emission of {\it perturbative} gluons.
Quite remarkably, the {\it medium-induced} spectrum \eq{our-spectrum-2} is parametrically similar to the radiation spectrum \eq{asym-spec} of an asymptotic parton of ``mass'' $M_\perp$. 

Finally, we stress that in the above calculation, the medium-induced spectrum has been defined with respect to an ``ideal" target for which $\ell_\perp =0$, see \eq{induced-def}. In practice, nuclear suppression is measured in a nucleus A with respect to that in a nucleus ${\rm B} < {\rm A}$, where soft rescatterings can also occur (even in the proton case ${\rm B}=1$). The medium-induced spectrum relevant to this situation should be defined with respect to the nucleus B and reads (as already announced in \eq{our-spectrum} for ${\rm B} =1$)
\be
\label{our-spectrum-3}
\omega \frac{\dd{I}}{\dd\omega}  = \frac{N_c \, \alpha_s}{\pi} \left\{ \ln{\left(1+\frac{\ell_{\perp {\rm A}}^2 E^2}{M_\perp^2 \omega^2}\right)} - \ln{\left(1+\frac{\Lambda_{\rm B}^2 E^2}{M_\perp^2 \omega^2}\right)} \right\} \, \Theta(\ell_{\perp {\rm A}}^2 - \Lambda_{\rm B}^2) \, , 
\ee
where $\Lambda_{\rm B}^2 \equiv {\rm max}(\Lambda_{_\mathrm{QCD}}^2, \ell_{\perp {\rm B}}^2)$. When the broadening in the nucleus ${\rm B}$ is too small, $\ell_{\perp {\rm B}}^2 < \Lambda_{_\mathrm{QCD}}^2$, the spectrum \eq{our-spectrum-3} coincides with the spectrum \eq{our-spectrum-2} defined with respect to the ``ideal" target with $\ell_\perp =0$. When $\ell_{\perp {\rm B}}^2 > \Lambda_{_\mathrm{QCD}}^2$, the induced radiation in ${\rm A}$ with respect to ${\rm B}$ becomes independent of $\Lambda_{_\mathrm{QCD}}$. 

\subsection{Application to quarkonium hadroproduction} 
\label{sec:appl-hadro}

As illustrated by the previous section, the behaviour $\Delta E \propto E$ for medium-induced parton energy loss is {\it not} forbidden by first principles. We expect such a behaviour in all \pA\ processes where some color charge is scattered at small angle in the nucleus rest fame, in particular in quarkonium hadroproduction at large $E$.

However, in order to apply the spectrum \eq{our-spectrum} to quarkonium hadroproduction, the underlying partonic subprocess should {\it effectively} look like the scattering of an energetic, pointlike color charge,  
at least within the formation time $t_{\mathrm{f}} \gg L$ of the radiated gluon. This is the case if 
\be
\label{hierarchy}
{\rm max}(L,\thard) \ll t_{\mathrm{f}} \ll t_{\rm octet} \lsim t_{\psi} \ \ \  {\rm and} \ \ \  r_\perp(t_{\mathrm{f}})  \ll 1/k_\perp  \, ,
\ee 
where $\thard$ is the hard process time scale, $t_{\rm octet}$ the lifetime of the color octet $Q \bar{Q}$ pair, and $t_{\psi}$ the quarkonium hadronization time (see the Introduction). The second condition states that the $Q \bar{Q}$ pair is effectively pointlike when the transverse wavelength $1/k_\perp$ of the radiated gluon is larger than the transverse size $r_\perp$ of the quark pair at a time $\sim t_{\mathrm{f}}$. In principle, the two conditions \eq{hierarchy} can be checked a posteriori, using the typical formation time contributing to the observable of interest (in our case nuclear attenuation). 

As an illustration, we show that the typical $t_{\mathrm{f}}$ contributing to the average loss \eq{mean-delta-E} formally satisfies \eq{hierarchy}. For simplicity we assume here  $L < \thard \sim E/M_\perp^2$ and $M_\perp \simeq M$, and denote $\ell_\perp^2 \sim \Delta q_\perp^2$. It is trivial to check that $\Delta E$ in \eq{mean-delta-E} arises from radiated energies $\omega \sim E\, (\ell_\perp /M) \ll E$ and transverse momenta $k_\perp^2 \sim \ell_\perp^2$ (see \eq{kperp-range}). The typical $t_{\mathrm{f}}$ thus satisfies
\be
\label{hierarchy-check}
\thard \sim \frac{E}{M^2} \ll t_{\mathrm{f}} \sim \frac{\omega}{k_\perp^2} \sim \frac{E}{M \ell_\perp} \ll t_{\psi} \sim \frac{E}{M} \, \tau_{\psi} \, ,
\ee
the inequalities arising from the nuclear broadening $\ell_\perp$ being soft compared to $M$, and hard (for large nuclei) compared to the non-perturbative scale $\tau_{\psi}^{-1}\simeq 0.6$~GeV (see Section~\ref{sec:absorption}). In this respect, let us recall that momentum broadening is related to the saturation scale $Q_s$ in the nucleus, $\ell_\perp^2 = Q_s^2$ \cite{Mueller:1999wm}, indeed considered as a {\it semi-hard} scale. The second condition in \eq{hierarchy} reads
\be 
k_\perp \,  r_\perp(t_{\mathrm{f}})  \sim k_\perp \,  v_\perp \, t_{\mathrm{f}}  \sim \ell_\perp \cdot \frac{\alpha_s M}{E} \cdot \frac{E}{M \ell_\perp} \sim \alpha_s \ll 1 \, ,
\ee
where the relative transverse velocity of the heavy quarks is estimated by $v_\perp \sim p_{B \perp} /E$, with $p_{B} \sim \alpha_s M$ the Bohr momentum of the quarkonium state.\footnote{The $Q \bar{Q}$ pair is produced perturbatively at the time $\thard$ with a relative momentum $\delta p_\perp \lsim M$ between the quark and antiquark. However, only those pairs with $\delta p_\perp \lsim \alpha_s M$ eventually have a non-negligible overlap with the quarkonium wave function.}
Thus, the conditions \eq{hierarchy} are fulfilled in the perturbative domain $\alpha_s \ll 1$ and provided nuclear broadening is a semi-hard scale. This defines the theoretical limit where the energy loss \eq{mean-delta-E} can be applied to quarkonium hadroproduction. 

In practice, $\ell_\perp \sim \sqrt{\hat{q} L} \gg \tau_{\psi}^{-1} $ might not be satisfied. Indeed, using $\hat{q} = 0.08 \, {\rm GeV}^2/{\rm fm}$ (see Section~\ref{sec:phenomenology}) and $L = 7 \,{\rm fm}$ we have $\sqrt{\hat{q} L} \simeq 0.7 \, {\rm GeV} \sim \tau_{\psi}^{-1} $, somewhat questioning the validity of \eq{mean-delta-E} for quarkonium hadroproduction. However, it should be stressed that the observable of interest in this paper -- nuclear attenuation -- does not directly depend on the average energy loss, but rather on the energy loss {\it probability distribution} ${\cal P}(\omega, E)$, see Section~\ref{sec:model}. As is well-known and generic to jet-quenching phenomenology~\cite{Baier:2001yt}, nuclear attenuation is dominated by the low energy tail of ${\cal P}(\omega, E)$, \ie, by $\omega$ much smaller than the typical $\omega$ contributing to the average loss.
This leads to smaller values of $t_{\mathrm{f}}$ to be used in \eq{hierarchy-check}, leading to the required condition $t_{\mathrm{f}} \ll t_{\psi}$. 

Under the conditions \eq{hierarchy} 
we thus expect the induced radiation spectrum 
to be given by Eq.~\eq{our-spectrum}, as derived in Section~\ref{sec:resolved-charge} for a fast color octet 
charge. It should be clear from Section~\ref{sec:resolved-charge} that the parametric dependence of the spectrum is uniquely determined, and should thus apply to other processes than quarkonium production (like open charm and light hadron hadroproduction), as discussed in the Introduction. 

Finally, we note that within our approximation \eq{hierarchy} the quarkonium bound state is formed far beyond the nucleus, $t_{\psi} \gg L$. This approximation may break down at low proton beam energy or at small values of $\xf$. We will comment on this when discussing the limits of applicability of the model in Section~\ref{sec:absorption}.

\section{Model}
\label{sec:model}

 \subsection{Shift in energy or ``medium-induced splitting"} 
\label{sec:xFshift}

The starting point of the model consists in expressing the $\jpsi$ differential production cross section 
$\dd\sigma/\dd E$ 
in \pA\ collisions simply as that in \pp\ collisions, with a shift in 
the quarkonium energy $E$ 
accounting for the energy loss $\varepsilon$ incurred by the octet $c \bar{c}$ pair propagating through the nucleus,
\be
\label{doubleshift-E}
\frac{1}{A}\frac{\dd\sigma_{\mathrm{pA}}^{\psi}}{\dd E}\left(E, \sqrt{s} \right)  = \int_0^{\varepsilon_{\rm max}} \dd \varepsilon \,{\cal P}(\varepsilon, E) \, \frac{\dd\sigma_{\mathrm{pp}}^{\psi}}{\dd E} \left( E+\varepsilon, \sqrt{s}\right)\ .
\ee
The energy loss $\varepsilon$ is more conveniently defined in the nucleus rest frame, and we thus denote $E$ and $E_\mathrm{p} \simeq s/(2 m_\mathrm{p})$ the $\jpsi$ and projectile proton energies in this frame (with $m_\mathrm{p}$ the proton mass). We have $\varepsilon \leq E_{\mathrm{p}} -E$ from energy conservation, and we impose $\varepsilon \leq E$ for consistency with the soft radiation approximation. Hence $\varepsilon_{\rm max} = {\rm min}(E_{\mathrm{p}} -E, E)$ in \eq{doubleshift-E}. The quantity ${\cal P}(\varepsilon, E)$ is the energy loss probability distribution or {\it quenching weight} associated to the radiation spectrum \eq{our-spectrum}, to be discussed in Section~\ref{sec:quenching-weight}.

The measured differential cross sections are usually expressed as a function of $\xf$ (or of the rapidity) rather than $E$. 
The variable $\xf$ is defined as the longitudinal momentum fraction between the $\jpsi$ and projectile proton in the c.m. frame of an elementary \pN\ collision (of energy $\sqrt{s}$). It can be related to the $\jpsi$ transverse mass $M_\perp$ and rapidity $y'$ in this frame. In the limit $\sqrt{s} \gg m_\mathrm{p}$,
\be
\label{xfdef}
\xf \equiv \frac{p'_\parallel}{p'_{\mathrm{p} \parallel}} = \frac{2 M_\perp\, \sinh{y'}}{\sqrt{s}}\, ; \ \ M_\perp \equiv \sqrt{M^2 +p_\perp^2}\, , \ \ y' \equiv \frac{1}{2} \ln{\left( \frac{E'+p'_\parallel}{E'-p'_\parallel} \right)} \, . 
\ee
Using $E = M_\perp \cosh{y} = M_\perp \cosh{(y' + \Delta y)}$, where $\cosh{\Delta y} = \sqrt{s}/(2 m_\mathrm{p})$ with $\Delta y$ the projectile proton rapidity in the c.m. frame, we obtain from \eq{xfdef} 
\be
E = E(\xf) = E_\mathrm{p} \cdot \left[ \frac{\xf}{2} + \sqrt{\left( \frac{\xf}{2} \right)^2 + \frac{M_\perp^2}{s}} \, \right] \, .
\label{Eofxf}
\ee
The relation \eq{Eofxf} can be inverted to give
\be
\label{xfofE}
\xf = \xf(E) = \frac{E}{E_{\mathrm{p}}} - \frac{E_{\mathrm{p}}}{E} \, \frac{M_\perp^2}{s} \, .
\ee
Introducing the variable $x^\prime$,
\be
\label{xprimedef}
x^\prime = x^\prime (E) = \frac{E}{E_{\mathrm{p}}} + \frac{E_{\mathrm{p}}}{E} \, \frac{M_\perp^2}{s} = \frac{2 M_\perp}{\sqrt{s}} \, \cosh{y'} = \sqrt{\xf^2  + 4 \mT^2/s} \ \, ,
\ee
we have ${\partial \xf}/{\partial E} = {x^\prime}/{E}$ and obtain from \eq{doubleshift-E} 
\be
\label{eq:xspA}
\frac{1}{A}\frac{\dd\sigma_{\mathrm{pA}}^{\psi}}{\dd \xf}\left(\xf, \sqrt{s} \right) = \int_0^{\varepsilon_{\rm max}} \dd\varepsilon \,{\cal P}(\varepsilon, E) \, \left[ \frac{E \, x^\prime(E+\varepsilon)}{(E+\varepsilon) \, x^\prime(E)} \right] \, \frac{\dd\sigma_{\mathrm{pp}}^{\psi}}{\dd\xf} \left( \xf(E+\varepsilon), \sqrt{s} \right) \, .
\ee

Note that at large $\xf\gg{M_\perp/\sqrt{s}}$, we have $\xf \simeq x^\prime \simeq E/E_{\mathrm{p}}$, the (Jacobian) factor in between brackets in \eq{eq:xspA} approaches unity, and the \pp\ cross section is evaluated at a shifted value of $\xf$, with the shift $\delta \xf(\varepsilon) \equiv  \xf(E+\varepsilon) - \xf(E) \simeq \varepsilon / E_{\mathrm{p}}$. At large $\xf$ the energy shift in \eq{doubleshift-E} is thus equivalent to a simple translation in $\xf$. This is not true at all values of $\xf$, due to the presence of the Jacobian. 

In the following we will use the expression \eq{eq:xspA}, where $E = E(\xf)$, and the relations \eq{Eofxf} and \eq{xfofE}, valid at all $\xf$. In a model where the $\jpsi$ is produced through a $2\to 1$ partonic subprocess, the expression \eq{Eofxf}, denoted as $E \equiv x_1 E_\mathrm{p}$, simply arises from the standard relations between parton momentum fractions, $x_1 x_2 =M_\perp^2/s$ and $x_1 - x_2 = \xf$ (note also that $x_1 + x_2 = x^\prime$). However, the kinematical relation \eq{Eofxf} is actually independent of the partonic subprocess. 

It is interesting to mention that the main equation of our model \eq{doubleshift-E} 
is equivalent to 
\be
\label{shift-frag}
\frac{1}{A}\frac{\dd\sigma_{\mathrm{pA}}^{\psi}}{\dd E}\left(E, \sqrt{s} \right) = \int^1_{z_{min}} \dd z  \,{\cal F}_{\mathrm{loss}}(z) \, \frac{\dd\sigma_{\mathrm{pp}}^{\psi}}{\dd E} \left( \frac{E}{z} , \sqrt{s} \right)\, , 
\ee
where $z \equiv E/(E+\varepsilon)$ is interpreted as a (medium-induced) splitting variable describing the energy loss process ($E$ is the energy of the charge {\it after} radiating the energy $\varepsilon$), and ${\cal  F}_{\mathrm{loss}}(z)$ as a  ``medium-induced splitting function". The expression \eq{shift-frag} follows from \eq{doubleshift-E} by changing variable from $\varepsilon$ to $z$ (giving $z_{min} = {\rm max}(\textstyle{E/E_\mathrm{p}},\textstyle{1/2})$), and using the fact that the quenching weight ${\cal P}(\varepsilon, E)$ is a scaling function of the ratio $\varepsilon/E$,\footnote{This can be trivially checked from \eq{Pstar} (Section~\ref{sec:quenching-weight}) and \eq{our-spectrum}.}
\be
\label{quenching-scaling}
E \cdot {\cal P}(\varepsilon, E) = \hat{\cal P} \left(\frac{\varepsilon}{E} \right) = \hat{\cal P} \left(\frac{1-z}{z} \right) \equiv  z^2 \, {\cal  F}_{\mathrm{loss}}(z) \, .
\ee

In writing \eq{doubleshift-E}, we implicitly assumed that the energy of the radiating octet $c \bar{c}$ pair is the same as the final $\jpsi$ energy $E$. However, the equivalent expression \eq{shift-frag} suggests that our nuclear suppression model based on a simple shift in 
$E$ 
might apply to more general situations where the final detected particle's energy arises from the fragmentation of some parent parton's energy, with a fragmentation variable $z <1$. Indeed, suppose that the observable \pp\ cross section is of the form\footnote{The dependence of \eq{fact1} on $\sqrt{s}$ and  on the projectile and target parton distribution functions is irrelevant to our discussion.}
\be
\label{fact1}
\frac{\dd\sigma_{\mathrm{pp}}}{\dd E}\left( E \right) = \int^1_{E/E_{\mathrm{p}}} \dd z  \, D(z) \, \frac{\dd\hat{\sigma}}{\dd E} \left( \frac{E}{z} \right) \, ,
\ee
and that medium-induced radiation and hadronization factorize,\footnote{This should be guaranteed by the separation of time scales, $t_{\mathrm{f}} \ll t_{\psi}$, see Section~\ref{sec:appl-hadro}.} 
\be
\label{fact2}
\frac{1}{A} \frac{\dd\sigma_{\mathrm{pA}}}{\dd E}\left(E \right) = \int^1_{E/E_{\mathrm{p}}} \dd z  \, D(z) \int_{z'_{min}}^1 \dd z' \, {\cal F}_{\mathrm{loss}}(z') \, \frac{\dd\hat{\sigma}}{\dd E} \left(\frac{E}{z\,z'} \right)\, , 
\ee
where $z'_{min} = {\rm max}(E/(z E_\mathrm{p}),1/2)$. Changing the order of the $z$ and $z'$ integrals in \eq{fact2} and using \eq{fact1}, we recover \eq{shift-frag}. 

The quarkonium \pA\ cross section is thus related to the {\it observable} \pp\ cross section according to \eq{shift-frag}, or equivalently \eq{doubleshift-E}, independently of the form of the fragmentation function $D(z)$. This result mostly follows from the scaling property \eq{quenching-scaling} of the quenching weight, and is thus expected to hold in all processes involving a {\it fractional} medium-induced energy loss ($\Delta E \propto E$), in particular in open charm and light hadron hadroproduction. 
Instead of describing the energy loss process as a shift in energy (see \eq{doubleshift-E} or \eq{eq:xspA}), one could alternatively describe it 
as in \eq{shift-frag} by a {\it medium-induced splitting function}\footnote{This designation is motivated by the fact that ${\cal F}_{\mathrm{loss}}$ is perturbatively calculable and can be Mellin convoluted with the (vacuum) fragmentation function $D(z)$ to give the ``medium modified fragmentation function'' $D_{\rm med}(z) = \int \dd{z'} D(z') \, {\cal F}_{\mathrm{loss}}(z/z')$. We however stress that ${\cal F}_{\mathrm{loss}}(z)$ is process-dependent (\eg, it is present in quarkonium but absent in DY production, see the Introduction). Thus, $D_{\rm med}(z)$ in the latter equation differs from the medium modified fragmentation functions assumed to be universal and discussed elsewhere, see for instance Ref.~\cite{Sassot:2009sh}.} ${\cal F}_{\mathrm{loss}}(z)$.

Finally, although in the above discussion we assumed the \pp\ cross section to obey 
factorization (see \eq{fact1}), we believe that \eq{doubleshift-E} might hold independently of this assumption. For instance, as long as the underlying partonic process is similar to the scattering of a color charge (as in Figs.~\ref{fig:generic-processes}a and \ref{fig:generic-processes}b), we may imagine the \pp\ cross section 
at large $E$  
to be affected by late comover rescattering \cite{Hoyer:1998ha} and thus to violate factorization, and nevertheless the \pA\ cross section to be given by \eq{doubleshift-E}. The only crucial assumption is that the partonic subprocess induces radiation, as dictated by perturbative QCD, independently of the precise mechanism fixing the quantum numbers of the final detected particle.

\subsection{Absolute production cross section}
\label{sec:xspp}

The dynamics of heavy-quarkonium production in hadronic collisions is still uncertain. 
In particular, none of the existing models proposed to describe heavy-quarkonium production is able to reproduce simultaneously all the features reported experimentally, 
at both $p_\perp \lsim M$ and $p_\perp \gg M$.

\begin{figure}
\begin{center}
    \includegraphics[width=7.5cm]{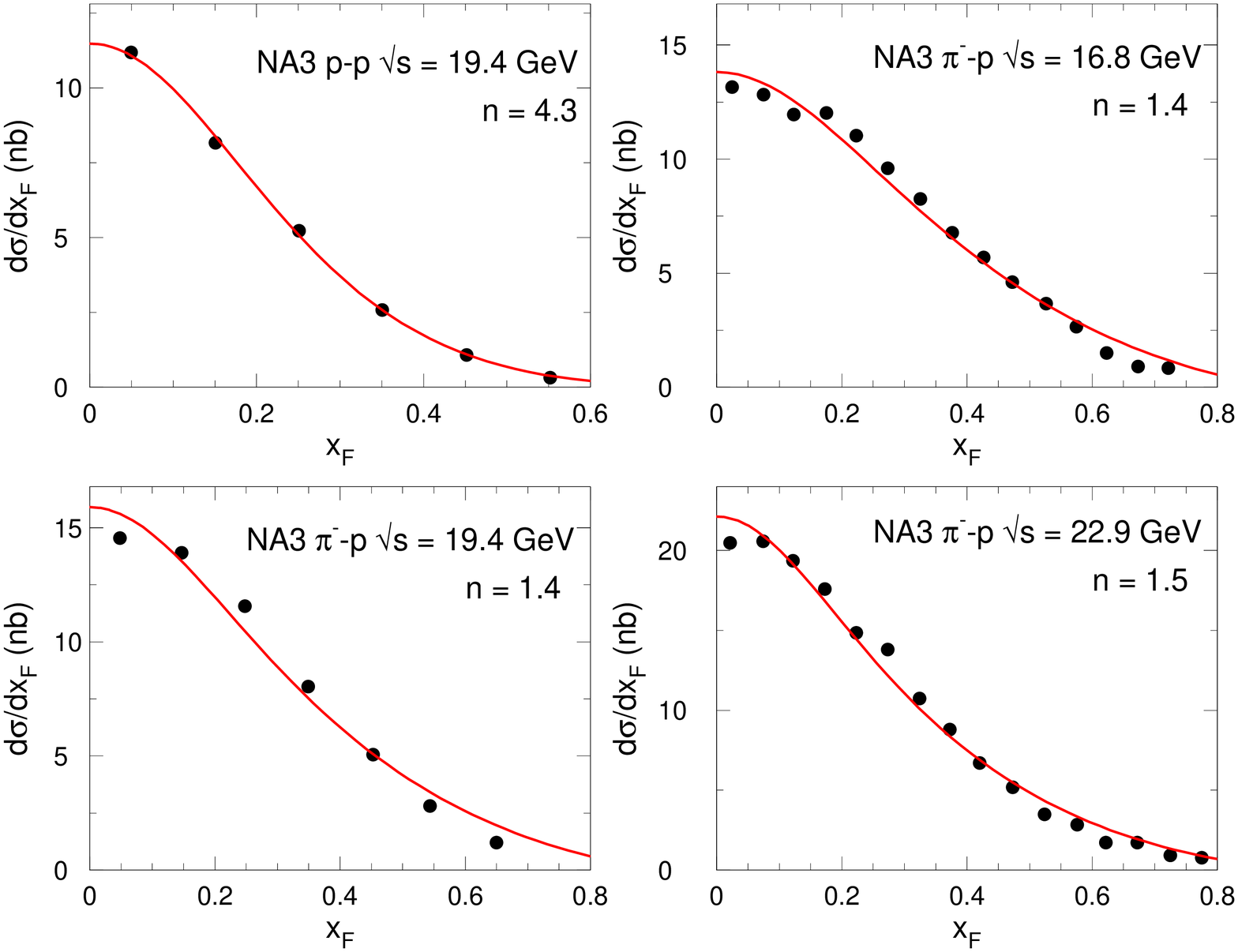}
    \includegraphics[width=7.5cm]{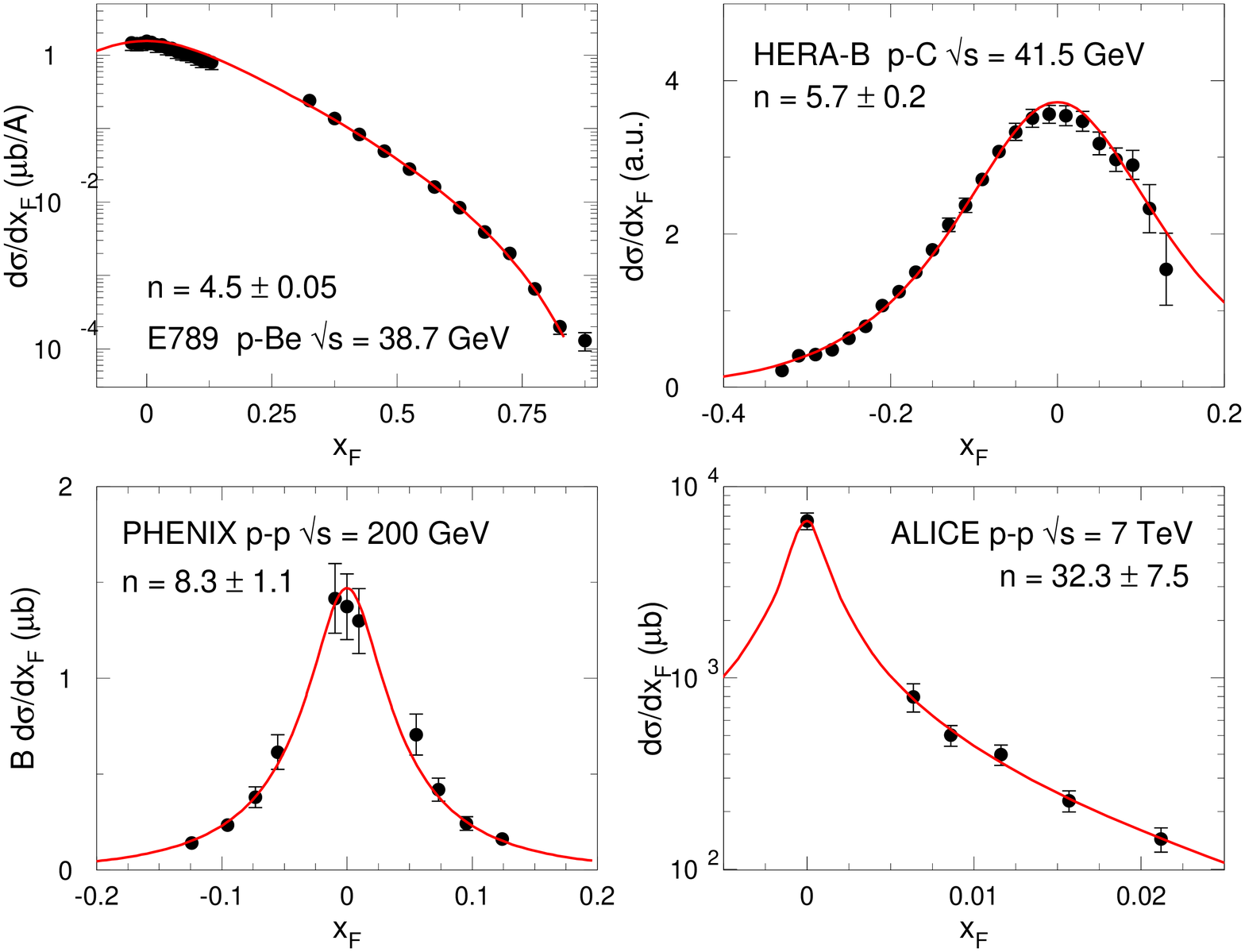}
  \end{center}
\vspace{-1cm}
\caption{Comparison between $\jpsi$ production data in p--p, 
$\pi^-$--p and p--A collisions and the fit \eq{pp-fit} (solid red line). The values of $n$ obtained from the fit are indicated in each panel and in Table~\ref{tab:expnjpsi}. Data are taken from~\cite{Badier:1983dg,Kowitt:1993ns,Abt:2008ya,Aamodt:2011gj,Adare:2006kf}.}
  \label{fig:fitjpsi}
\begin{center}
    \includegraphics[width=4.6cm]{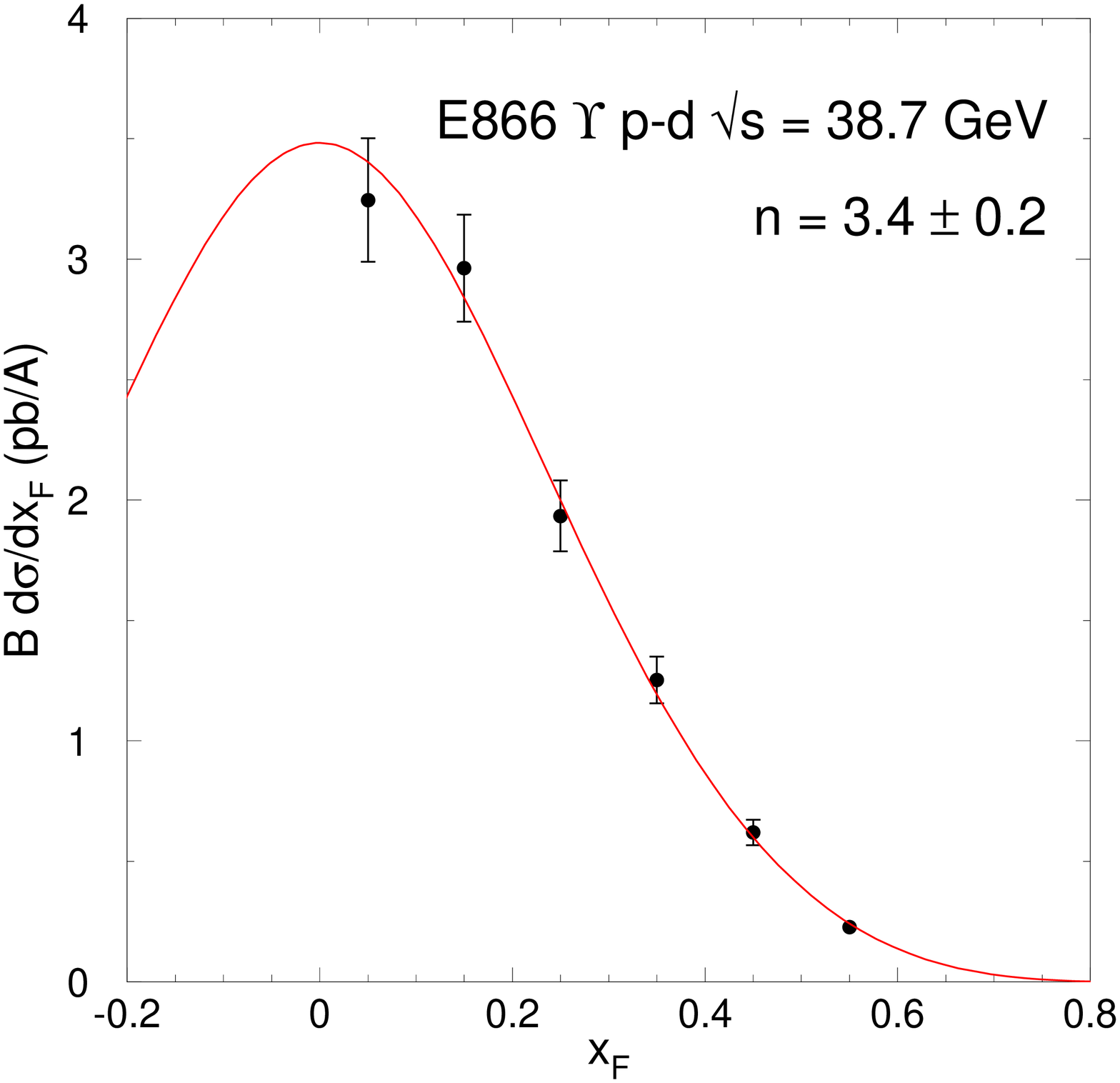}
    \includegraphics[width=4.6cm]{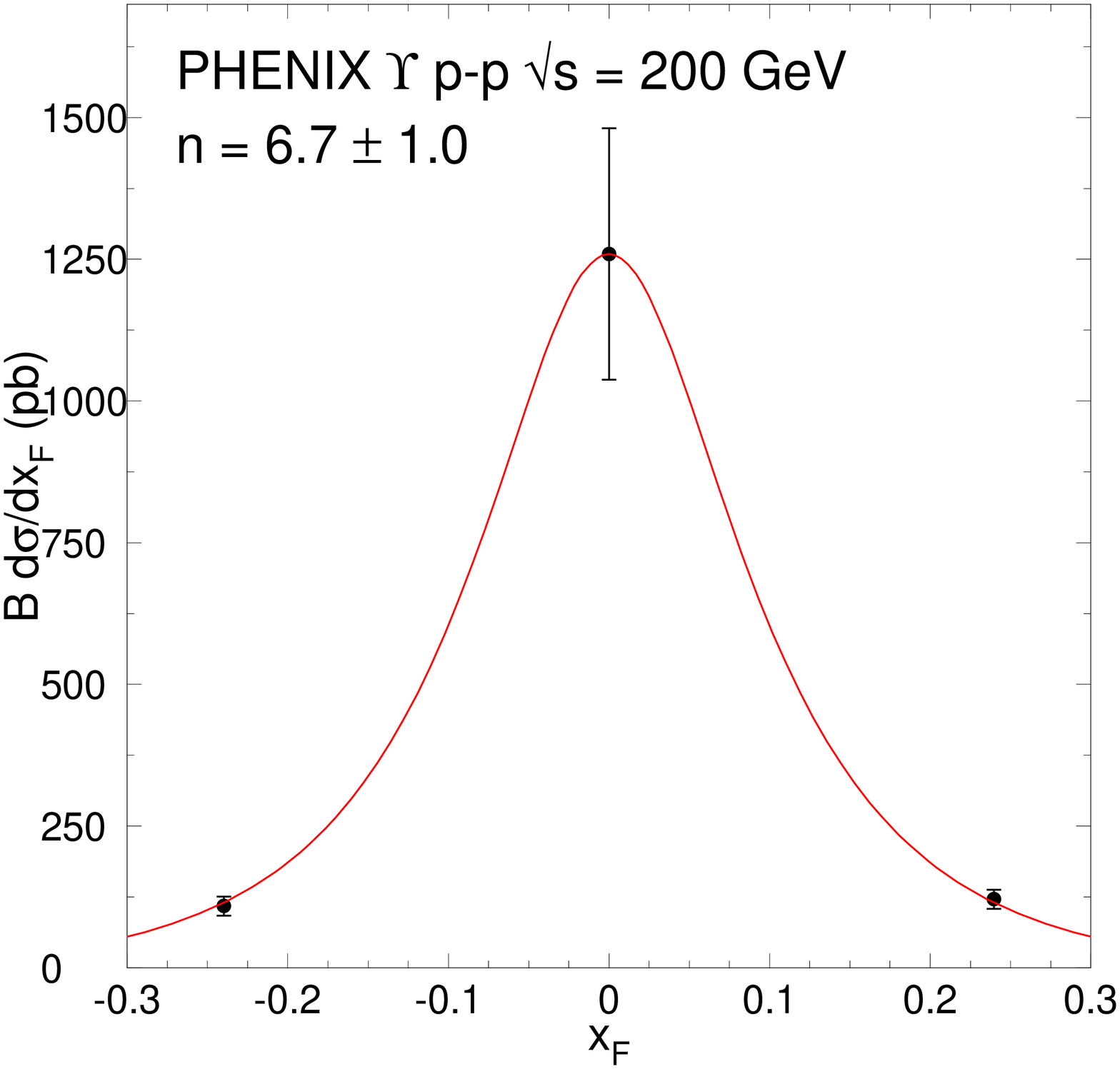}
    \includegraphics[width=4.6cm]{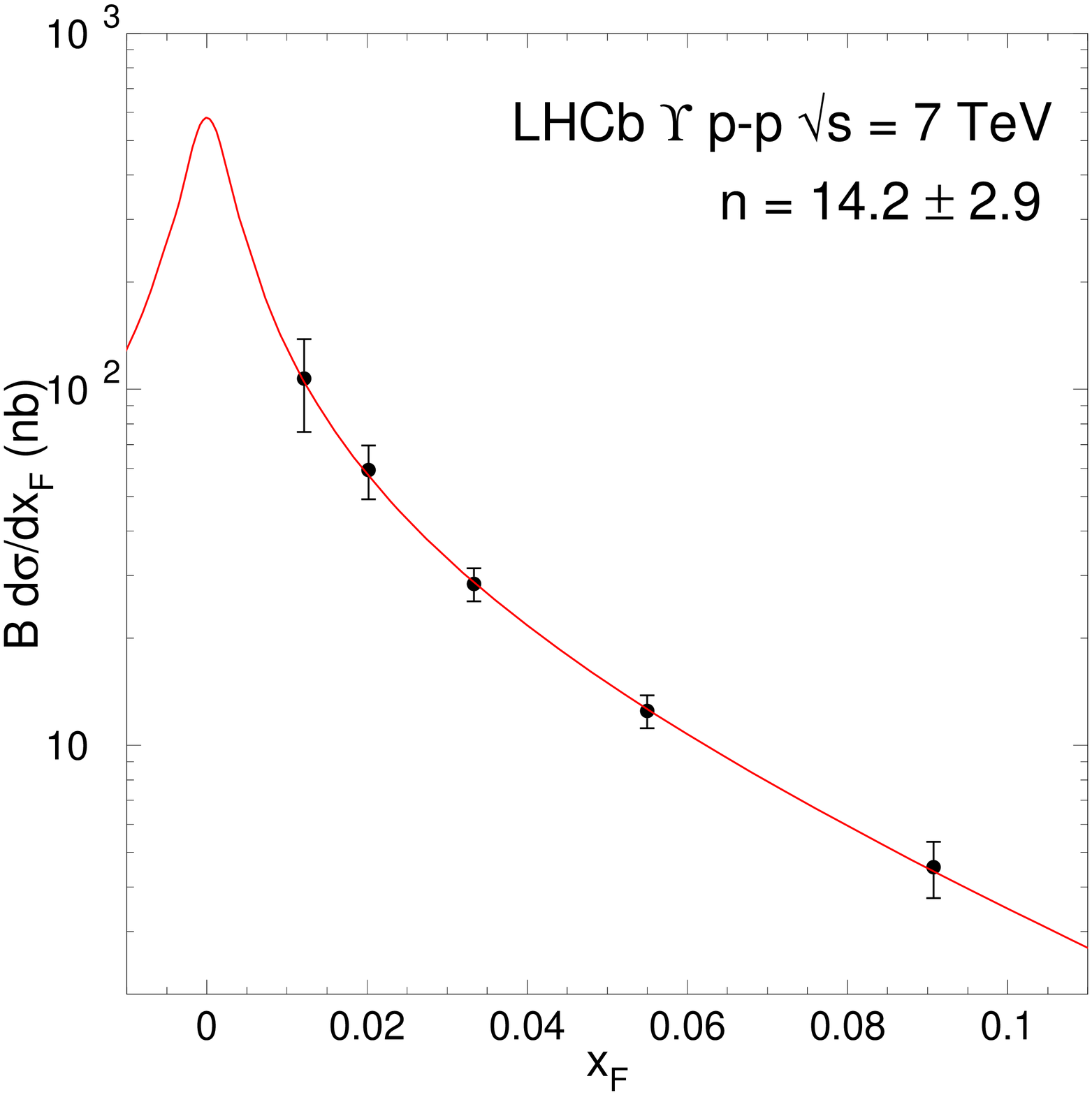}
  \end{center}
\vspace{-1cm}
\caption{Comparison between $\Upsilon$ production data in p--p and p--d collisions and the fit \eq{pp-fit} (solid red line). The values of $n$ obtained from the fit are indicated in each panel and in Table~\ref{tab:expnups}. Data are taken from~\cite{Zhu:2007aa,Adare:2012bv,LHCb:2012aa}.}
  \label{fig:fitups}
\end{figure}

In the present approach, a crucial ingredient entering \eq{eq:xspA} is the $\xf$ single differential absolute cross section, $\dsigpp$, of $\jpsi$ and $\Upsilon$ production in \pp\ collisions at 
a given center-of-mass energy. In order to be as model-independent as possible, $\dsigpp$ used in \eq{eq:xspA} is not taken from theory but determined from a fit to the data. We found that it can be conveniently parametrised as
\be
\label{pp-fit}
\frac{\dd\sigma_{\mathrm{hp}}^{\psi}}{\dd\xf} \left(\xf\right) \propto \frac{(1-x^\prime)^{n}}{x^\prime} \, ,
\ee
where $x^\prime$ is defined in \eq{xprimedef} and  
the exponent $n$ is extracted from p--p\footnote{p--A data, where A is a light nucleus such as Be or C,
were also used, see Fig.~\ref{fig:fitjpsi}.} and $\pi^-$--p data taken at the same center-of-mass energy - whenever possible - as the p--A and $\pi^-$--A measurements discussed in this paper.\footnote{These data fits have been made much easier thanks to the Quarkonii database which can be found at~\url{http://hepdata.cedar.ac.uk/review/quarkonii/}.} Note that the normalization parameter in (\ref{pp-fit}), or equivalently the total production cross section, is irrelevant for our purpose since only nuclear production ratios are considered in this paper, see Eq.~\eq{RpA}.

\begin{table}[t]
{\footnotesize
 \centering
 \begin{tabular}[c]{p{2.3cm}ccccc|ccc}
   \hline
   \hline
  \multicolumn{9}{c}{}  \\
 & \multicolumn{5}{c}{p beam}  & \multicolumn{3}{c}{$\pi^-$ beam}\\
  \multicolumn{9}{c}{}  \\
Experiment   & NA3 & E789  &  HERA-B & PHENIX & ALICE& NA3& NA3& NA3\\
$\sqrt{s}$ (GeV) &19.4& 38.7 & 41.5  & 200  & 7000 &16.8& 19.4 & 22.9 \\
$n$ & $4.3$ & $4.5\pm0.05$ &  $5.7\pm0.2$ &  $8.3\pm1.1$     &  $32.3\pm7.5$& 1.4 & 1.4 &  1.5\\
 & & & & &  & & & \\
\hline
\hline
\end{tabular}
 \caption{Values of $n$ extracted from $\jpsi$ production in p--p 
 (or p--Be, p--C) 
 and $\pi^-$--p collisions; see text for details.}
 \label{tab:expnjpsi}
 }
{\footnotesize
 \centering
 \begin{tabular}[c]{p{2.3cm}ccc}
   \hline
   \hline
 & & &    \\
Experiment    &  E866 & PHENIX & LHCb \\
$\sqrt{s}$ (GeV) &38.7  & 200  & 7000 \\
$n$ &  $3.4\pm0.2$ &  $6.7\pm1.0$     &  $14.2\pm 2.9$\\
 & & &  \\
\hline
\hline
\end{tabular}
 \caption{Values of $n$ extracted from $\Upsilon$ production in p--p 
 (or p--d)
 collisions; see text for details.}
 \label{tab:expnups}
 }
\end{table}

The values of $n$ for $\jpsi$ production from SPS to LHC energy are summarized in Table~\ref{tab:expnjpsi}.\footnote{We were unable to estimate the experimental errors when retrieving NA3 data from~\cite{Badier:1983dg}. Therefore we do not quote the uncertainties on the exponent $n$ in this case.} The index for \pp\ production
grows smoothly from $n\simeq 4$--5 at low energies ($\sqrt{s}\lesssim 40$~GeV) up to $n\simeq 8$ at RHIC and $n\simeq30$ at the LHC. At LHC, the relative uncertainty on $n$ is as large as $\delta{n} / n \sim 20\%$ ($n=32.3\pm7.5$) because of the too small $\xf$ domain, $|\xf| \lesssim 0.02$, covered by the data and the fact that around mid-rapidity, $x^\prime \ll 1$, the parametrization \eq{pp-fit} becomes independent of $n$. However we checked that the resulting uncertainty on $R_{\mathrm{pA}}^{\jpsi}$ at LHC (and similarly at RHIC) is marginal because of these very reasons. In the $\Upsilon$ channel, the exponents are slightly smaller than for $\jpsi$ production; see Table~\ref{tab:expnups}.

The comparison between the fits and $\jpsi$ (respectively, $\Upsilon$) production data is shown in Fig.~\ref{fig:fitjpsi} (respectively, Fig.~\ref{fig:fitups}). An excellent agreement is observed on a very wide range of center-of-mass energies, spanning from $\sqrts=16.8$~GeV to $\sqrts=7$~TeV, on the full $\xf$ domain covered experimentally, and for both $\jpsi$ and $\Upsilon$ production. These results indicate that the parametrization \eq{pp-fit} can be safely used in \eq{eq:xspA} in order to compute heavy-quarkonium nuclear suppression. 

Independently of the present work, this parametrization can prove useful in future phenomenological studies for which a data-driven knowledge of the $\xf$ single differential cross sections is necessary.

\subsection{Transport coefficient and nuclear broadening}
\label{sec:qhat}

The amount of medium-induced gluon radiation, and hence the strength of $\psi$ suppression in \pA\ collisions, is controlled by the nuclear broadening $\Delta q_\perp^2 \equiv \ell_\perp^2$ in Eq.~\eq{our-spectrum}. For a path length $L$ crossed in the target, the broadening reads
\be
\ell_\perp^2  = \hat{q} \, L \  ,
\label{broadening}
\ee
where the transport coefficient $\hat{q}$ in the target nucleus is related to the gluon distribution $G(x)$ in a target nucleon as~\cite{Baier:1996sk}\footnote{We assume that the relevant transport coefficient in our approach is that of an energetic {\it gluon}, hence the factor $N_c$ in the numerator of \eq{qhat-gluondensity}.}
\be
\hat{q} = \frac{4 \pi^2 \alpha_s(\hat{q} L) N_c}{N_c^2-1} \, \rho \, x G(x,\hat{q} L) \simeq  \frac{4 \pi^2 \alpha_s  N_c}{N_c^2-1} \, \rho \, x G(x)\, .
\label{qhat-gluondensity}
\ee
In the latter equation $\rho$ is the target nucleon number density, and the scaling violations in the running of $\alpha_s$ and in the evolution of the gluon density are neglected since $\hat{q} L \lesssim 1$~GeV$^2$.

The typical value of $x$ at which $xG(x)$ should be evaluated in the r.h.s. of \eq{qhat-gluondensity} is discussed in Appendix~\ref{appendix-qhat}. The result depends on whether the hard subprocess is incoherent, $\thard \ll L$, or coherent, $\thard \gg L$, in the nucleus. Assuming $2\to1$ subprocess kinematics we estimate $\thard \sim E/M^2 \sim x_1 E_{\mathrm{p}} / (x_1 x_2 s) \sim 1/(2 m_{\mathrm{p}} x_2)$, so that the incoherent and coherent limits correspond respectively to $x_2 \gg x_0$ and $x_2 \ll x_0$, with $x_0 = x_0(L) \equiv (2 m_{\mathrm{p}} L)^{-1}$. In the incoherent case $xG(x)$ should be evaluated at $x \sim x_0$~\cite{Baier:1996sk}, whereas in the coherent case $x \sim x_2$ (see Appendix \ref{appendix-qhat}), \ie, $x = \min(x_0, x_2)$. 

Using the power-law behavior $xG(x) \sim x^{-0.3}$ suggested by small-$x$ ($x<10^{-2}$) fits to HERA data~\cite{GolecBiernat:1998js}, we can thus extract the $x$ and $\rho$ dependence of $\hat{q}$ by writing
\be
\label{qhat-model}
\hat{q} =  \hat{q}(x) \, \frac{\rho}{\rho_0} \ ; \ \ \ \hat{q}(x) \equiv \hat{q}_0 \left( \frac{10^{-2}}{x} \right)^{0.3}\ ;\ \ \ x = \min(x_0, x_2) \ ; \ \ \  x_0 \equiv \frac{1}{2 m_\mathrm{p} L}\, ,
\ee
where $\rho_0$ is in principle an arbitrary constant density, and $\hat{q}_0 \equiv \hat{q}(x=10^{-2},\rho=\rho_0)$.

In Ref.~\cite{Arleo:2012hn} we used the hard sphere (HS) approximation\footnote{This approximation is defined by $\rho(r) = (A/V)\,\Theta(R_A - r)$, with $R_A=r_0 \, A^{1/3}$ ($r_0 =1.12 \, {\rm fm}$).} and thus the same uniform density $\rho = \rho_{_{\rm HS}} = [(4/3)\pi r_0^3]^{-1} \simeq 0.17 \, {\rm fm}^{-3}$ for all nuclei. Within this approximation, $\hat{q} =  \hat{q}(x)$ with $\hat{q}_0 \equiv \hat{q}(x=10^{-2},\rho=\rho_{_{\rm HS}})$, the average of $L = \int \dd z$ is found to be $L_{_{\rm HS}} = 3 R_A/2$, and the broadening is directly obtained from \eq{broadening}. In the present study we will use more realistic (non-uniform) nuclear density profiles $\rho(r)$ extracted from electron--proton scattering experiments~\cite{DeJager:1987qc}. In order to formally recover the situation considered in Ref.~\cite{Arleo:2012hn} when $\rho(r)$ is constant, we thus choose $\rho_0 = \rho_{_{\rm HS}}$ in \eq{qhat-model}. 

When $\rho(r)$ is not constant, the parameter $L = \int \dd z$ entering \eq{broadening} is badly defined, but the broadening is still well-defined, since it is proportional to $\int \rho \, \dd z$ rather than to $\int \dd z$. Using \eq{qhat-model} we can write
\be
\label{broadening2}
\ell_\perp^2  = \int  \dd \ell_\perp^2 = \int \hat{q} \, \dd z = \hat{q}(x) \cdot \Leff \ \ ; \ \ \ \Leff \equiv \frac{1}{\rho_0}  \int \rho \, \dd z \, .
\ee
The effective path length $\Leff$ is mathematically well-defined and can be related to the number $N_{\rm part}$ of nucleons participating to the broadening of the fast 
color charge. 
Using $\dd N_{\rm part} = \rho \, \sigma \, \dd z$, where $\sigma$ is interpreted as the cross section for having non-zero broadening in parton-nucleon scattering, we obtain\footnote{The integration constant in \eq{Leff} follows from the fact that for a proton target, $N_{\rm part} =1$, see \eq{Npart-ave}.}
\be
\Leff - L_{\rm p}  = \frac{N_{\rm part}-1}{\rho_0 \, \sigma} \, .
\label{Leff}
\ee

For minimum bias \pA\ collisions, the average of $N_{\rm part}$ in the events with $\jpsi$ production can be calculated within Glauber theory and reads 
\be
\label{Npart-ave}
 \langle N_{\rm part} \rangle_{_{\jpsi}} = 1+ \sigma \, \frac{(A-1)}{A^2} \int \dd^2\vec{b} \, \,  T_A(b)^2 \, ,
\ee
where we used the normalization $\int \dd^3\vec{r} \, \rho(r) =  \int \dd^2\vec{b}\,\, T_A(b) = A$. The effective path length becomes
\be
\label{Leff2}
\Leff - L_{\rm p} =  \frac{(A-1)}{A^2 \rho_0} \int \dd^2\vec{b} \,\, T_A(b)^2 \, .
\ee

As can be seen from Eq.~\eq{Leff2}, the additional (effective) path length in a nucleus with respect to a proton is independent of $\sigma$, and can thus be uniquely determined knowing the nuclear density profile. 
For the effective length in the proton, we take $L_{\rm p} = 3 R_{\rm p}/2 = 1.5 \, {\rm fm}$, using $R_{\rm p} =  1 \, {\rm fm}$ for a generic proton length scale.\footnote{We checked that varying the proton effective length in the range $L_{\rm p}=1.3$--$1.7$~fm only marginally affects our results.} In summary, $\Leff$ is given by 
\be
\label{Leff3}
\Leff = 1.5 \, {\rm fm}  + \frac{(A-1)}{A^2 \rho_0} \int \dd^2\vec{b} \,\, T_A(b)^2  \, .
\ee
The values of $\Leff$ obtained from \eq{Leff3} using realistic nuclear density profiles~\cite{DeJager:1987qc} are listed in Table~\ref{tab:Leff} for various nuclei, and compared to the values $L_{_{\rm HS}} = 3 R_A/2$ corresponding to the hard sphere approximation previously used in Ref.~\cite{Arleo:2012hn}.

Using \eq{Leff3} in \eq{broadening2} fully determines the nuclear broadening and hence the induced gluon spectrum \eq{our-spectrum}. The transport coefficient $\hat{q}_0\equiv\hat{q}(x=10^{-2},\rho=\rho_0)$ is the {\it only free parameter} of the model.

\begin{table}[t]
{\footnotesize
 \centering
 \begin{tabular}[c]{p{2.3cm}cccccccccc}
   \hline
   \hline
  \multicolumn{10}{c}{}  \\
Nucleus           & {\rm p} & {\rm Be} & {\rm C}  & {\rm Ca} & {\rm Fe} & {\rm Cu} & {\rm W} & {\rm Pt} & {\rm Au} &  {\rm Pb} \\
Atomic mass       & 1       &    9     &   12     &     40   &     56   &     63   &  184    &   196    &  197     &    208  \\
$\Leff$ (fm) & 1.5     & 3.24     & 3.94     & 5.69     & 6.62     &  6.67    & 9.35    & 10.85    &  10.21   &    10.11   \\
$L_{_{\rm HS}}$ (fm)    & 1.68    & 3.49     & 3.85     & 5.75     & 6.43     &  6.68    & 9.56    & 9.76     &  9.78    &    9.95   \\
 & & & & &  & & & & & \\
\hline
\hline
\end{tabular}
 \caption{Values of $\Leff$ for various nuclei, using \eq{Leff3} and realistic nuclear densities~\cite{DeJager:1987qc}, or within the hard sphere approximation.}
 \label{tab:Leff}
 }
\end{table}

\subsection{Energy loss probability distribution}
\label{sec:quenching-weight}

In this section we discuss the quenching weight ${\cal P}(\varepsilon, E)$ entering \eq{eq:xspA}. A well-known procedure (used for instance in the case of large $\pt$ jet-quenching in A--A collisions \cite{Baier:2001yt}) to construct a normalized ${\cal P}(\varepsilon, E)$ from the single gluon emission spectrum $\dd I/\dd\omega$, consists in assuming {\it independent} emissions of {\it soft} gluons. In this so-called Poisson approximation, 
\be
\label{eq:quenchingweight}
{\cal P}(\varepsilon, E) = \sum^\infty_{n=0} \, \frac{1}{n!}\ \left[ \prod^n_{i=1} \, \int_0^{\varepsilon} \, \dd\omega_i \, \frac{\dd{I}(\omega_i)}{\dd\omega} \right] 
\times  \, \delta \left(\varepsilon- \sum_{i=1}^n  \omega_i\right) \, \exp{ \left\{- \int_0^{\infty} \dd\omega \, \frac{\dd{I}}{\dd\omega} \right\} }\, . 
\ee
Let us mention that ${\cal P}(\varepsilon, E)$ is a solution of the equation
\be
\label{landau-kinetic}
\frac{\partial {\cal P}(\varepsilon, E)}{\partial L} = \int_0^{\infty} \dd\omega \left[  {\cal P}(\varepsilon - \omega, E) - {\cal P}(\varepsilon, E)\right] \frac{\dd I}{\dd\omega \dd{L}} \, ,
\ee
where by convention ${\cal P}(\varepsilon <0, E)=0$, and $L$ stands for any parameter entering the expression of $\dd{I}/\dd\omega$. If $L$ is the medium length crossed by the fast charge, \eq{landau-kinetic} is formally identical to the kinetic equation used by Landau to study ionization losses in normal matter \cite{Landau:1944if}. 
An important feature of the Poisson approximation \eq{eq:quenchingweight} is that not only each $\omega_i$, but also the accumulated loss $\sum \omega_i$, is supposed soft as compared to $E$. In other words, the ``energy degradation'' of the fast particle during the multiple emission process is neglected. 

Whether the Poisson approximation is appropriate or not obviously depends on each energy loss process and on the specific properties of $\dd I/\dd\omega$, in particular on the multiplicity $N_{\omega}$ of radiated gluons with energy $\sim \morder{\omega}$. In the present case, $\dd I/\dd\omega$ given in \eq{our-spectrum-3} is a scaling function of $\omega/\hat{\omega}_{\rm A}$, where $\hat{\omega}_{\rm A} \equiv (\ell_{\perp {\rm A}}/M_\perp) \, E$,\footnote{For the present discussion we neglect the second term of the spectrum \eq{our-spectrum-3}.} and $N_{\omega}$ is estimated as
\be
\label{gluon-mult}
N_{\omega} \sim \int_{\omega/2}^{2\omega} \dd\omega' \, \frac{\dd{I}}{\dd\omega'} \sim \omega \, \frac{\dd{I}}{\dd\omega} \sim \alpha_s \ln{\left(1+\frac{\hat{\omega}_{\rm A}^2}{\omega^2}\right)} \, .
\ee
When $\omega \ll \hat{\omega}_{\rm A}$, $N_{\omega}$ becomes potentially large, $\alpha_s \ln{(\hat{\omega}_{\rm A}/\omega)} \sim \morder{1}$, questioning the assumption of {\it independent} multiple emissions.\footnote{In the case of large $\pt$ jet-quenching \cite{Baier:2001yt}, a large multiplicity at small $\omega$ is compensated by a factor $t_{\rm f}/L \ll 1$, resulting in a small gluon {\it occupation number} $\sim (t_{\rm f}/L)\, N_{\omega} \ll 1$, thus supporting the assumption of independent emissions. In our case the spectrum \eq{our-spectrum-3} arises from large formation times $t_{\rm f} \gg L$, and the estimates of gluon occupation number and gluon multiplicity coincide.} When $\omega \gsim \hat{\omega}_{\rm A}$, $N_{\omega}$ is small, $N_{\omega} \lsim \morder{\alpha_s} \ll 1$. However in this region each emitted gluon carries away a fixed fraction $\sim \ell_{\perp {\rm A}}/M_\perp$ of the energy $E$, and the fast particle energy degradation cannot be neglected.\footnote{This problem was previously addressed in Ref.~\cite{Arleo:2002kh} in the context of large $p_\perp$ jet-quenching.} Thus, in our context the Poisson approximation proves fishy. 

A simple way to deal with this problem is to supplement \eq{eq:quenchingweight} with the condition that the energy $\varepsilon$ is carried away by a {\it single} gluon, \ie, $\delta \left(\varepsilon- \sum \omega_i \right) \to n \, \delta \left(\varepsilon- \omega_j \right)$ in \eq{eq:quenchingweight}. This yields the (normalized) quenching weight
\be
\label{Pstar}
{\cal P}(\varepsilon, E) = \frac{\dd I}{\dd\varepsilon} \, \exp \left\{ - \int_{\varepsilon}^{\infty} \dd\omega  \frac{\dd{I}}{\dd\omega} \right\} = \frac{\partial}{\partial \varepsilon} \, \exp \left\{ - \int_{\varepsilon}^{\infty} \dd\omega  \frac{\dd{I}}{\dd\omega} \right\} \, .
\ee 
The latter is simply interpreted as the product between the ``probability'' $\dd{I}/\dd\varepsilon$ to radiate a gluon with $\omega_j = \varepsilon$ and the probability (given by the exponential Sudakov factor) to have no extra radiation carrying $\omega_k \gsim \varepsilon$. In our context, the expression \eq{Pstar} of the quenching weight is better founded than the Poisson expression \eq{eq:quenchingweight}, and will be used in \eq{eq:xspA}.

\subsection{Other nuclear effects}
\label{sec:othereffects}

Besides energy loss effects, other mechanisms might affect $\psi$ suppression in nuclei. In this section, the role of nuclear absorption and saturation on $\psi$ suppression in p--A collisions is discussed. 

\subsubsection{Nuclear absorption}
\label{sec:absorption}

At small $\psi$ energy in the nucleus rest frame, the hadronization time $t_\psi=\tau_\psi \cdot (E/M)$ (where $\tau_{\psi}$ 
is the proper hadronization time) becomes comparable to the typical nuclear size, $t_\psi \lesssim L$. Consequently, $\psi$ states are produced on average {\it within} the target nucleus and might suffer inelastic interaction with nuclear matter, the so-called nuclear absorption process. From \eq{Eofxf}, this should be the case at low proton beam energy $\Ep$ (\ie\ at low $\sqrt{s}\simeq \sqrt{2\mprot \Ep}$) or at small values of $\xf$. The $\jpsi$ suppression in p--A collisions at the SPS (\eg\ by the NA60 experiment at $\Ep=158$ and $450$~GeV~\cite{Arnaldi:2010ky}) has in particular often been attributed to nuclear absorption effects. 

Nuclear absorption effects are not included in this analysis for two reasons. First of all, the strength of nuclear absorption strongly depends on the (effective) absorption cross section $\sigma_{\mathrm{abs}}^{\psi}$ which is poorly constrained from data~\cite{Arleo:2006qk,Lourenco:2008sk}. Moreover, when $t_\psi \sim L$, the hierarchy \eq{hierarchy} upon which the medium-induced spectrum \eq{our-spectrum-3} relies is no longer valid and hence the use of the latter becomes dubious. In this paper we will therefore focus on the region $t_\psi \gg L$ (the region of validity of \eq{our-spectrum-3}), where the effect of nuclear absorption is irrelevant.

In the calculations presented in Section~\ref{sec:phenomenology}, we shall indicate by an arrow the value of $\xfcrit$ (or $\ycrit$), defined as $E(\xfcrit)/M \times \tau_\psi = L$, below which nuclear absorption might play a role. For the numerical values of $\xfcrit$, the $\jpsi$ hadronization time is given by the mass splitting between 1S and 2S states, 
$\tau_{\jpsi} = {(M_{\psi^\prime} - M_{\jpsi})}^{-1} \simeq (0.6 \,{\rm GeV})^{-1} \simeq 0.3$~fm (a similar estimate is obtained in the $\Upsilon$ channel). Note that $\xfcrit$ becomes negative at large collision energy, in which case the model should not only apply at large positive $\xf$ but also down to $\xf <0$.

\subsubsection{Saturation and nuclear PDF effects}
\label{sec:saturation}

At small values of $x$, partons inside the nucleus wavefunction start to overlap, leading to the phenomenon of saturation (see for instance~\cite{Gelis:2007kn} for a review). Although saturation effects should also occur in a proton, they are expected to scale roughly like the nucleus transverse density, $V/S\sim A^{1/3}$, therefore being stronger in large nuclei at a fixed value of $x$. As a consequence, the $\psi$ normalized yield in p--A collisions is likely to be suppressed with respect to that in p--p collisions --~independently of the energy loss effects discussed above~-- either at large $\xf$ and/or at high energies (RHIC, LHC) where small values of $x$ are probed in the target nucleus.

The effects of (gluon) saturation on $\jpsi$ suppression in p--A and A--A collisions have been addressed by many authors, see \eg~\cite{Kharzeev:2005zr,Fujii:2006ab}. In the present paper, we shall implement the physics of saturation following the work of Fujii, Gelis and Venugopalan~\cite{Fujii:2006ab}, where $\jpsi$ suppression has been computed within the Color Glass Condensate assuming 
$2\to1$ kinematics for the production process. The nuclear suppression is a scaling function of the saturation scale $Q_s$ and can be simply parametrised as~\cite{Fujii:2006ab}
\be
{\cal S}_{\mathrm A}^{\jpsi}(x_2, L) \simeq \left(1+ \frac{Q_s^2(x_2, L)}{b} \right)^{-\alpha},
\label{RpAsat}
\ee
with $b=2.65$~GeV$^2$ and $\alpha=0.417$. Unfortunately, no equivalent parametrization has been given in the $\Upsilon$ channel. We will assume in the present approach that saturation effects on heavy-quarkonium production are a scaling function of $Q_s / M_\perp$.\footnote{Ideally this {\it ansatz} should be checked numerically. Indeed, it can only be approximate, since running coupling effects will explicitly spoil this scaling hypothesis.} Therefore, the $\Upsilon$ suppression due to saturation reads
\be
{\cal S}_{\mathrm A}^{\Upsilon}(Q_s) = {\cal S}_{\mathrm A}^{\jpsi}(Q_s \times M_\perp^{\jpsi} \big/ M_\perp^{\Upsilon}).
\ee
In order to make reliable predictions at RHIC and LHC, the $\jpsi$ and $\Upsilon$ nuclear production ratio is determined assuming energy loss effects, $R_{\mathrm{pA}}^{\rm{E.loss}}$ from Eq.~(\ref{eq:xspA}), with and without saturation effects,
\bea
{\rm (i)}\qquad R_{\mathrm{pA}} &=& R_{\mathrm{pA}}^{\rm{E.loss}} \ \, ,\nonumber\\
{\rm (ii)}\qquad R^{\rm sat}_{\mathrm{pA}} &=& R_{\mathrm{pA}}^{\rm{E.loss}} \times {\cal S}_{\mathrm A} / {\cal S}_{\mathrm p} \ \,. \nonumber
\eea

The saturation scale appearing in (\ref{RpAsat}) is closely related to the transport coefficient $\hat{q}$ given by \eq{qhat-gluondensity}, namely \cite{Mueller:1999wm,Baier:2002tc}
\be
\label{Qs-qhat}
Q_s^2(x, L) = \hat{q} \, L \, .
\ee
In other words $Q_s^2$ is nothing but the transverse momentum broadening discussed in Section~\ref{sec:qhat}, see Eqs.~\eq{broadening} and \eq{broadening2}.
The inclusion of saturation effects thus does not require any additional parameter once the parametrization \eq{qhat-model} for $\hat{q}$ is employed and $\qzero$ is determined. 

Let us mention that $\hat{q}_0$ is related to the saturation momentum in a proton at $x=10^{-2}$, $Q_{s0}$. We have, from \eq{qhat-model} and \eq{Qs-qhat},
\be
Q_{s0}^2 = Q_s^2(x_2=10^{-2}, L_{\rm p}) = \hat{q}_0\ L_{\rm p} \simeq 0.1 \, {\rm GeV}^2  \left( \frac{\hat{q}_0}{0.06} \right) \, ,
\label{Qs-estimate}
\ee
with $\hat{q}_0$ in ${\rm GeV}^2/{\rm fm}$. Comparing the value of $\hat{q}_0$ obtained in our model from a fit to the E866 $\jpsi$ nuclear suppression data and the current estimates of $Q_{s0}^2$ obtained from a fit to small-$x_2$ DIS data \cite{Albacete:2010sy} should provide a non-trivial (though not conclusive) test of our model. 

Another, earlier approach in order to model the modifications of parton densities in nuclei is the use of leading-twist nuclear PDF (nPDF) which have been determined from global fit analyses of e--A DIS or p--A Drell-Yan data for more than a decade (see \eg~\cite{Armesto:2006ph}). In this framework, $\psi$ production in p--A collisions is proportional to the gluon distribution in the nucleus $G^{\rm A}(x_2, \mT)$. Therefore, $\psi$ suppression can be modelled as
\ben
{\rm (iii)}\qquad R_{\mathrm{pA}}^{\rm nPDF} = R_{\mathrm{pA}}^{\rm{E.loss}} \times G_{\rm A}^{\rm nPDF}(x_2, \mT) / G_{\rm p}(x_2, \mT)\ .
\een
The predictions to be discussed in the next section will be performed assuming energy loss effects, supplemented with predictions including saturation effects at RHIC and LHC energies where these are expected to play a role. For completeness, we will also critically compare in Section~\ref{sec:npdfs} these results with those obtained using nPDF.

\section{Phenomenology}
\label{sec:phenomenology}

After the description of the energy loss model, the phenomenology of $\psi$ suppression in hadron--nucleus collisions is investigated in this section. In the practical applications, we take $\Lambda_{_{\rm QCD}}=0.25\,{\rm GeV}$, $\pt=1\,{\rm GeV}$ in the transverse mass $M_\perp=\sqrt{M^2+p_\perp^2}$, and $M=3\,{\rm GeV}$ ($M=9\,{\rm GeV}$) for the mass of a compact $c\bar{c}$ ($b\bar{b}$) pair. 
As we can easily verify a posteriori, the typical scale entering the running coupling constant is not too large, $\hat{q} L\sim1\,{\rm GeV}^2$, which justifies the assumption of a frozen coupling, $\alpha_s=1/2$, at such semi-hard scales.

\subsection{Fitting procedure}
\label{sec:procedure}

The only parameter of the model, the transport coefficient $\qzero$, is determined by fitting the $\jpsi$ suppression measured by E866~\cite{Leitch:1999ea} in p--W over p--Be collisions ($\sqrt{s}=38.7$~GeV). This choice is motivated by the fact that the E866 measurements are the most precise performed so far and cover a wide range in $\xf$. We choose to perform the fit in the [0.2--0.8] $\xf$-range for the following reasons: at $\xf \lesssim 0.2$ $\jpsi$ suppression might be affected by nuclear absorption (see Section~\ref{sec:absorption}) while at $\xf\gtrsim0.8$, we expect 
quark-induced subprocesses 
to come into play, possibly modifying the overall normalization of the medium-induced spectrum \eq{our-spectrum}. Note also that this $\xf$-range at E866 energy corresponds to values of $\xtwo \gtrsim 10^{-2}$ for which saturation effects are expected to be small, of the order of 5\% at most on the W/Be ratio.

The fit gives $\qzero=0.075\pm0.005$~\gevsqfm, where the quoted uncertainty is determined from the $\chi^2$ minimization procedure. A systematic uncertainty on the value of $\qzero$ can be roughly estimated by restricting the $\xf$-range used for the fit
to the interval [0.3--0.7]; we found that it would increase the value of $\qzero$ 
to $\qzero \simeq 0.087$~\gevsqfm.

\begin{figure}[htbp]
\begin{center}
    \includegraphics[width=7.8cm]{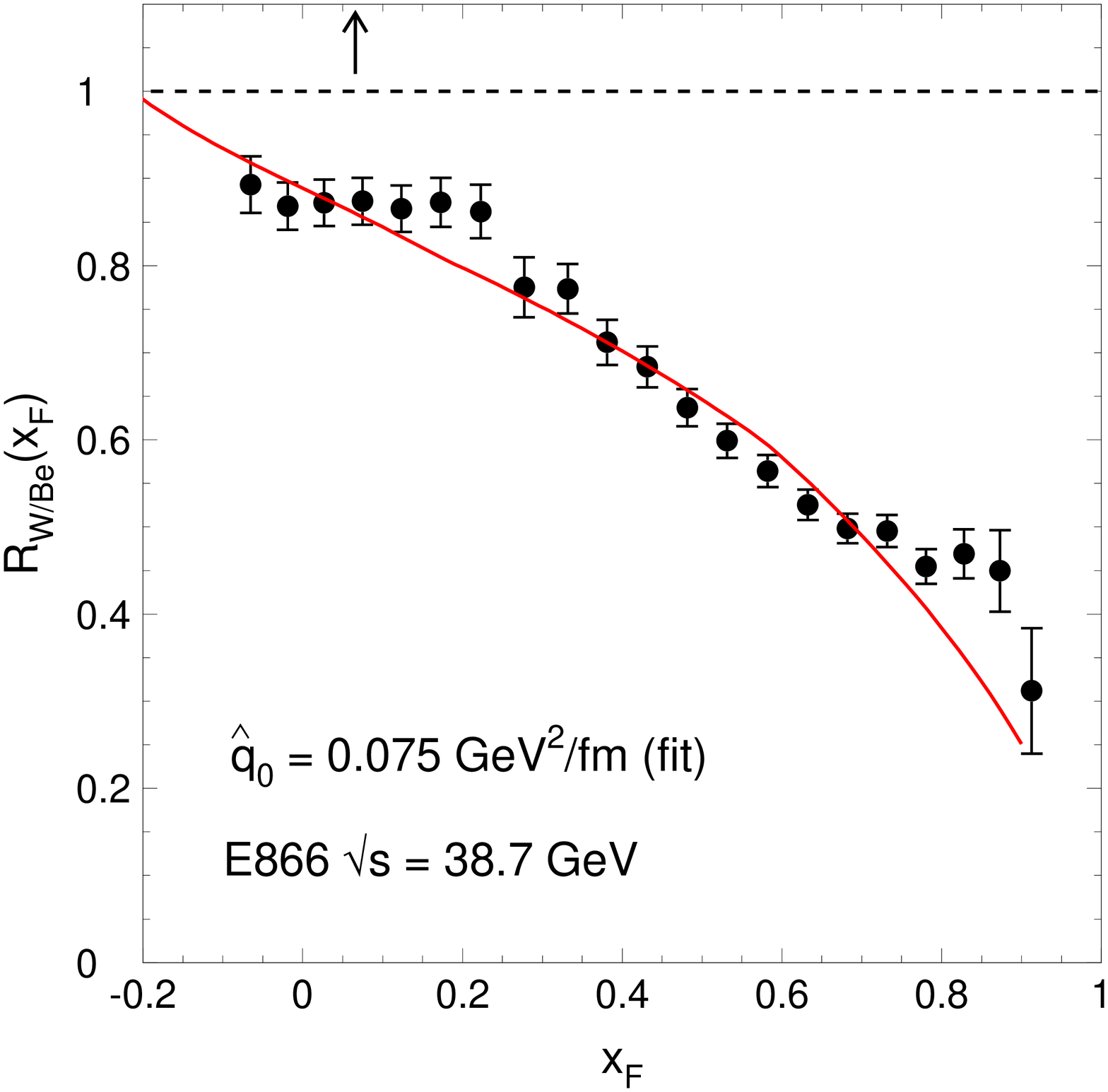}
  \end{center}
  \vspace{-1.cm}
\caption{E866 $\jpsi$ suppression data~\cite{Leitch:1999ea} in p--W collisions compared to the energy loss model.}
  \label{fig:wbe_e866}
\end{figure}

The result of the fit is shown in Fig.~\ref{fig:wbe_e866} where excellent agreement is observed in the whole fit range. 
 Note however that for $\xf\lesssim0.1$, nuclear absorption is expected to play a role; see the vertical arrow at $\xfcrit\simeq0.07$ below which the $\jpsi$ formation time becomes smaller than the size of the target tungsten nucleus (see Section~\ref{sec:absorption}).

It is worth mentioning that the fitted transport coefficient, $\qzero=0.07$--$0.09$~\gevsqfm, would correspond to the saturation scale in a proton $Q_s^2(x=10^{-2})=0.11$--$0.14 \ \rm{GeV}^2$ using \eq{Qs-estimate}, which is consistent with (yet slightly smaller than) estimates based from fits to $F_2$ DIS data~\cite{Albacete:2010sy}. Note that the saturation scale in large nuclei and at smaller $x$ considerably exceeds that in a proton, yielding $\hat{q} L \sim$1~GeV$^2$, where the use of perturbative techniques is commonly assumed to be legitimate.

\subsection{Scaling properties of heavy-quarkonium suppression}
\label{sec:scaling}

Before comparing the model predictions to the other available data, we discuss in this section the expected scaling properties of $\psi$ suppression in the present model. 

Let us first mention that the nuclear dependence of quarkonium suppression is often pa\-ra\-me\-tri\-sed as a power law,
\be
\label{standard-fit}
\frac{\dd\sigma_{\mathrm{pA}}^{\psi}}{\dd \xf} = A^{\alpha} \, \frac{\dd\sigma_{\mathrm{pp}}^{\psi}}{\dd \xf}\ \  \Rightarrow \ \ R_\mathrm{pA}^{\psi} = \, A^{\alpha -1},
\ee
where $\alpha$ is assumed to be independent of $A$. The power law is empirical. It can be
inferred in the Glauber picture of $\psi$ {\it absorption} in the nucleus, $S^{\rm abs} \simeq \exp{\left( - {\rm cst}\cdot A^{1/3} \right)}$, and using the approximation $A^{1/3} \simeq \log{A}$, which is accurate to the 10\% level for $5 \leq A \leq 200$. However, the Glauber picture of nuclear absorption is expected to hold when the $\psi$ energy $E$ in the nucleus rest frame is small enough, see Section \ref{sec:absorption}. The heuristic law \eq{standard-fit} has no reason to be valid at high $E$ where the compact color octet $Q \bar{Q}$ pair crosses the nucleus and hadronizes far beyond.

We checked that the $\jpsi$ suppression expected in our model does not follow the pa\-ra\-me\-tri\-za\-tion \eq{standard-fit}. To illustrate this, the typical values of $\alpha$ are found to vary by up to 10\% depending on whether $\jpsi$ suppression in p--W collisions (in the E866 kinematics) is normalized either to p--p or p--Be collisions. Clearly the attenuation factor $R_{\mathrm{pA}}$ should be preferred to the effective power $\alpha$ when discussing nuclear suppression. We thus focus on $R_{\mathrm{pA}}$ rather than on $\alpha$ throughout our study, and now discuss its scaling properties. 

In the energy loss model of Gavin and Milana~\cite{Gavin:1991qk}, quarkonium suppression exhibits an approximate $\xf$ scaling, \ie, $R_{\rm pA}$ is a function of $\xf$ but independent of $\sqrt{s}$. 
Indeed, assuming that the shape of $\dsigpp$ is independent of $\sqrt{s}$, and considering the limit $\xf\gg\mT/\sqrt{s}$ where $E \simeq \xf E_\mathrm{p}$, we obtain from \eq{shift-frag} 
\be
\label{shift-frag-approx}
\frac{1}{A}\frac{\dd\sigma_{\mathrm{pA}}^{\psi}}{\dd \xf}\left(\xf \right) \simeq \int^1_{z_{min}} \dd z  \,{\cal F}_{\mathrm{loss}}(z) \, \frac{\dd\sigma_{\mathrm{pp}}^{\psi}}{\dd\xf} \left(\frac{\xf}{z}\right)\, .
\ee
In the present approach, however, the approximate $\xf$ scaling of quarkonium suppression is broken for several reasons:
\begin{enumerate}
\item The transport coefficient $\hat{q}$, and therefore the function ${\cal F}_{\mathrm{loss}}$ in \eq{shift-frag-approx}, depends ex\-pli\-citly on $x_2$ at small $x_2<x_0$, see \eq{qhat-model}. As we shall see, this effect is particularly important at LHC energies;
\item As discussed in Section~\ref{sec:xspp}, the slope of the p--p production cross section does depend on $\sqrt{s}$ (see Tables~\ref{tab:expnjpsi} and \ref{tab:expnups});
\item Finally, the saturation (or nPDF) effects also scale with $x_2$  yet this effect is actually rather small.
\end{enumerate}

\begin{figure}[t]
\begin{center}
    \includegraphics[width=4.5cm]{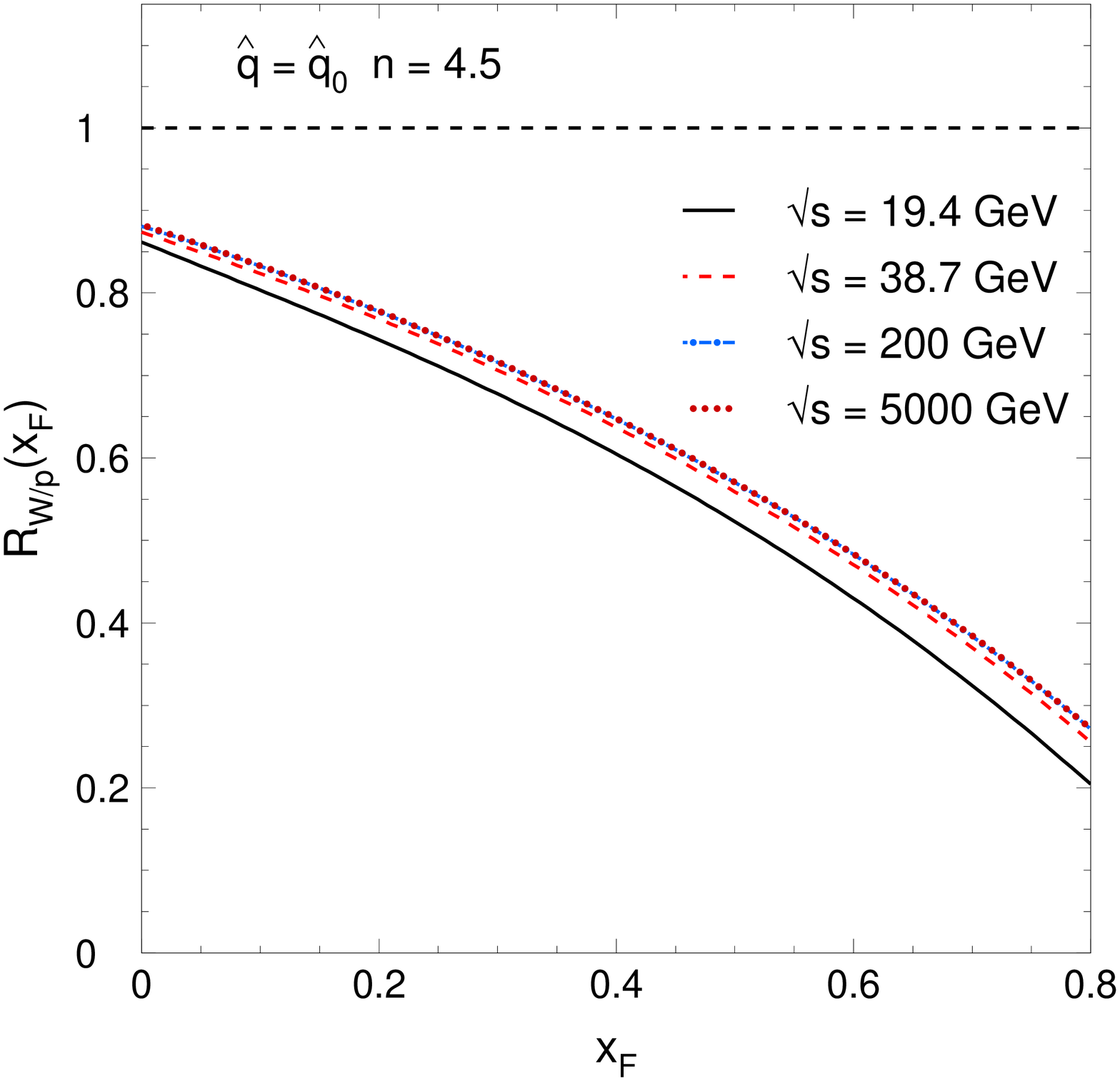} \ \ \ 
    \includegraphics[width=4.5cm]{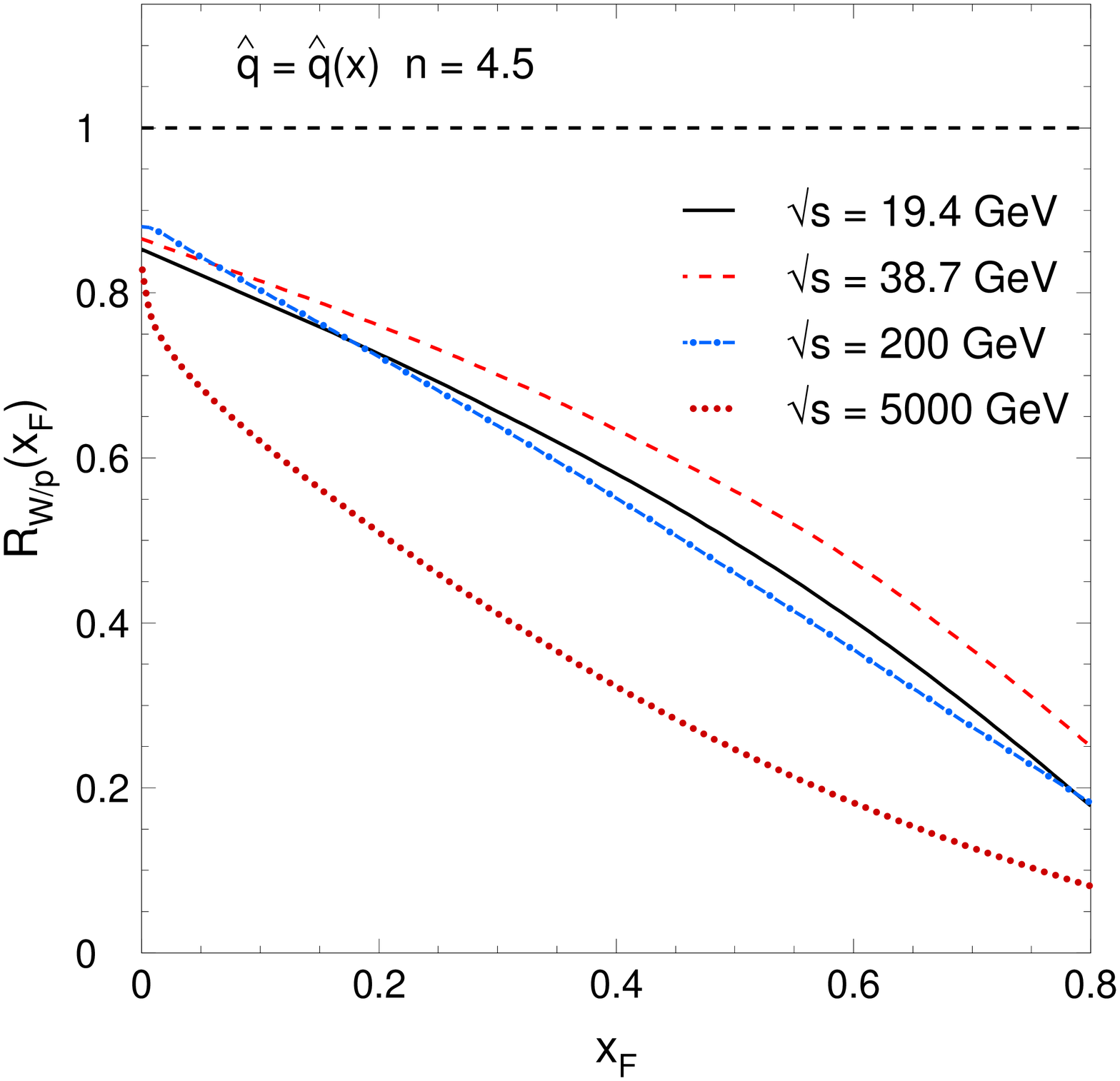} \ \ \ 
    \includegraphics[width=4.5cm]{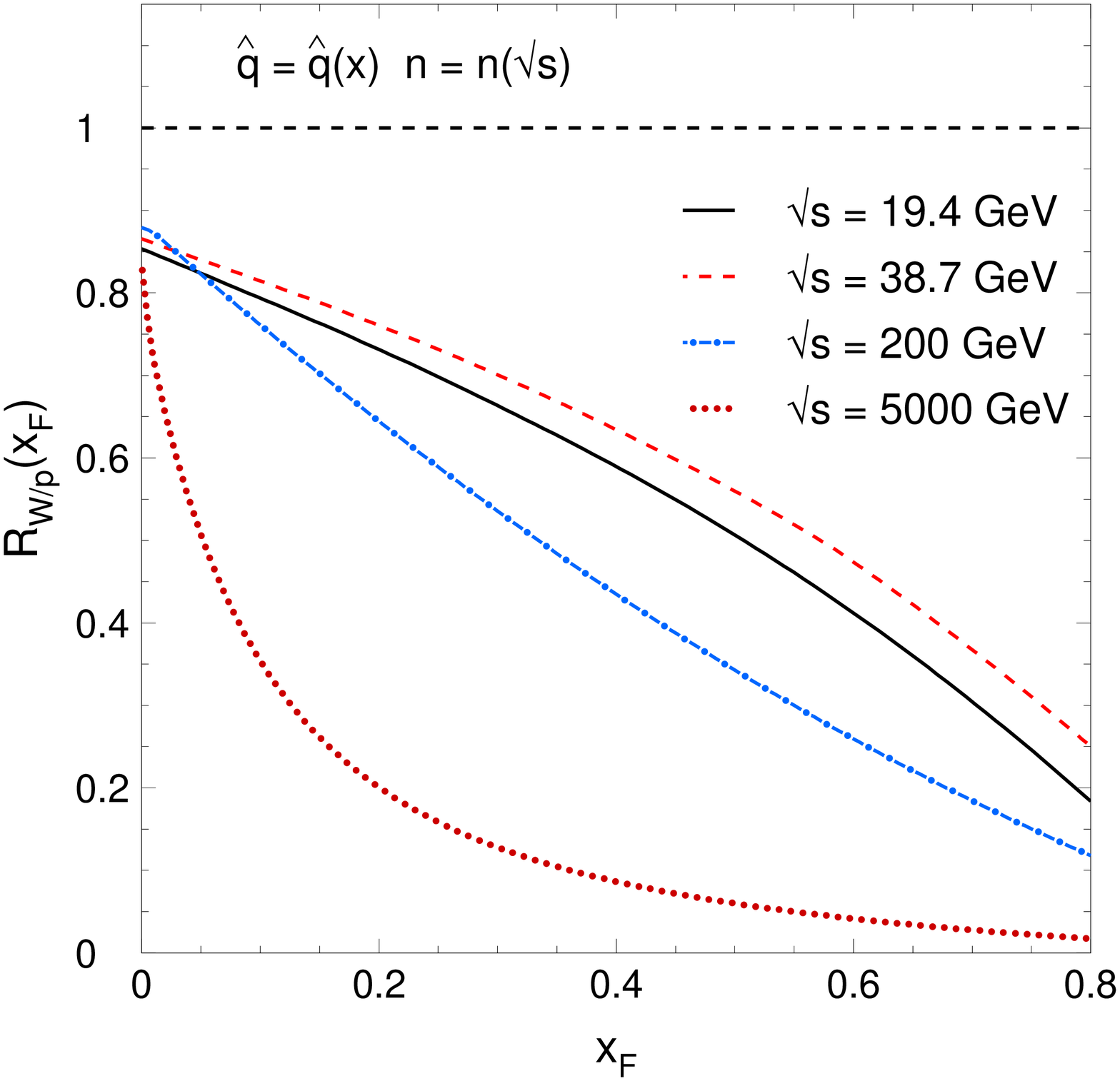} \ \ \ 
  \end{center}
\vspace{-1.cm}
\caption{Scaling of $\jpsi$ suppression predicted in p--W collisions in the range $\sqrt{s}=19.4$--$5000$~GeV for various assumptions regarding the transport coefficient $\hat{q}$ and the value of the exponent $n$. See text for details.}
  \label{fig:scaling}
\end{figure}

In order to illustrate this, $\jpsi$ suppression has been computed in p--W collisions as a function of $\xf$ at NA3 ($\sqrt{s}=19.4$~GeV), E866 ($\sqrt{s}=38.7$~GeV), RHIC 
($\sqrt{s}=200$~GeV) and LHC ($\sqrt{s}=5$~TeV) energies in Fig.~\ref{fig:scaling}, without saturation effects but under various assumptions. In the left panel, the transport coefficient is frozen, $\hat{q}(x)=\qzero=0.075$~\gevsqfm, and the exponent of the p--p cross section is fixed to $n=4.5$. With no surprise the $\xf$ scaling is observed to a very good accuracy, except at the lowest $\sqrt{s}$ for which the approximation $E\simeq\xf E_{\rm p}$ is no longer valid. When taking explicitly into account the $x$ dependence of $\hat{q}(x)$ but keeping a fixed exponent $n=4.5$ (central panel), the $\xf$ scaling is strongly violated at LHC energy, but still approximately verified from NA3 up to RHIC energies. Finally, the deviations from $\xf$ scaling are even more pronounced (right panel) when considering the actual exponents $n$ extracted at each $\sqrt{s}$ in Section~\ref{sec:xspp}, with a stronger suppression at RHIC ($n=8.3$) and LHC ($n=32.3$).

Note that at LHC, the variation of $R_\mathrm{pA}^{\jpsi}$ with $\xf$ is extremely fast
at very small $\xf$.
This strong dependence comes from the small-$x$ behavior of the transport coefficient $\hat{q}(x)$, Eq.~\eq{qhat-model}, together with the fast variation of $x=\xtwo$ with $\xf$ at $|\xf|\lesssim 10^{-2}$.
As we shall see in Section~\ref{sec:lhc} the variation of $R_\mathrm{pA}^{\jpsi}$ with $y\sim\ln\xf$ is naturally much smoother.

In order to check experimentally whether $\jpsi$ suppression scales with $\xf$, it would be crucial to measure $\jpsi$ production in p--A collisions at RHIC and LHC at large $\xf$, say $\xf\gtrsim0.1$, which is out of reach with the present apparatus. Such measurements could in particular shed light on the $x$ dependence of the transport coefficient $\hat{q}(x)$.


\subsection{Predictions and comparison to $\jpsi$ data}
\label{sec:jpsidata}

Once $\qzero$ is determined from the fitting procedure described in Section~\ref{sec:procedure}, the $\xf$ dependence of the $\jpsi$ quenching factor $R_{\mathrm{pA}}^{\psi}$ can be predicted in any target nucleus and at any center-of-mass energy for which the absolute p--p differential cross section $\dsigpp$ has been measured. In this section we systematically compare the model predictions with all available data. 

\subsubsection{E866, NA3, E537, NA60, HERA-B}\label{se:E866}

\begin{figure}[ht]
\begin{center}
    \includegraphics[width=7.8cm]{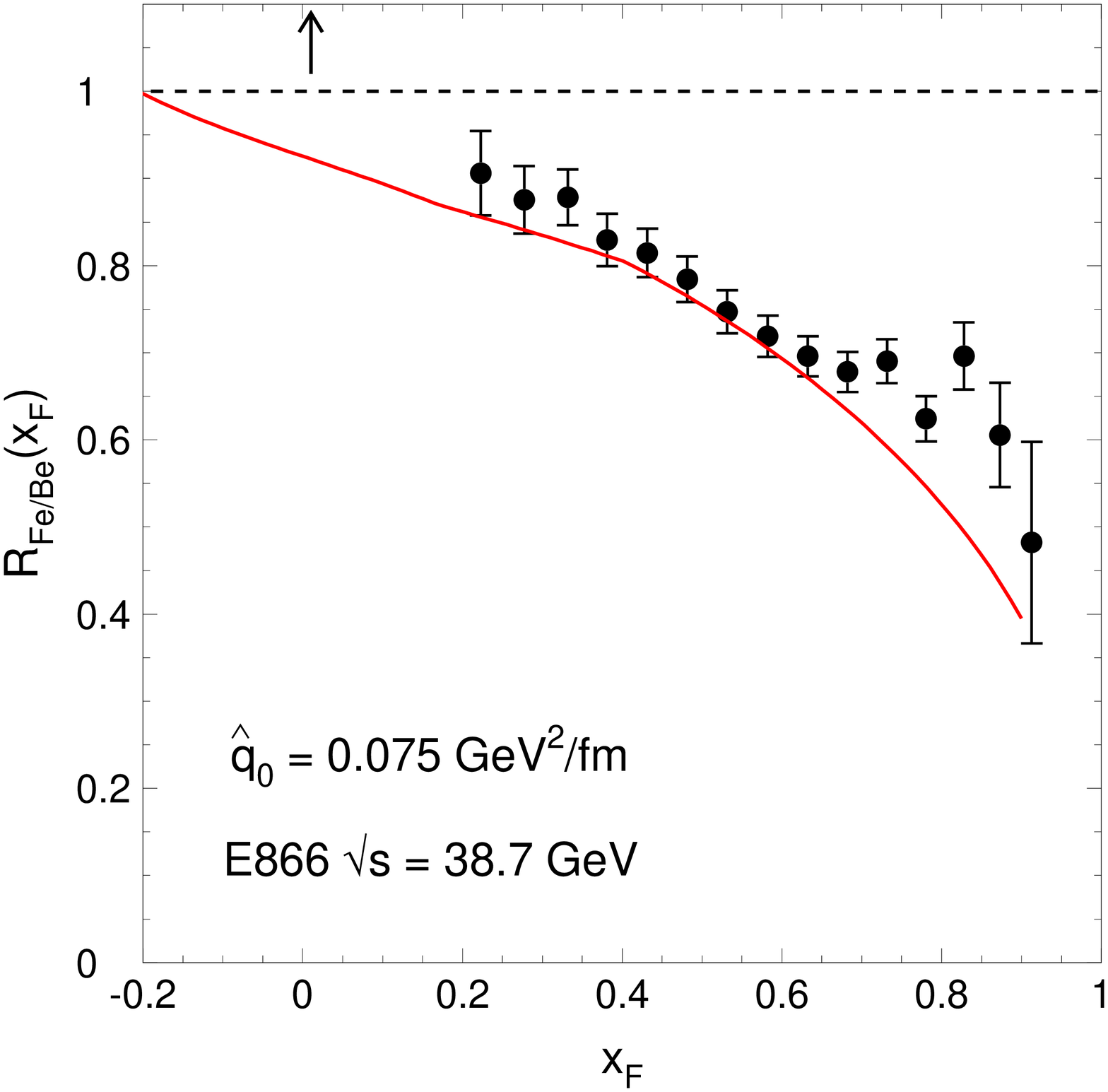}
  \end{center}
  \vspace{-1.cm}
\caption{E866 $\jpsi$ suppression data~\cite{Leitch:1999ea} in p--Fe collisions compared to the energy loss model.}
\label{fig:febe_e866}
\begin{center}
    \includegraphics[width=14.cm]{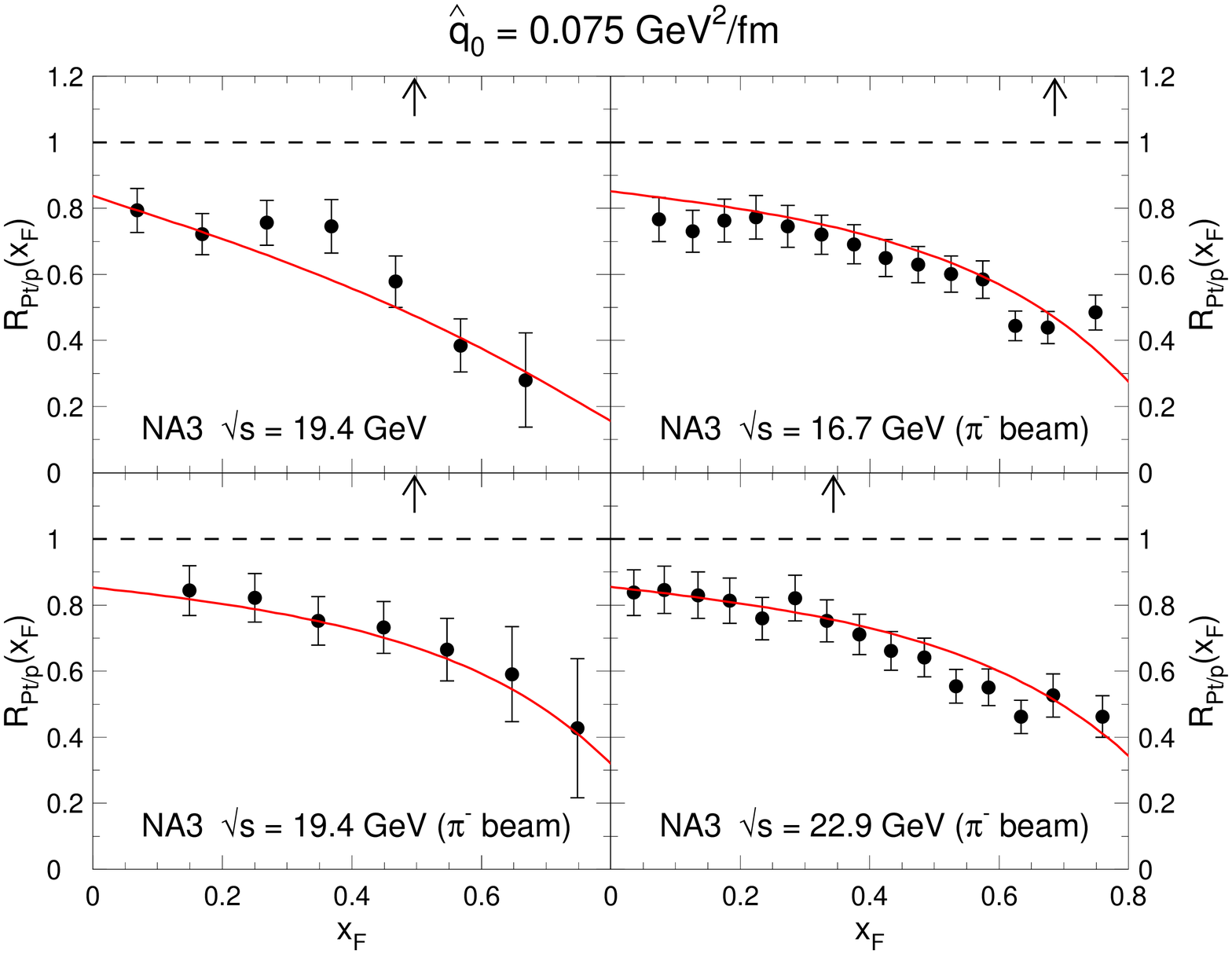}
  \end{center}
  \vspace{-1.cm}
\caption{NA3 $\jpsi$ suppression data~\cite{Badier:1983dg} in p--A and $\pi^-$--A collisions compared to the energy loss model.}
  \label{fig:na3}
\end{figure}

Let us start with the comparison of $\jpsi$ suppression expected in an iron target
and the E866 data for $R_{\rm{Fe/Be}}$, 
\ie, taken at the same energy as the fitted ratio $R_{\rm{W/Be}}$. The excellent agreement reported in Fig.~\ref{fig:febe_e866} fully supports the atomic mass dependence of the model. This is at variance with the calculations by Gavin and Milana~\cite{Gavin:1991qk} which overestimated the ratio $R_{\rm{Fe/Be}}$, at that time measured by E772.\footnote{We do not show the agreement between our model predictions and $\jpsi$ E772 data~\cite{Alde:1990wa} since those measurements were superseded by E866~\cite{Leitch:1999ea}.} It is therefore a hint that the $L$-dependence expected here, $\Delta E \propto \sqrt{L}$ 
(see \eq{mean-delta-E}), is probably more appropriate than the {\it ad hoc} assumption of Ref.~\cite{Gavin:1991qk}, $\Delta E \propto L$.
 
Data taken at lower $\sqrt{s}$ or smaller $\xf$ are also compared to the model. As can be seen in Fig.~\ref{fig:na3} the agreement with NA3 p--A and $\pi^-$--A data is excellent, both in shape and magnitude, over a very wide range in $\xf$. It is also remarkable that the model is able to reproduce the different magnitude of suppression in \pA\ and $\pi^-$--A collisions reported by NA3~\cite{Badier:1983dg}. This difference cannot be understood within nuclear absorption models, where nuclear suppression is a purely final state effect, thus independent of the projectile type. It cannot either be explained by nPDF
effects, unless the nPDF to proton PDF ratios for valence quarks and for gluons, probed respectively in $\pi^-$--A and \pA\ collisions, prove completely different.\footnote{On top of this, nPDF effects in the NA3 kinematical domain, $x_2\sim0.1$--$0.2$, are known to be small for both valence quarks and gluons, see for instance the discussion in~\cite{Arleo:2002ph}.} In our picture, the smaller $\jpsi$ suppression in $\pi^-$--A collisions naturally arises from the flatter differential cross section, $n_{\pi{{\mathrm{p}}}}=1.4$ vs. $n_{{\mathrm{p}}{\mathrm{p}}}=4.3$ at $\sqrt{s}=19.4$~GeV, see~Table~\ref{tab:expnjpsi}. Although no prediction of the exponent $n$ is made in our model, it is clear that this feature can be explained from the larger slope, at large $x$, of the gluon PDF in a proton, 
$xG(x)\sim(1-x)^3$~\cite{Martin:2009iq}, when compared to that of a valence antiquark PDF in a pion, $x\bar{q}(x)\sim(1-x)$~\cite{Sutton:1991ay}.
In this respect, let us mention that our assumption of an incoming {\it gluon} in quarkonium hadroproduction (see the Introduction) does not hold for NA3 {\it pion}-nucleus collisions, where subprocesses with an incoming valence antiquark dominate. In spite of this, a very good agreement between the model and the NA3 $\pi^-$--A data is found, suggesting a mild dependence of the energy loss on the incoming parton type.  A similar remark applies to the case of the $\pi^-$--A E537 data discussed below.

\begin{figure}[h]
  \begin{center}
    \includegraphics[width=7.cm]{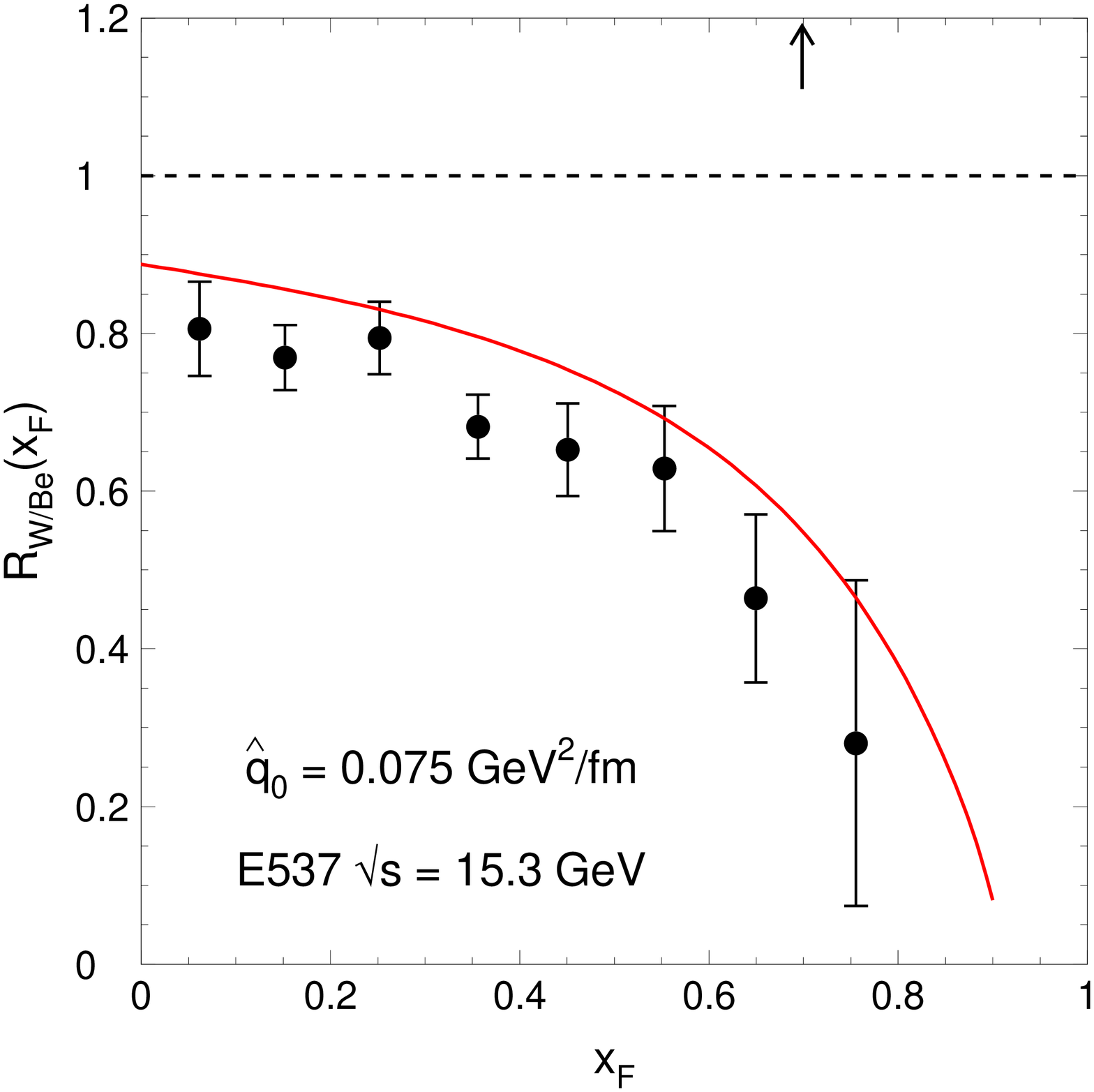}
    \includegraphics[width=7.cm]{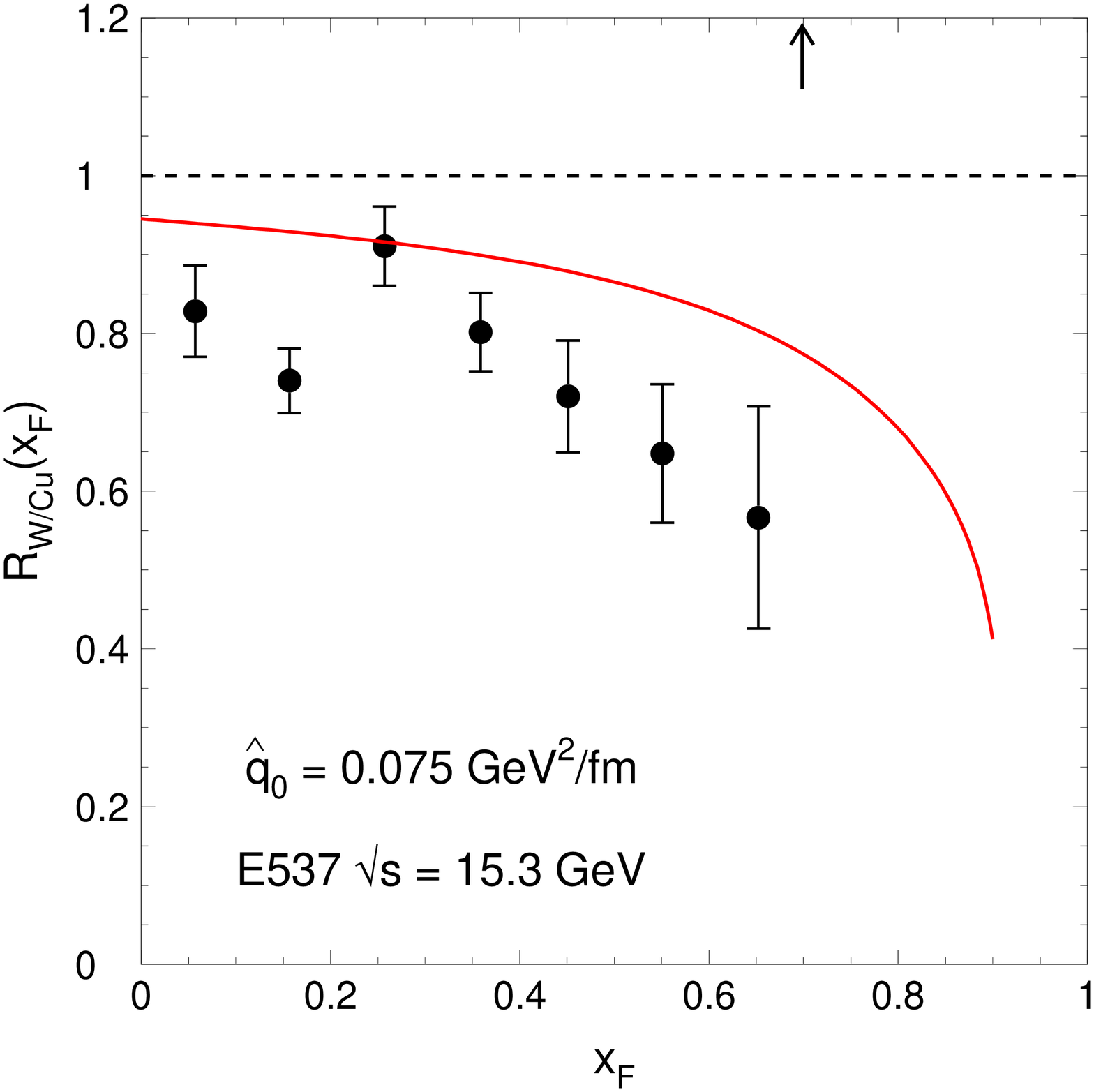}
  \end{center}
\vspace{-0.8cm}
\caption{E537 $\jpsi$ suppression data~\cite{Katsanevas:1987pt} in $\pi^-$--A collisions compared to the energy loss model.}
  \label{fig:e537}
\end{figure}

The E537 experiment also reported on measurements of $\jpsi$ production in $\pi^-$ induced collisions on various nuclear targets (Be, Cu, W) at $\sqrt{s}=15.3$~GeV~\cite{Katsanevas:1987pt}. 
Our results\footnote{Lacking $\pi^-$--p data at E537 energy, we choose the exponent $n=1.4$ (see Table~\ref{tab:expnjpsi}).} are found in reasonable agreement with the measured ratio $R_{\mathrm{W/Be}}$ (Fig.~\ref{fig:e537}, left); the 
slight underestimation of the suppression by the model  
might be attributed to nuclear absorption. 
Indeed, at this energy $\xfcrit\simeq0.7$, and all E537 data lie in the $\xf \leq \xfcrit$ domain. This might also explain 
the (more pronounced) difference between the observed and predicted magnitudes of the 
ratio $R_{\mathrm{W/Cu}}$ (Fig.~\ref{fig:e537}, right).

\begin{figure}[t]
  \begin{center}
    \includegraphics[width=7.cm]{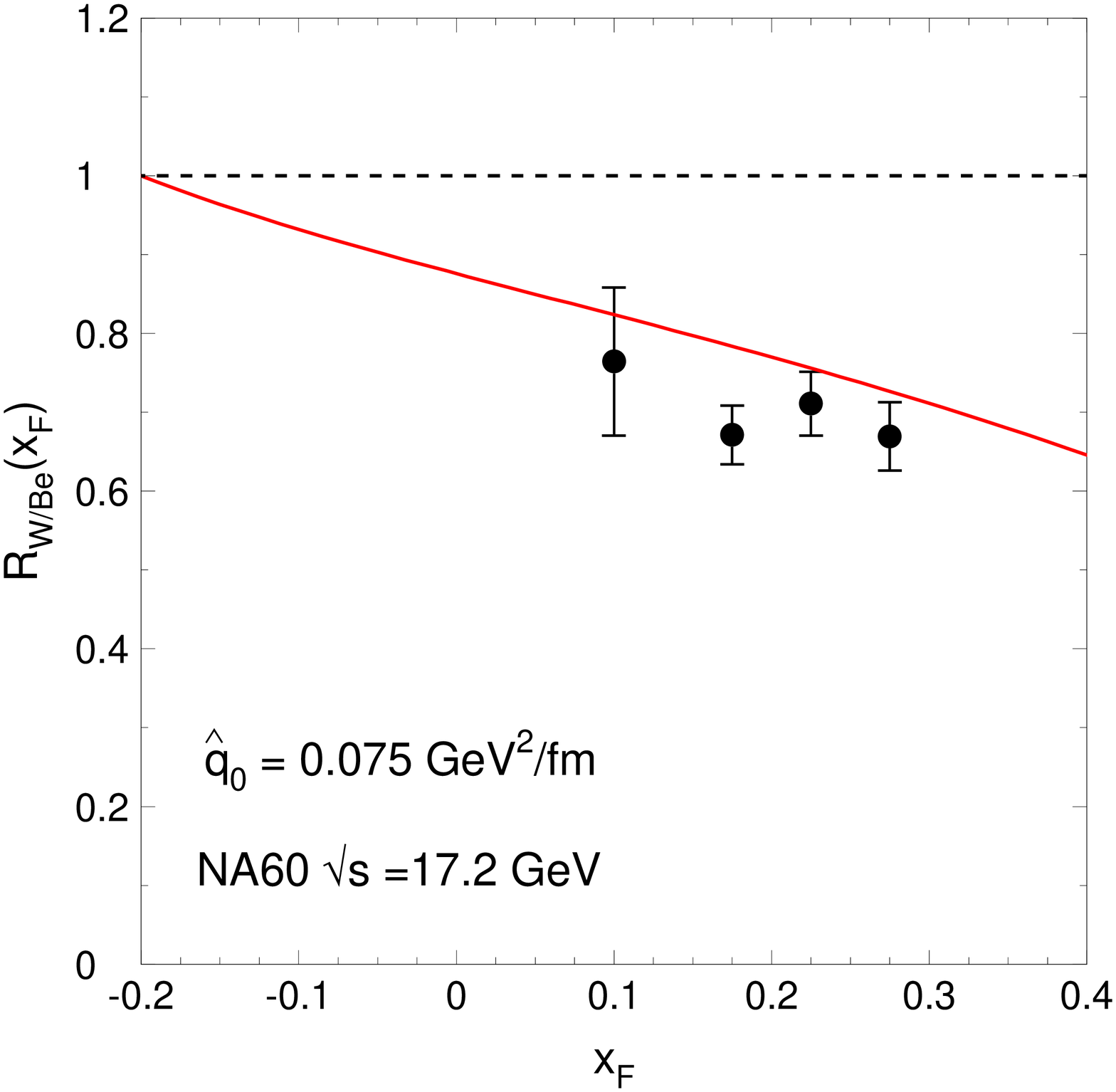}
    \includegraphics[width=7.cm]{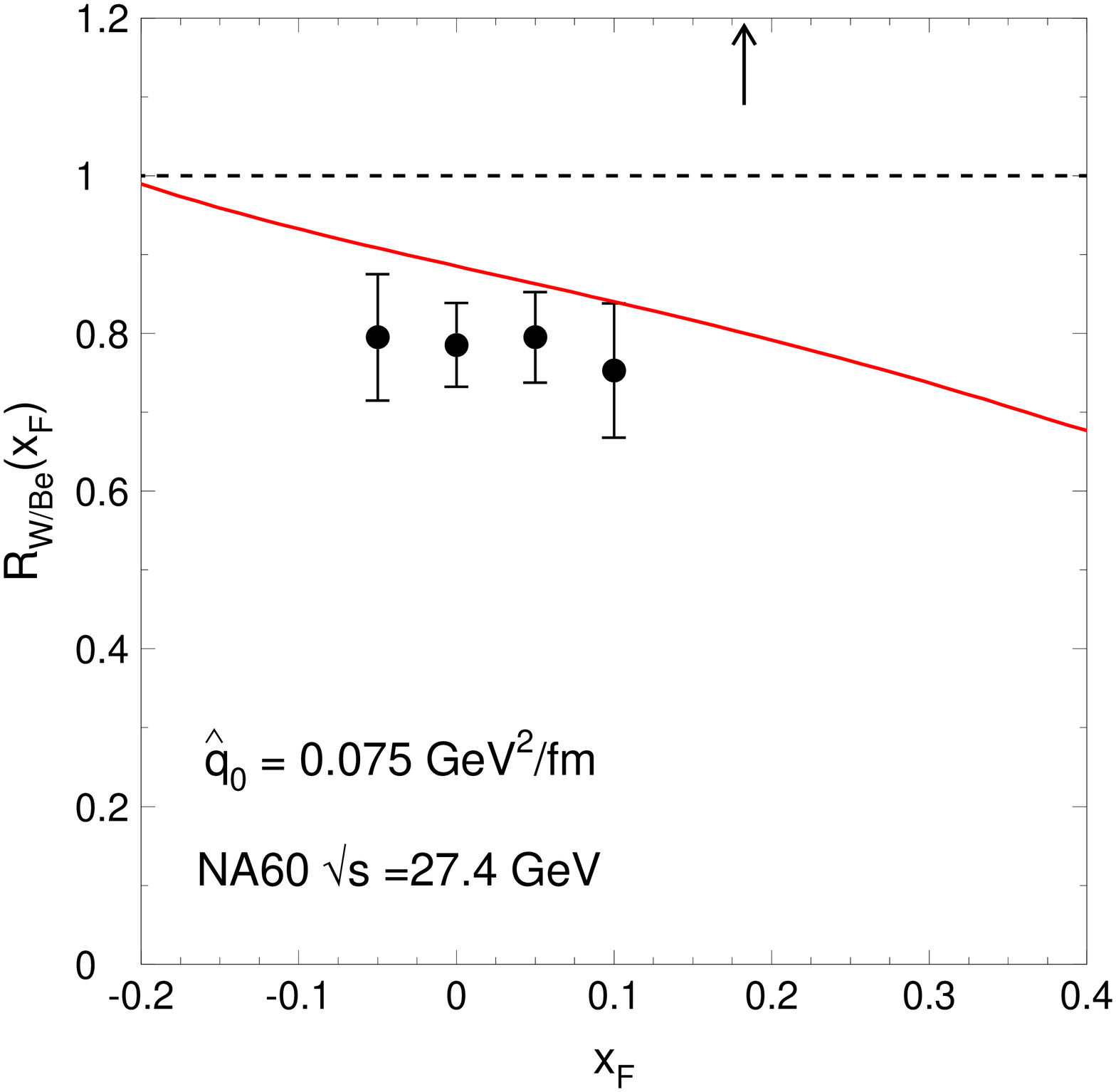}
  \end{center}
\vspace{-0.8cm}
\caption{NA60 $\jpsi$ suppression data~\cite{Arnaldi:2010ky} in p--A collisions compared to the energy loss model.}
  \label{fig:na60}
\end{figure}

\begin{figure}[ht]
  \begin{center}
    \includegraphics[width=7.2cm]{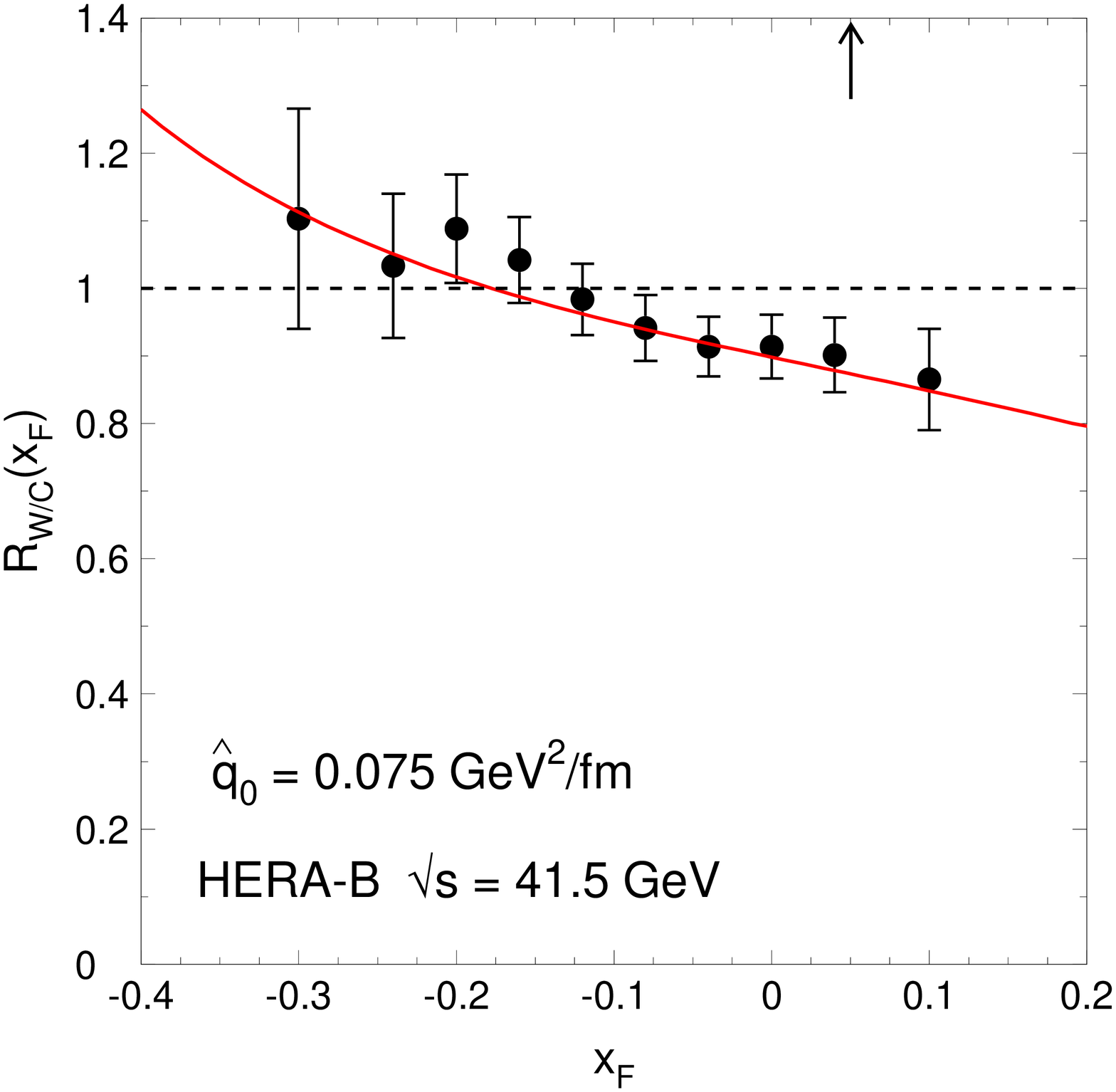}
  \end{center}
\caption{HERA-B $\jpsi$ suppression data~\cite{Abt:2008ya} in p--A collisions compared to the energy loss model.}
  \label{fig:herab}
\end{figure}

In Figs.~\ref{fig:na60} and \ref{fig:herab} we compare our predictions with NA60~\cite{Arnaldi:2010ky}\footnote{Lacking p--p data at NA60 energies, we choose the exponent $n=4.3$ (see Table~\ref{tab:expnjpsi}).} and HERA-B~\cite{Abt:2008ya} \pA\ 
measurements. Although the center-of-mass energy is larger than those of NA3 and E537, 
the typical $\jpsi$ energy range covered by NA60 and HERA-B is actually {\it lower} 
because of the smaller $\xf$ values probed by these experiments. As a consequence, $\jpsi$ suppression can be affected more strongly by nuclear absorption effects, as can be inferred by the position of the $\xfcrit$ arrows in Figs.~\ref{fig:na60} and~\ref{fig:herab}, below which hadronization typically takes places inside the nuclear target.\footnote{In the left panel of Fig.~\ref{fig:na60}, the arrow is not visible as $\xfcrit > 0.4$.} 


Nevertheless, the model predictions prove in very good agreement with data. In particular, the enhancement of $\jpsi$ production observed by HERA-B at very negative $\xf$, $\xf \lesssim -0.2$ (see Fig.~\ref{fig:herab}) is well reproduced by the model. 
(The origin of $R_{\rm pA}>1$ can be simply understood from the {\it positive} slope of $\dsigpp$ in the target fragmentation region, $\xf<0$, see the HERA-B data in Fig.~\ref{fig:fitjpsi}.)
There is however room for $\jpsi$ absorption with a cross section of a few millibarns, 
as suggested by the slight overprediction of $R_{\rm pA}$ 
at NA60 precisely in the region where $\jpsi$ absorption can no longer be neglected.

\subsubsection{PHENIX}
\label{sec:RHIC}

\begin{figure}[htbp]
  \begin{center}
    \includegraphics[width=7.8cm]{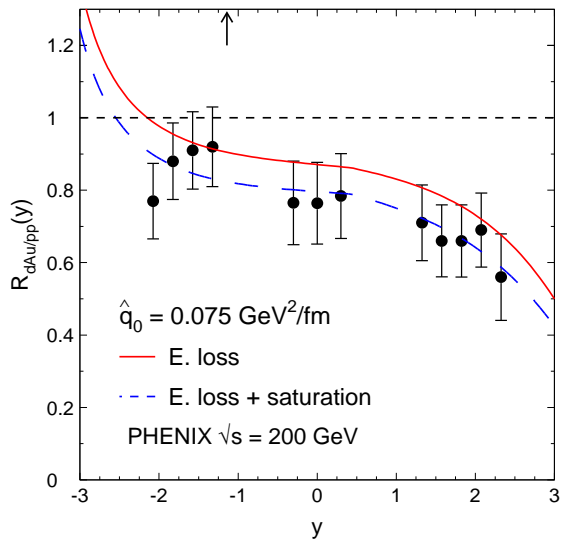}
   \end{center}
 \caption{PHENIX $\jpsi$ suppression data~\cite{Adare:2010fn} in d--Au collisions compared to the energy loss model, with (dashed line) and without (solid line) saturation effects.}
   \label{fig:rhic}
\end{figure}

The predictions in d--Au collisions at RHIC, $\sqrt{s}=200$~GeV, are shown in Fig.~\ref{fig:rhic} in comparison with PHENIX data~\cite{Adare:2010fn}, with (dashed line) and without (solid line) saturation effects. The energy loss model is able to reproduce nicely the $\jpsi$ suppression at all rapidities. Note that in several phenomenological analyses, the suppression observed in the most forward rapidity bins has often been attributed to gluon saturation effects or to strong small-$x$ shadowing in the nuclear PDF (see \eg~\cite{Vogt:2004dh,Kharzeev:2005zr}). Here, the sole energy loss effects might be responsible for the observed suppression, although saturation might play a role as well. As a matter of fact, 
the agreement is better when saturation is included. Remarkably, an excellent agreement is also observed in some 
negative $y$ bins, for which nuclear absorption might also play a role (at least for $y<\ycrit=-1.1$). We shall discuss further these data in Section~\ref{sec:npdfs} when comparing to the predictions including nPDF effects.

The $\jpsi$ suppression has also been measured by PHENIX for various d--Au centrality classes~\cite{Adare:2010fn} and more recently as a function of its transverse momentum~\cite{Adare:2012qf}. Discussing these data would go beyond the scope of the present article and is left for future work.

\subsection{Predictions and comparison to $\Upsilon$ data}
\label{sec:upsilondata}

The above comparison between $\jpsi$ suppression data and our model predictions supports both the medium length and energy dependence of the model. The mass dependence of heavy-quarkonium suppression can be studied by investigating the suppression of $\Upsilon$ states in p--A collisions. Unfortunately the data are rather scarce; to our knowledge, the measurements have only been performed by E772 at Fermilab~\cite{Alde:1991sw} and PHENIX and STAR~\cite{Adare:2012bv,Reed:2011zza} at RHIC.

\begin{figure}[h]
  \begin{center}
    \includegraphics[width=4.9cm]{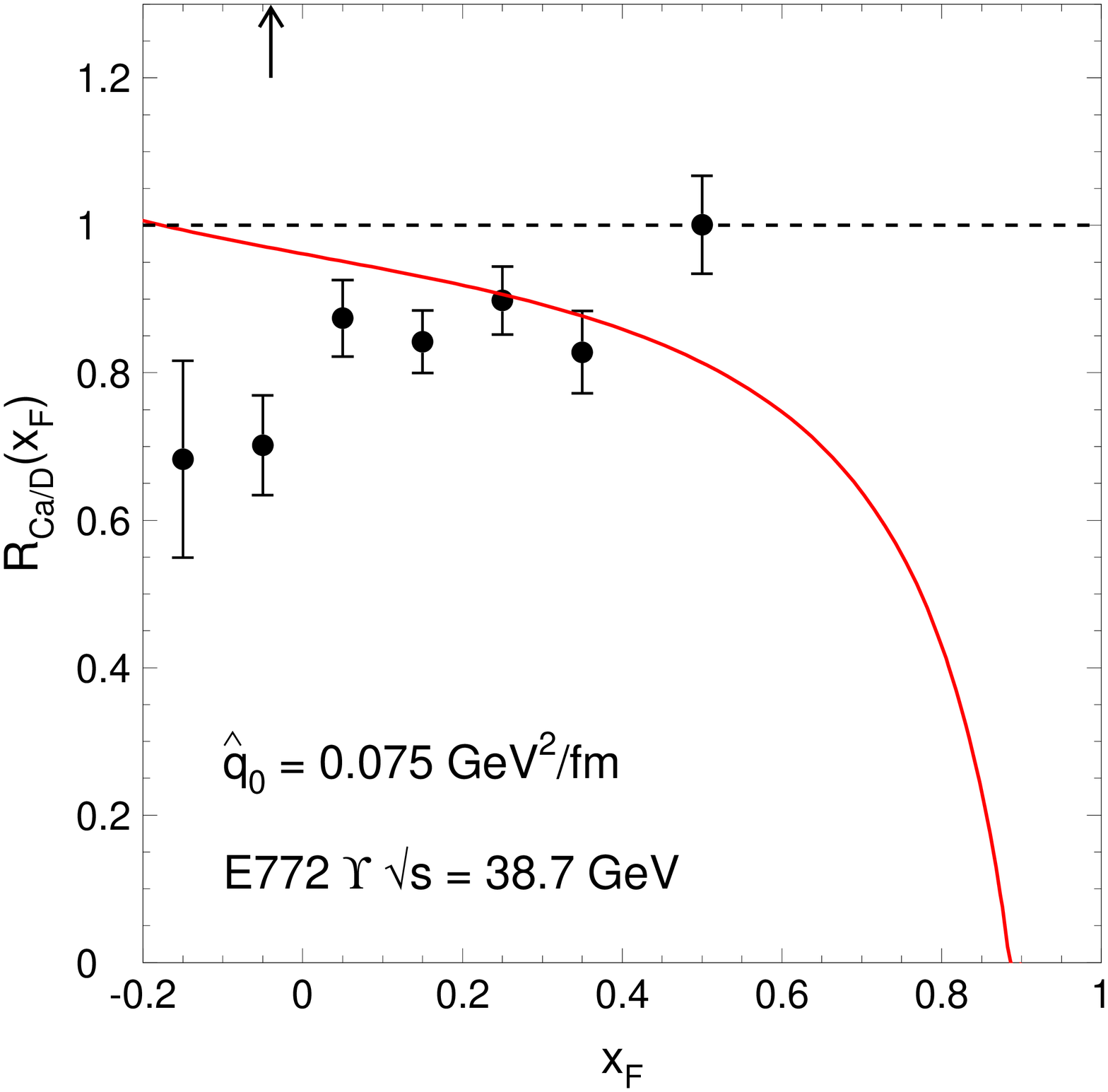}
    \includegraphics[width=4.9cm]{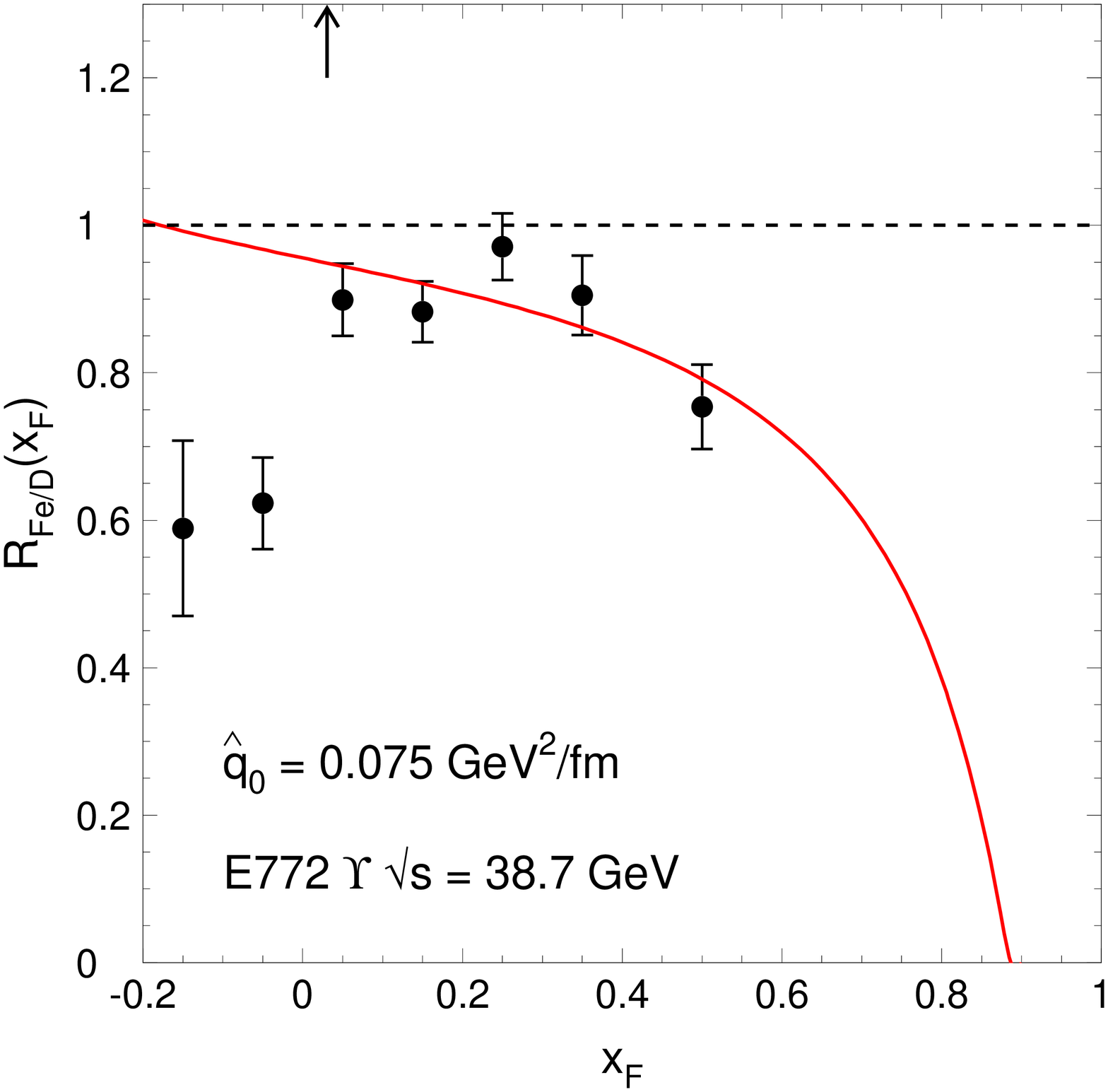}
    \includegraphics[width=4.9cm]{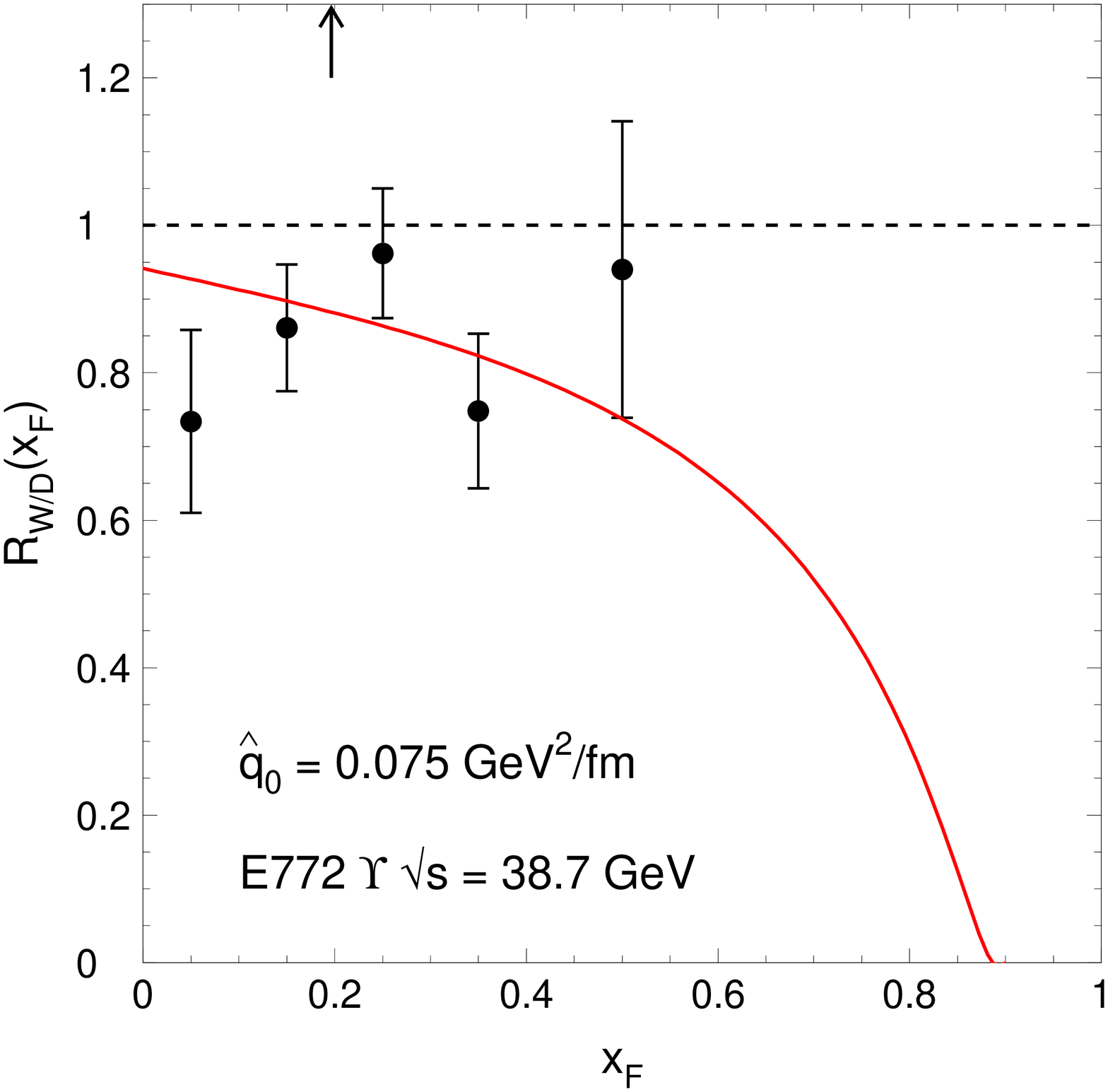}
  \end{center}
\vspace{-0.8cm}
\caption{E772 $\Upsilon$ suppression data~\cite{Alde:1991sw} in p--A collisions compared to the energy loss model.}
  \label{fig:e772ups}
\end{figure}

The E772 data are shown in Fig.~\ref{fig:e772ups} for various nuclear targets (Ca, Fe, W) and in comparison to the model predictions. A rather good agreement between data and theory is found for $\xf > \xfcrit$, although smaller experimental uncertainties would be necessary to further check the $M$ 
dependence of the model. At small $\xf < \xfcrit$,  the measured nuclear production ratio $R_{\rm pA}^{\Upsilon}$ lies much  below our predictions, probably too much to be accommodated by $\Upsilon$ nuclear absorption. However, let us mention that the E772 measurements at low $\xf$ might be affected by uncorrected acceptance effects due to the correlation between $\xf$ and $\pt$ (see the discussion in~\cite{Leitch:1999ea}); it is therefore difficult to draw any conclusion from the significant disagreement observed at negative $\xf$.

These $\Upsilon$ data nevertheless allow the mass dependence of the energy loss to be constrained. In their paper~\cite{Gavin:1991qk}, Gavin and Milana assumed that the mean energy loss scales as $\Delta E \propto M^{-n}$, and considered explicitely the cases $n=2$ (``power suppressed'') and $n=0$. From the comparison of their calculations with E772 data, these authors concluded that neither of these two choices were satisfactory: assuming $\Delta E \propto M^{-2}$ led to too little $\Upsilon$ attenuation while 
a too strong suppression was predicted with the hypothesis $\Delta E \propto M^{0}$. It is therefore interesting to note that the scaling $\Delta E \propto M^{-1}$ predicted in~\cite{Arleo:2010rb} and used here 
(see \eq{mean-delta-E}) 
supports this empirical observation, as the agreement in Fig.~\ref{fig:e772ups} indicates.

\begin{figure}[t]
  \begin{center}
    \includegraphics[width=7.8cm]{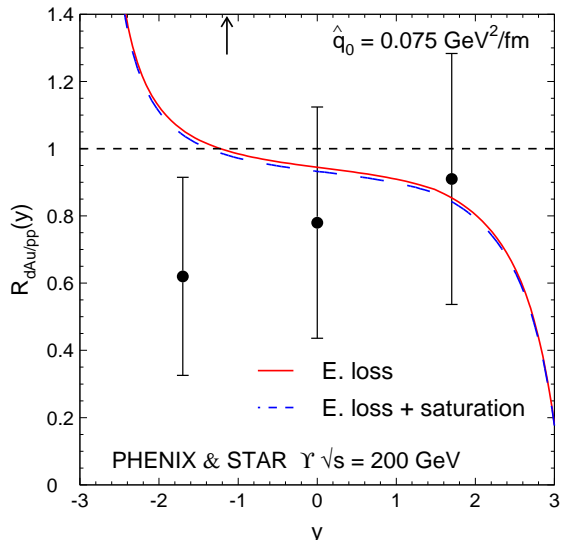}
  \end{center}
  \vspace{-1.cm}
\caption{PHENIX ($|y|=1.7$) and STAR ($y=0$) $\Upsilon$ suppression data~\cite{Reed:2011zza,Adare:2012bv} in d--Au collisions compared to the energy loss model.}
  \label{fig:rhicups}
\end{figure}

The predictions at RHIC are shown in Fig.~\ref{fig:rhicups}. As expected, the suppression is less pronounced than for $\jpsi$ production, compare to Fig.~\ref{fig:rhic}. Since saturation effects are very small in the $\Upsilon$ channel, the predictions including saturation or not are virtually indistinguishable. 
The PHENIX and STAR experiments reported on preliminary measurements of $\Upsilon$ suppression\footnote{The PHENIX and STAR data correspond to the suppression of $\Upsilon$(1S)$+\Upsilon$(2S)$+\Upsilon$(3S) states.} in d--Au collisions~\cite{Reed:2011zza,Adare:2012bv}, see Fig.~\ref{fig:rhicups}.
Hopefully more precise data will soon allow for clarifying the strength of $\Upsilon$ suppression in cold nuclear matter.

\subsection{LHC predictions}
\label{sec:lhc}

\begin{figure}[htbp]
  \begin{center}
    \includegraphics[width=7.8cm]{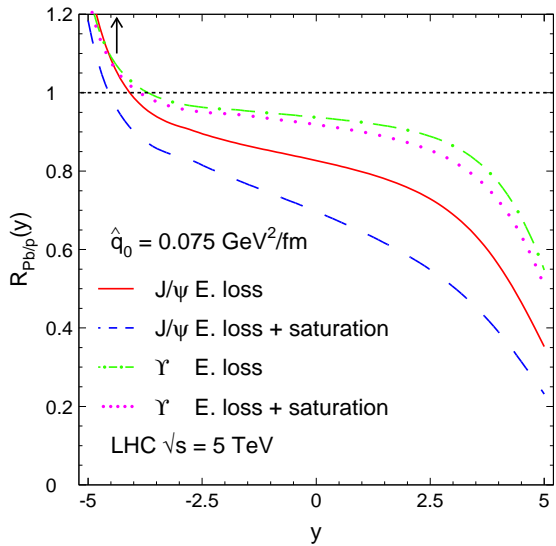}
  \end{center}
  \vspace{-1.cm}
\caption{$\jpsi$ and $\Upsilon$ suppression expected in p--Pb collisions at the LHC, with and without saturation effects (see legend).}
  \label{fig:lhc}
\end{figure}

We discuss in this section the  $\jpsi$ and $\Upsilon$ suppression expected in p--Pb collisions at the LHC ($\sqrt{s}=5$~TeV). 
In Fig.~\ref{fig:lhc} we show the $R_{\rm pPb}$ ratios for both states as a function of the rapidity in the center-of-mass frame.\footnote{Note that in p--Pb collisions at the LHC, the laboratory frame is shifted by $\Delta y \simeq 0.47$ with respect to the center-of-mass frame.} Interestingly, $\jpsi$ production is significantly suppressed at large positive rapidity, \eg\ $R_{\rm pPb}^{\jpsi}\simeq0.7$--$0.8$ at $y=1$ and down to $R_{\rm{pPb}}\lsim 0.5$ at $y\gtrsim4$.  Because of the high center-of-mass energy of the collision at the LHC, saturation effects in the $\jpsi$ channel are significant: in addition to energy loss, the suppression due to saturation ranges from ${\cal S}_{\mathrm A}^{\jpsi}\simeq 0.9$ in the most negative rapidity bins down to ${\cal S}_{\mathrm A}^{\jpsi}\simeq 0.65$ at $y=5$. In the target fragmentation region ($y<0$), the suppression is moderate ($\sim10$--$20\%$) while a possible $\jpsi$ enhancement might be seen at very backward rapidities, $y\lesssim -5$. Predictions using EPS09~\cite{Eskola:2009uj} and DSSZ~\cite{deFlorian:2011fp} nPDF sets are also discussed in Section~\ref{sec:npdfs}.

In the $\Upsilon$ channel, the suppression 
is more moderate because of the mass dependence of energy loss,\footnote{Another effect, yet rather marginal, comes from the flatter $\xf$ distributions in p--p collisions (see Table~\ref{tab:expnups} in Section~\ref{sec:xspp}).} $\Delta E \propto {\mT}^{-1}$, \eg\  $R^{\Upsilon}_{\rm{pPb}}\simeq0.85$ at $y=3$.  At the LHC the saturation effects in the $\Upsilon$ channel prove quite small, although more pronounced than at RHIC.
As can be seen from the arrow in Fig.~\ref{fig:lhc}, $\jpsi$ and $\Upsilon$ hadronization should take place outside the nuclear medium for $y>\ycrit\simeq-5$; nuclear absorption should thus play little or no role at the LHC. 

These predictions can be compared to the future measurements during the p--Pb run scheduled at the LHC in January 2013. 
In order to test the model, the nuclear dependence of $\psi$ production should ideally be measured for various rapidity bins and on a rather broad range, which hopefully should be possible with the ALICE or LHCb experiments.\footnote{We recall that the calculations are made using $\pt=1$~GeV in the calculation of the transverse mass. Therefore 
our predictions on $\jpsi$ suppression should be adapted for
the CMS acceptance which requires a transverse momentum cut, $\pt\gtrsim6$~GeV, in the $\jpsi$ channel.}

\subsection{E906 predictions}
\label{sec:e906}

The E906 ``SeaQuest'' collaboration~\cite{e906} aims at measuring Drell-Yan production in p--p and p--A collisions at $E_{\rm p}=120$~GeV ($\sqrt{s}=15$~GeV) at Fermilab. Although the first goal of this experiment is to study the sea quark asymmetry in the nucleon, it will also be able to measure the nuclear dependence of both Drell-Yan and $\jpsi$ production on various nuclear targets and on a wide range in $\xf$. In this section we present our model predictions on $\jpsi$ suppression in p--A collisions at E906 energy, to be compared to the measurements that might already be available in 2013.

\begin{figure}[ht]
  \begin{center}
    \includegraphics[width=7.cm]{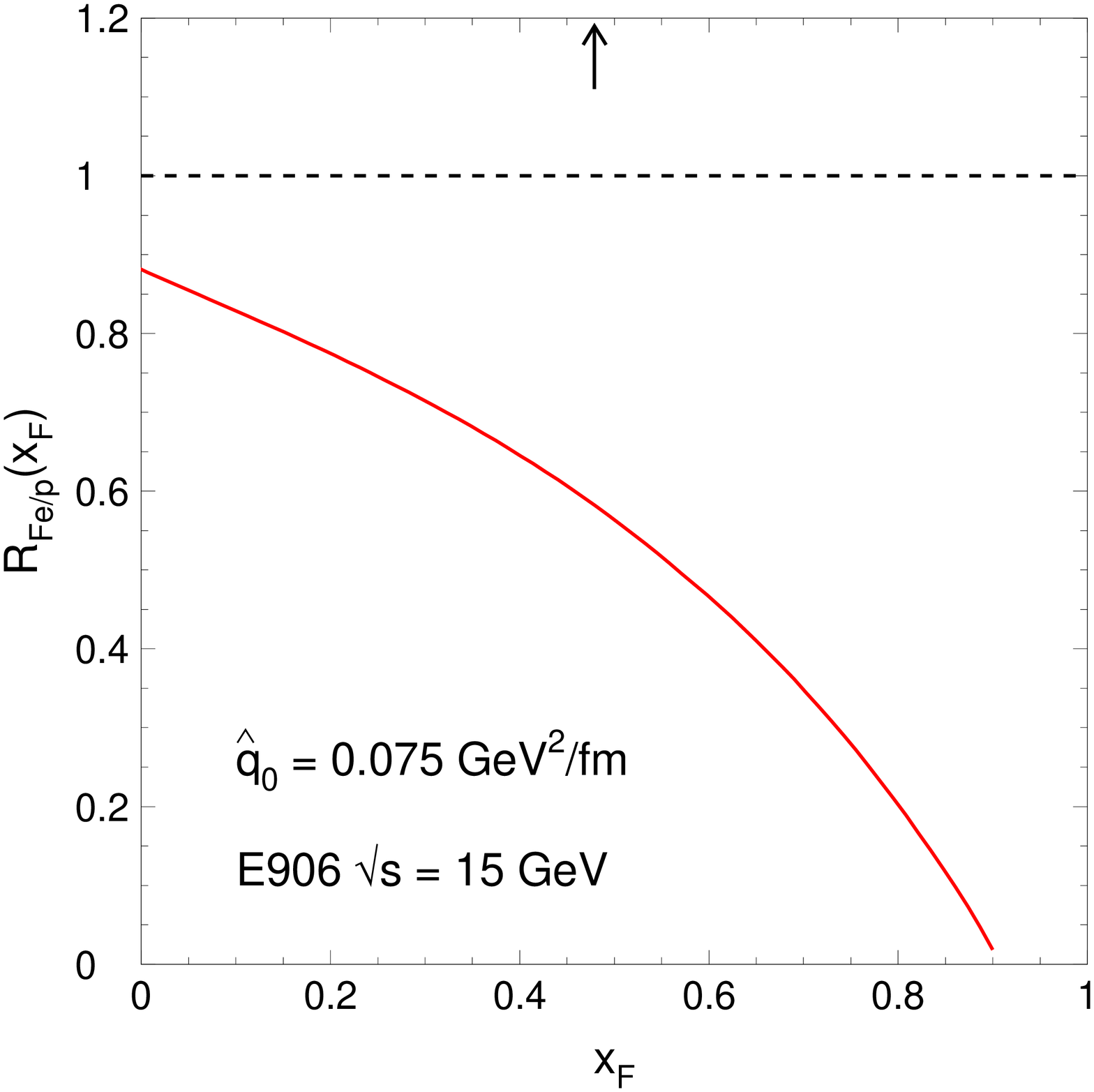}
    \includegraphics[width=7.cm]{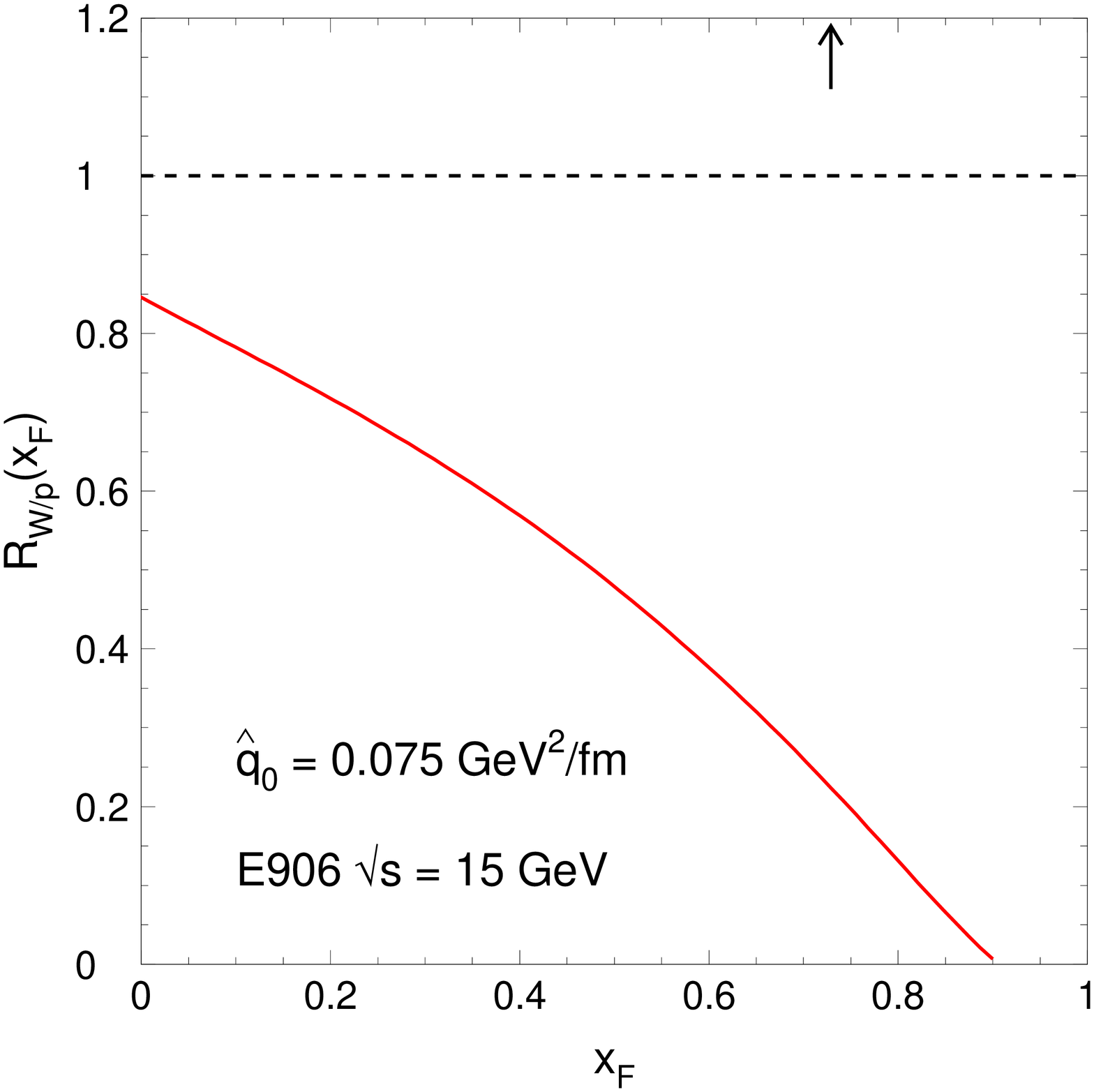}
  \end{center}
\vspace{-0.8cm}
\caption{$\jpsi$ suppression in p--Fe and p--W collisions in the E906 kinematics.}
  \label{fig:e906}
\end{figure}

In Fig.~\ref{fig:e906} we plot the predictions in p--Fe (left) and p--W (right) collisions.\footnote{Lacking p--p data at this energy, we choose the exponent $n=4$ to be consistent with the systematics discussed in Section~\ref{sec:xspp}.} The suppression is very pronounced especially at large $\xf$, for which however the $\jpsi$ production cross section should be extremely small. 

\subsection{Comparing predictions using saturation vs. nPDF}
\label{sec:npdfs}

For completeness, we compare in this section the former results on $\jpsi$ suppression at RHIC and LHC obtained in the ``energy loss + saturation'' model with the predictions using the EPS09~\cite{Eskola:2009uj} and DSSZ~\cite{deFlorian:2011fp} nPDF leading-order sets instead of saturation. Unlike saturation effects, nPDF corrections should be valid (and possibly non-negligible) even at not too small values of $\xtwo$, and in particular at E866 energy. Therefore, the transport coefficient $\qzero$ using each of the two nPDF sets has been consistently refitted to E866 data. The corresponding values, used for the RHIC and LHC predictions with these two sets, are indicated in Table~\ref{tab:rhic}.

\begin{figure}[h]
\begin{center}
    \includegraphics[width=7.cm]{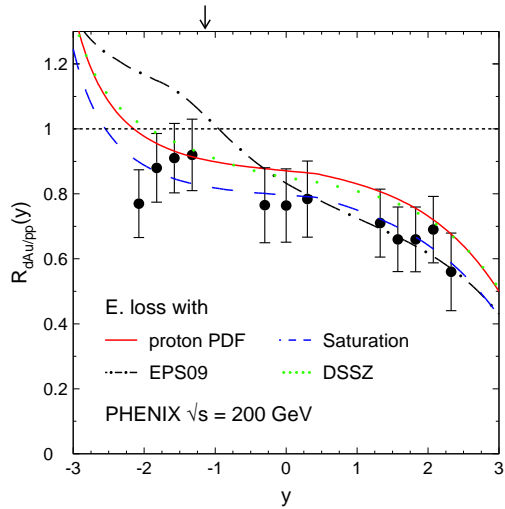}
  \end{center}
\caption{$\jpsi$ suppression predicted in d--Au collisions at RHIC  in the energy loss model,
for various assumptions regarding the nuclear modifications of gluon distributions in nuclei. PHENIX data are from~\cite{Leitch:1999ea}.}
  \label{fig:npdfrhic}
\end{figure}

The comparison is shown in Fig.~\ref{fig:npdfrhic} at RHIC. The predictions using saturation, EPS09 and DSSZ somehow differ in the rapidity dependence of $R_{\rm dAu}^{\jpsi}$. The DSSZ modifications are rather small, leading to a suppression similar to the one assuming energy loss effects only. On the contrary, the EPS09 set exhibits larger modifications to the gluon nPDF (and in particular a slightly faster variation in this $x$ domain) increasing the slope of $R_{\rm dAu}^{\jpsi}$ versus $y$. At mid- and forward rapidity, the various predictions are similar; in particular all of them reproduce nicely the data at $y=1$--$2$ with a slightly better description with saturation or using the EPS09 set as compared to DSSZ (yet this is not statistically significant). On the other hand, the predictions are different at backward rapidity. The best agreement is obtained assuming saturation effects (in addition to parton energy loss) or DSSZ nPDF instead of EPS09. This observation obviously depends on the present energy loss model, preventing us from drawing a firmer conclusion. 

\begin{figure}[h]
\begin{minipage}{7.cm}
\begin{center}
    \includegraphics[width=7.cm]{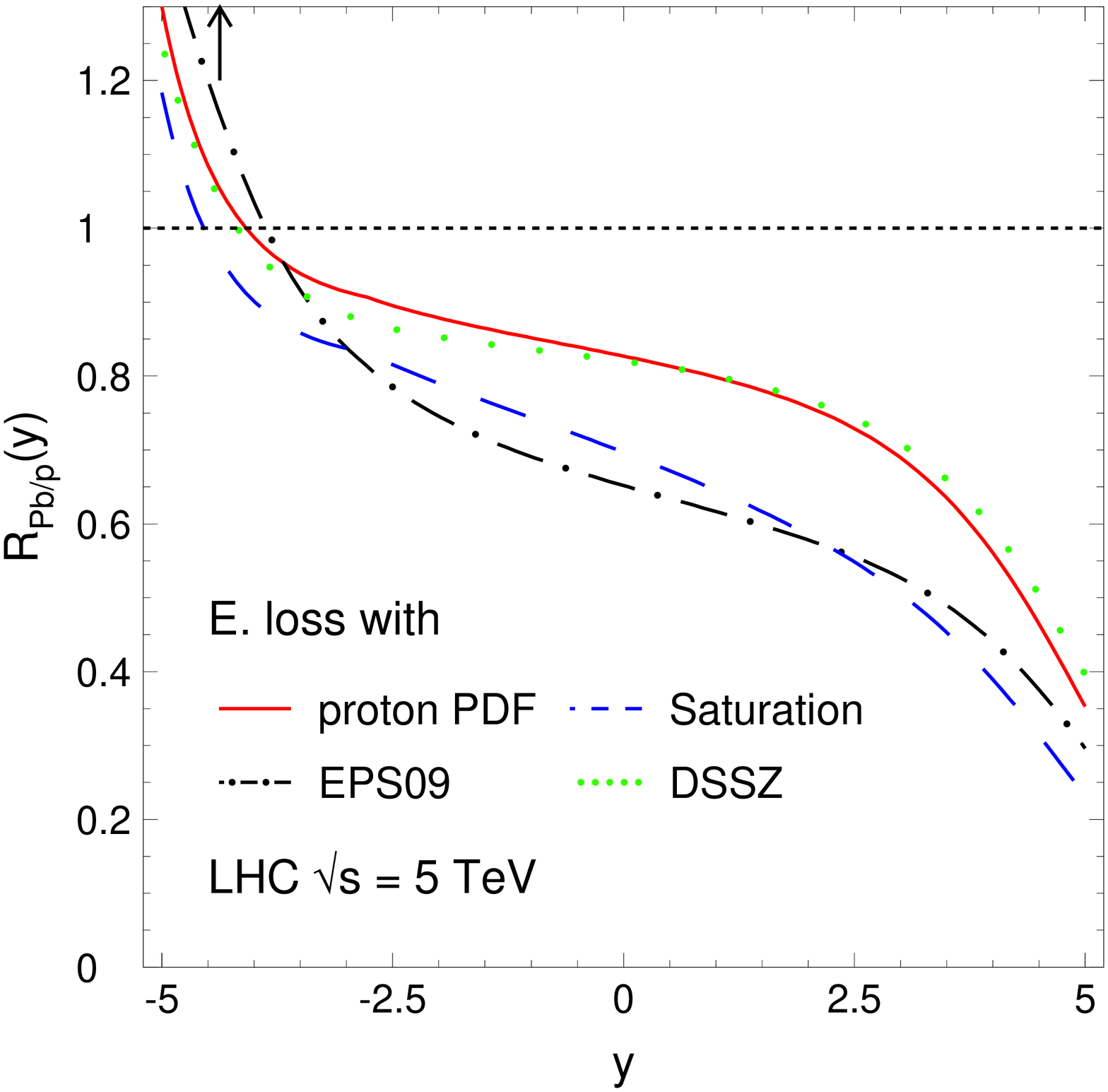}
  \end{center}
\end{minipage}
\hfill
\begin{minipage}{7.cm}
\begin{center}
    \includegraphics[width=7.cm]{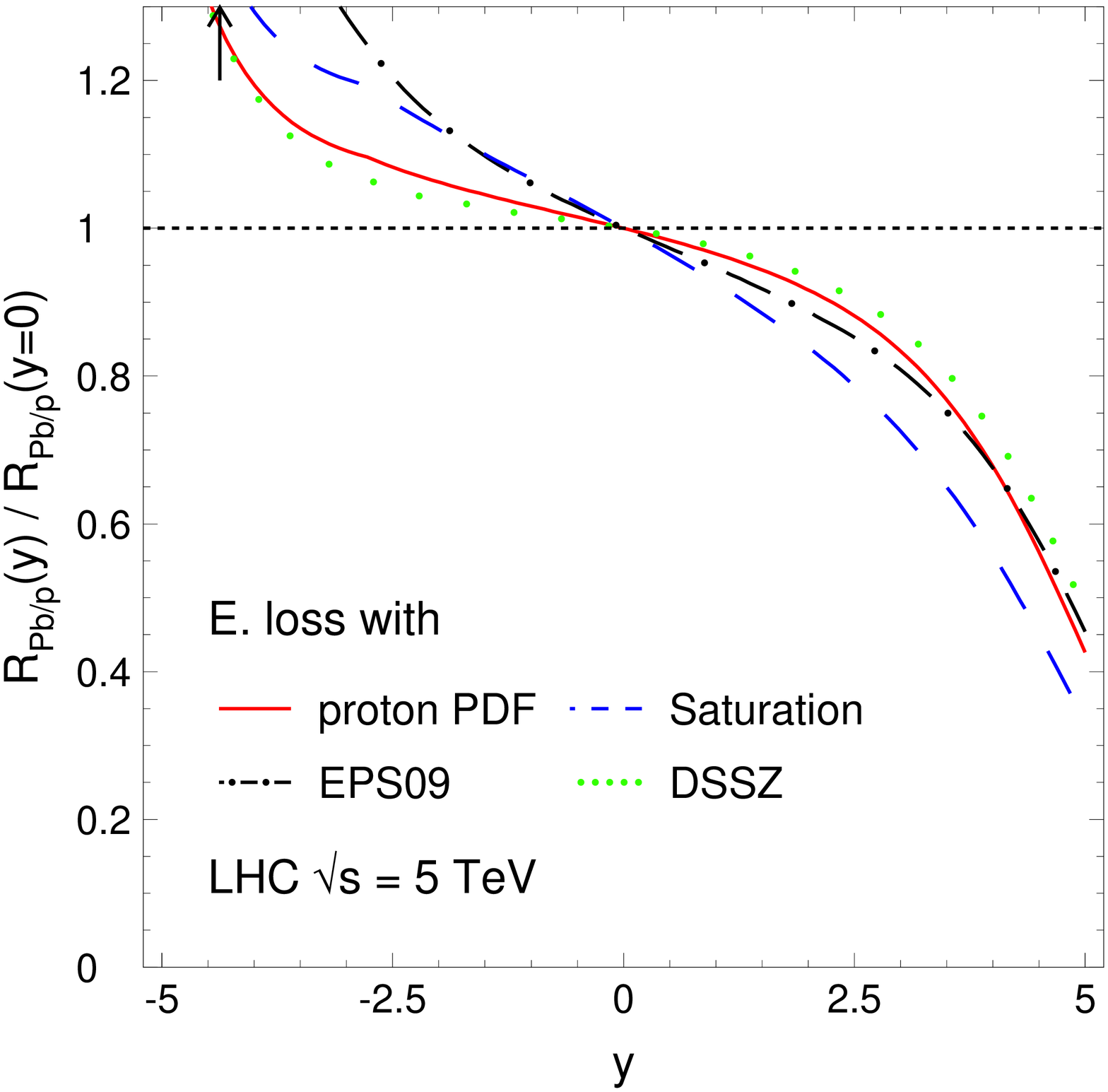}
  \end{center}
\end{minipage}
\caption{Left: $\jpsi$ suppression predicted in p--Pb collisions at the LHC in the energy loss model,
for various assumptions regarding the nuclear modifications of gluon distributions in nuclei. Right: same when normalized to its expected suppression at mid-rapidity, $R_{\mathrm{pPb}}(y)/R_{\mathrm{pPb}}(y=0)$.}
  \label{fig:npdflhc}
\end{figure}

\begin{table}[h]
 \centering
 \begin{tabular}[c]{p{3.6cm}ccc}
   \hline
   \hline
     & $\qzero$ (GeV$^2$/fm) & $\left({\chi^2/\ndf}\right)_{_{\rm E866}}$ & $\left({\chi^2/\ndf}\right)_{_{\rm PHENIX}}$
\\ 
\hline 
Energy loss   & 0.075  & 2.3  & 1.1 \\[0.2cm]
    \hline\\[-0.4cm]
E. loss + saturation & (0.075) &  --- &  0.3       \\
E. loss + EPS09 & 0.046 &  1.7 &  2.7       \\
E. loss + DSSZ & 0.064 &  2.2 &  1.1       \\[0.2cm]
    \hline\\[-0.4cm]
Saturation 	&  (0.075)  &  --- & 2.7 \\
EPS09    	&  --- & 285  &    4.7     \\
DSSZ		&  --- & 384  &    5.7      \\[0.1cm]
\hline
\end{tabular}
 \caption{$\qzero$ and ${\chi^2/\ndf}$ of E866 and PHENIX data, with (upper rows) and without (lower) energy loss effects, for various assumptions regarding the nuclear modifications of gluon distributions in nuclei.}
 \label{tab:rhic}
\end{table}

In order to analyze a bit more quantitatively these results, the values of ${\chi^2/\ndf}$ obtained for the E866 and PHENIX data sets are quoted in Table~\ref{tab:rhic}. As can be seen, the E866 data do not allow one to differentiate the various energy loss predictions including (or not) EPS09/DSSZ nPDF corrections. On the contrary, the agreement at RHIC is considerably better when energy loss is supplemented by saturation effects (${\chi^2/\ndf}=0.3$) rather than by nPDF ($\chi_{_{\rm EPS09}}^2/\ndf=2.7$, $\chi_{_{\rm DSSZ}}^2/\ndf=1.1$), as mentioned above. For completeness we also quote the values of  ${\chi^2/\ndf}$ assuming {\it no} energy loss but only saturation\footnote{The value of $\qzero=0.075$~\gevsqfm\ quoted in Table~\ref{tab:rhic} is here to determine the saturation scale.} or nPDF corrections. As can be seen from Table~\ref{tab:rhic}, saturation without energy loss still gives a fair description of PHENIX data, ${\chi^2/\ndf}=2.7$ (as well as EPS09 and DSSZ to a lesser extent), although saturation and nPDF effects alone would totally fail to reproduce E866 data.

The predictions at the LHC are shown in Fig.~\ref{fig:npdflhc} (left). The expected suppression with saturation effects or using the EPS09 set prove remarkably similar. On the contrary, the nPDF corrections given by DSSZ are tiny in the forward rapidity bins, despite the small values of $x$ probed in the Pb nucleus. Although the absolute magnitude of the $\jpsi$ suppression differs depending on the assumption regarding saturation/nPDF effects, the rapidity dependence (especially at $y>0$) is mostly governed by energy loss effects. This could be tested experimentally. In the present model, energy loss effects are moderate 
at mid-rapidity which corresponds to the maximum of the p--p production cross section. As a consequence, the expected suppression at $y=0$, $R_{\mathrm{pPb}}(y=0)$, is 
more sensitive to 
saturation/nPDF effects. Moreover, since the rapidity dependence is essentially due to energy loss effects, the double ratio $R_{\mathrm{pPb}}(y)/R_{\mathrm{pPb}}(y=0)$ is rather independent of the strength of saturation/nPDF effects. This is illustrated in Fig.~\ref{fig:npdflhc} (right) where $R_{\mathrm{pPb}}(y)/R_{\mathrm{pPb}}(y=0)$ is plotted. As can be seen this double ratio proves remarkably similar whether or not energy loss is supplemented with nPDF 
or saturation effects.

\section{Discussion}
\label{sec:discussion}

The agreement between our model and the \pA\ data for quarkonium nuclear suppression is quite remarkable. With a single free parameter $\hat{q}_0$, both the slope and normalization of $R_{\mathrm{pA}}$ (or $R_{\mathrm{pA}}/R_{\mathrm{pB}}$ and also $R_{\mathrm{\pi{A}}}$) are accurately described, for various collision energies, various target nuclei and different masses ($\jpsi$, $\Upsilon$), and over a broad range in $\xf$ (or rapidity). We also stressed that the effect of saturation alone fails in describing $\jpsi$ nuclear suppression at different collision energies. This strongly supports parton energy loss as a dominant effect in \pA\ quarkonium nuclear suppression, the main conclusion of our study. The successful description of the data is mostly due to the (medium-induced) energy loss scaling as $\Delta E \propto E$, where $E$ is the energy of the $Q \bar{Q}$ pair in the nucleus rest frame. This behaviour arises when the partonic subprocess looks like small angle scattering of an asymptotic charge, and thus holds within our 
assumption of a {\it long-lived}, {\it color-octet} $Q \bar{Q}$ pair.  Our 
results support the parametric ($E$, $M$ and $L$) dependence of the induced radiation spectrum \eq{our-spectrum}, which is derived from first principles 
in Section \ref{sec2}.

These results also give some hint on the mechanism for low $p_\perp$ 
($p_\perp \lsim M$)
heavy-quarkonium hadroproduction. We argued in the Introduction that at large $\xf$, the octet $Q \bar{Q}$ pair should be long-lived in any quarkonium production model, including the Color Singlet Model (CSM). The agreement of the energy loss model with the large $\xf$ suppression data thus cannot distinguish between production models. But we found that the agreement extends to small values of $\xf$ (see in particular the comparison to RHIC data at $y \sim 1$--$2$, corresponding to $\xf \sim 0.04$--$0.1$, in Fig.~\ref{fig:rhic}), where assuming a long-lived color-octet $Q \bar{Q}$ pair becomes inaccurate in the CSM. The CSM mechanism thus seems somewhat disfavoured by our results, at least as a dominant contribution to inclusive (\ie, low $p_\perp$ and low $\xf$) $\jpsi$ production. The future measurements in p--Pb collisions at the LHC will probe small values of $\xf$ ($|\xf|<0.1$), yet in a rather large rapidity interval ($|y|<5$), and might thus further clarify the underlying dynamics of heavy-quarkonium production. In fact our energy loss explanation of $\jpsi$ suppression is consistent with any $\jpsi$ production model where $t_{\rm hard} \ll t_{\rm octet}$, leaving room for gluon radiation with 
$t_{\rm hard} \ll t_{\mathrm{f}} \ll t_{\rm octet}$, see \eq{hierarchy}. It was argued in Ref.~\cite{Hoyer:1998ha} that a qualitative analysis of the quarkonium {\it production} data suggests a mechanism for quarkonium hadroproduction, named ``Comover Enhancement Scenario'',
where color neutralization is realized at the time $t_{\rm octet}$ by a semi-hard scattering between the $Q \bar{Q}$ pair and the comoving {\it radiation field} of the incoming parton. It is intriguing that the condition on $t_{\rm octet}$ inferred from global {\it production} features, $t_{\rm hard} \ll t_{\rm octet} \ll t_{\psi}$~\cite{Hoyer:1998ha}, is consistent with the condition \eq{hierarchy} necessary to explain nuclear {\it suppression} from radiative parton energy loss. 

The induced radiation spectrum was derived assuming a given hierarchy between the nuclear size $L$, gluon formation time $t_{\mathrm{f}}$ and quarkonium hadronization time $t_{\psi}$. Thus, as we emphasized several times, the model should in principle be valid only when the quarkonium state hadronizes outside the nucleus, \ie, when $E$ is large enough or $\xf > \xfcrit$. It is quite striking that the extrapolation of the model to the region $\xf < \xfcrit$ is either consistent with the data (within error bars, see e.g. the NA3 data in Fig~\ref{fig:na3}, HERA-B data in Fig.~\ref{fig:herab} and PHENIX data in Fig.~\ref{fig:rhic}), or systematically {\it underestimates} quarkonium nuclear suppression (NA60 data in Fig.~\ref{fig:na60}). This suggests parton energy loss to play a role in a broader domain than expected, the possible  
additional suppression required at $\xf < \xfcrit$ being due to nuclear {\it absorption} of the fully formed quarkonium state. 

In our study we also assumed quarkonium production in {\it proton}--nucleus collisions
to arise from the splitting of an incoming gluon into an octet $Q \bar{Q}$ pair.
This assumption becomes inaccurate at very large $\xf$, where 
quark-induced processes (such as $q \bar{q} \to Q \bar{Q}$) 
come into play. Although we expect the parametric dependence of the associated radiation spectrum to be unchanged, the overall color factor might 
be changed in this case. 
A possibly {\it smaller} effective color factor at very large $\xf$ might explain the milder $\jpsi$ suppression observed by E866 at $\xf \gsim 0.8$ (see Fig.~\ref{fig:wbe_e866}) than predicted in our model.
However, as already mentioned in Section \ref{se:E866}, the very good agreement between the model and the NA3 {\it pion}--nucleus data, for which the $q\bar{q}$ annihilation channel is dominant at all $\xf$, suggests a relatively weak dependence of the energy loss on the incoming parton type. 

We might envisage refinements of the parton energy loss model presented here, such as including quarkonium absorption at $\xf < \xfcrit$ and quark-induced processes, in order to extend the domain of validity of our approach. However, we find it more important to first confirm the dominant role of parton energy loss in \pA\ collisions, where gluon-induced processes dominate and our model assumptions apply.

First, the energy loss model can be tested in forthcoming p--Pb collisions at the LHC, for which our predictions for the $y$-dependence of $\jpsi$ and $\Upsilon$ suppression are shown in Fig.~\ref{fig:lhc}, and in p--A collisions in the E906 fixed-target experiment at Fermilab (Fig.~\ref{fig:e906}). The model should as well be confronted to the existing RHIC d--Au data on the $p_\perp$ and centrality dependence of $\jpsi$ suppression, measured at various rapidities~\cite{Adare:2010fn,Adare:2012qf}. This requires generalizing \eq{eq:xspA} to double differential (in $\xf$ and $p_\perp$) cross sections, and will be the subject of a future work. It will be interesting to check whether the $L$-dependence of the energy loss predicted in \eq{mean-delta-E} is consistent with the centrality dependence of the RHIC d--Au data. Our study might also help interpreting quantitatively quarkonium measurements performed in heavy-ion collisions at RHIC~\cite{Adare:2006ns,Abelev:2009qaa} and LHC~\cite{Aad:2010aa,Chatrchyan:2012np,Abelev:2012rv}. Indeed, parton energy loss through the incoming cold nuclei is expected to combine with hot effects (such as Debye screening or final state energy loss in a QGP). The evaluation of $\jpsi$ suppression in \hi\ collisions 
expected from cold parton energy loss alone 
will be presented in a future study. Finally, as discussed in the Introduction other processes than quarkonium production should be sensitive to a parametrically similar parton energy loss, such as open charm and light hadron production in \pA\ collisions. This work is also in progress.

\acknowledgments
We thank Rodion Kolevatov for contributing to the evaluation of the effective path length within Glauber theory, and Elena Ferreiro and Jian-Wei Qiu for useful discussions. FA thanks CERN PH-TH division for hospitality. This work is funded by ``Agence Nationale de la Recherche'' under grant ANR-PARTONPROP.

\appendix

\section{$x$ dependence of $\hat{q}$}
\label{appendix-qhat}

In Ref.~\cite{Baier:1996sk}, the transport coefficient $\hat{q}$ was related to the gluon distribution $G(x)$ in a target nucleon, see the expression \eq{qhat-gluondensity}. The value of $x$ to be used in $x G(x)$ in \eq{qhat-gluondensity} can be estimated by considering the kinematics of the rescattering process. 

Following Ref.~\cite{Baier:1996sk}, let us consider the specific case of DIS, where an energetic light quark of momentum $p$ is produced and then rescatters with transfer $q$ on a target nucleon of momentum $P$. Working in the target nucleus rest frame, we have $P = (m_\mathrm{N}, \vec{0})$, with $m_\mathrm{N}$ the nucleon mass. Choosing light-cone coordinates $p^\pm = p^0 \pm p^z$ and $p = (p^+,p^-, \vec{0}_\perp)$ along the negative $z$-direction, the condition for the final quark to be on-shell reads
\be
\label{on-shell-cond}
(p+q)^2 = (p^{+} + q^{+})(p^{-} + q^{-}) - q_\perp^2 = 0 \Rightarrow  p^{+} + q^{+} \simeq \frac{q_\perp^2}{p^-} \, ,
\ee
where we neglected $q^{-}$ as compared to $p^{-} = 2 E$. 

\subsubsection*{parton produced inside the target}

When the hard production time $\thard \sim E/Q^2 \ll L$, or equivalently when the Bjorken variable $x_B \equiv Q^2/(2m_\mathrm{N} E) \gg x_0 \equiv 1/(2 m_\mathrm{N} L)$, the parton $p$ is effectively produced incoherently inside the nucleus. This is the situation considered in Ref.~\cite{Baier:1996sk}, which we now briefly review. If the rescattering occurs at a distance $z$ from the production point, from the uncertainty principle we have $|p^{+}| \sim 1/z$ just before the scattering. From the constraint $z \leq L$, we obtain $|p^{+}| \gsim 1/L$. For $p^-$ large enough Eq.~\eq{on-shell-cond} gives $q^{+} \simeq |p^{+}|$. The momentum fraction of the rescattering gluon thus satisfies \cite{Baier:1996sk} 
\be
x \equiv \frac{q^+}{P^+} = \frac{q^+}{m_\mathrm{N}} \simeq  \frac{|p^{+}|}{m_\mathrm{N}} \sim \frac{1}{2 m_\mathrm{N} L} = x_0\ \ \ \  (x_B \gg x_0) \, .
\ee

\subsubsection*{parton produced far before the target}

When $\thard \gg L \Leftrightarrow x_B \ll x_0$, the virtual photon splits into a light quark-antiquark pair far before the nucleus, and the DIS process is coherent over the whole nucleus. In this case, the quark virtuality $|p^2| = |p^{+} p^{-}| \sim Q^2$, and $|p^{+}| \sim Q^2/p^{-}$ is not bounded by $1/L$ any longer. From Eq.~\eq{on-shell-cond} we obtain (using $q_\perp^2 \ll Q^2$)
\be
x = \frac{q^+}{m_\mathrm{N}} \simeq  \frac{|p^{+}|}{m_\mathrm{N}} \sim \frac{Q^2}{m_\mathrm{N}\, p^-} = x_B \ \ \ \  (x_B \ll x_0) \, .
\ee

For a generic hard process (for instance in a \pA\ collision) of coherence length $\thard \sim E/M^2 \sim 1/(2 m_\mathrm{N} x_2)$, the above DIS example supports the following estimate for the value of $x$ to be used in Eq.~\eq{qhat-gluondensity},
\be
x = x_0 \, \Theta(x_2 > x_0) + x_2 \, \Theta(x_2 < x_0) = \min(x_0, x_2) \ ; \ \ \ x_0 \equiv \frac{1}{2 m_\mathrm{N} L} \, ,
\label{x-estimate}
\ee
thus specifying the $x_2$-dependence of the transport coefficient $\hat{q}$.

\providecommand{\href}[2]{#2}\begingroup\raggedright\endgroup

\end{document}